\documentclass[preprintnumbers,nofootinbib,showpacs,floatfix]{revtex4}

\usepackage{graphicx}
\usepackage{latexsym}
\usepackage[english]{babel}
\usepackage{amsmath}
\usepackage{amsthm}
\usepackage{epsfig}
\usepackage{color}
\usepackage{amssymb}
\usepackage{epsfig}
\usepackage{verbatim}
\usepackage{psfrag}
\usepackage{bm}
\usepackage{bbm}
\usepackage{amssymb}

\begin{document}

\title{\Large Decoherence in an Interacting Quantum Field Theory: \\ The Vacuum Case}

\preprint{ITP-UU-09/50, SPIN-09/41}

\preprint{HD-THEP-09-23}


\author{Jurjen F. Koksma}
\email[]{J.F.Koksma@uu.nl, T.Prokopec@uu.nl,
M.G.Schmidt@thphys.uni-heidelberg.de} \affiliation{Institute for
Theoretical Physics (ITP) \& Spinoza Institute, Utrecht
University, Postbus 80195, 3508 TD Utrecht, The Netherlands}

\author{Tomislav Prokopec}
\email[]{J.F.Koksma@uu.nl, T.Prokopec@uu.nl,
M.G.Schmidt@thphys.uni-heidelberg.de} \affiliation{Institute for
Theoretical Physics (ITP) \& Spinoza Institute, Utrecht
University, Postbus 80195, 3508 TD Utrecht, The Netherlands}

\author{Michael G. Schmidt}
\email[]{J.F.Koksma@uu.nl, T.Prokopec@uu.nl,
M.G.Schmidt@thphys.uni-heidelberg.de}\affiliation{ Institut f\"ur
Theoretische Physik, Heidelberg University, Philosophenweg 16,
D-69120 Heidelberg, Germany}

\begin{abstract}
We apply the decoherence formalism to an interacting scalar field
theory. In the spirit of the decoherence literature, we consider a
``system field'' and an ``environment field'' that interact via a
cubic coupling. We solve for the propagator of the system field,
where we include the self-energy corrections due to the
interaction with the environment field. In this paper, we consider
an environment in the vacuum state ($T=0$). We show that
neglecting inaccessible non-Gaussian correlators increases the
entropy of the system as perceived by the observer. Moreover, we
consider the effect of a changing mass of the system field in the
adiabatic regime, and we find that at late times no additional
entropy has been generated.
\end{abstract}

\maketitle

\section{Introduction}
\label{Introduction}

\subsection{Outline}
\label{Outline}

We aim to apply the decoherence formalism to an interacting
quantum field theoretical model. The main idea in the framework of
decoherence (\cite{Zeh:1970,Zurek:1981xq, Joos:1984uk,
Zurek:1991vd}, for reviews see \cite{Hartle:1992as,
Giulini:1996nw, Paz:2000le, Zurek:2003zz, Schlosshauer}) is that a
macroscopic system cannot be separated from its environment. The
conventional strategy is to assume the existence of a distinct
system, environment and observer. If the observer and the
environment are weakly coupled, we are allowed to integrate the
environment out to study its average effect on the system
perceived by the observer. Alternatively one could say that the
environmental degrees of freedom are inaccessible to the observer.
This averaging process is an intrinsically non-unitary operation,
which consequently gives rise to an increase in entropy of the
system. A quantum system with a large entropy, in turn,
corresponds to an effectively classical system.

It is however difficult to realistically apply the decoherence
machinery to quantum field theory: this requires in general
involved out-of-equilibrium, finite temperature, interacting
quantum field theoretical computations. It is of course widely
appreciated that entropy can be generated as a result of an
incomplete knowledge of a system. We thus need to keep in mind
what quantities are actually measured in quantum field theory: all
information in a system that can in principle be observed by a
``perfect observer'' is contained in the $n$-point correlators of
the system. However, realistic observers are not capable of
measuring irreducible $n$-point functions of arbitrary order as
they are limited by the sensitivity of their apparatus. Therefore,
it is important to realise that inaccessible higher order
correlators, from the observer's perspective, yield an increase in
entropy of the system. We thus propose the following viewpoint
when applying the decoherence program to quantum field
theory\footnote{Older work can already be interpreted in a similar
spirit \cite{Prokopec:1992ia, Brandenberger:1992jh}. Here, we
propose it as a strict procedure \cite{Talk, Talk2} of how to study
entropy generation in the 2PI formalism, also see \cite{JFKTPMGS}.
Recently, during the last stages of writing this paper, we found
an interesting article by Giraud and Serreau \cite{Giraud:2009tn}
addressing the question of entropy production in an interacting
field theory from a similar perspective.}:

\begin{center}
\begin{minipage}{.55\textwidth}
Neglecting observationally inaccessible correlators will give rise
to an increase in entropy of the system as perceived by the
observer.
\end{minipage}
\end{center}

As an example, consider some interacting quantum field theory
where information is stored in either two-point or Gaussian
correlators or in higher order, non-Gaussian correlators. The
latter are generated generically in any interacting field theory.
If we assume that the information stored in these non-Gaussian
correlators is barely accessible in experiments, then neglecting
this information will give rise to an increase in the entropy.
From the Gaussian correlators, we can fix the entropy uniquely
\cite{Campo:2008ij, JFKTPMGS}. As before, a quantum system with a
considerable amount of entropy corresponds to a classical
system\footnote{Our definition of classicality differs from the
approach used in some of the literature, where for example
coherent states with large occupation numbers are also considered
to have many classical properties, even though these states have
vanishing entropy.}. We emphasise that this definition can be
improved if e.g. three- or four-point correlators are accessible
through experiments such that knowledge of these correlators is
included in the definition of the entropy \cite{JFKTPMGS}. It is
important to stress that this procedure does not require a
non-unitary process of tracing out environmental degrees of
freedom.

In order to apply these rather abstract ideas to a scalar field
toy model, let us outline our paper. We will consider the
following interacting scalar field theory:
\begin{equation}\label{action:tree1}
S[\phi,\chi] = \int \mathrm{d}^{\scriptscriptstyle{D}}\!x {\cal
L}[\phi,\chi] = \int \mathrm{d}^{\scriptscriptstyle{D}}\! x {\cal
L}_{0}[\phi] + {\cal L}_{0}[\chi]+ {\cal
L}_{\mathrm{int}}[\phi,\chi] \,,
\end{equation}
where:
\begin{subequations}
\label{action:tree2}
\begin{eqnarray}
{\cal L}_{0}[\phi] &=& -\frac{1}{2} \partial_\mu\phi(x)
\partial_\nu \phi(x) \eta^{\mu\nu} - \frac{1}{2} m^{2}_{\phi}(t)
\phi^{2}(x)
\label{action:tree2a}\\
{\cal L}_{0}[\chi] &=& -\frac{1}{2} \partial_\mu\chi(x)
\partial_\nu \chi(x)\eta^{\mu\nu} - \frac{1}{2} m^{2}_{\chi} \chi^{2}(x)
\label{action:tree2b}\\
{\cal L}_{\mathrm{int}}[\phi,\chi] &=& -
\frac{\lambda}{3!}\chi^{3}(x) -\frac{1}{2}h \chi^{2}(x) \phi(x)
\,, \label{action:tree2c}
\end{eqnarray}
\end{subequations}
where $\eta_{\mu\nu}=\mathrm{diag}(-1,1,1,\cdots)$ is the
$D$-dimensional Minkowski metric. Here, $\phi(x)$ will play the
role of the system, interacting with an environment $\chi(x)$,
where we assume that $\lambda \gg h$ such that the environment is
in thermal equilibrium at temperature $T$. In this paper, we study
an environment of temperature $T=0$, i.e.: an environment in its
vacuum state and we postpone the finite temperature corrections to
a future publication. We assume that $\langle\hat{\phi}\rangle = 0
= \langle\hat{\chi}\rangle$, which can be realised by suitably
renormalising the tadpoles.

Another application of our calculation is baryogenesis in an early
Universe setting where the system is driven out-of-equilibrium by
a changing mass term $m_{\phi}^{2}(t)$ generated by a time
dependent Higgs-like scalar field during a symmetry breaking. For
electroweak baryogenesis, we can neglect the Universe's expansion
during the phase transition and our assumption to work in
Minkowski spacetime is well justified.

As we study out-of-equilibrium quantum field theory, we work in
the Schwinger-Keldysh formalism. The 2PI (two particle
irreducible) effective action then captures the effect of
perturbative loop corrections to the various propagators $\imath
\Delta_{\phi}$ and $\imath \Delta_{\chi}$. Of course we will
discuss these equations of motion in greater detail in the main
text, but when we omit all indices and arguments, they have the
following structure:
\begin{subequations}
\label{introeom}
\begin{eqnarray}
(\partial^{2}-m^{2}_{\phi})\imath \Delta_{\phi} -\int M_{\phi}
\imath \Delta_{\phi} &=& \imath\delta^{\scriptscriptstyle{D}}
\label{introeoma}\\
(\partial^{2}-m^{2}_{\chi})\imath \Delta_{\chi} -\int
M_{\chi}\imath \Delta_{\chi} &=&
\imath\delta^{\scriptscriptstyle{D}} \label{introeomb}\,,
\end{eqnarray}
\end{subequations}
where $M_{\phi}$ and $M_{\chi}$ are the corresponding self-masses.
These two equations are non-Gaussian due to the coupling of the
two fields with coupling constant $h$. Multiplying equation
(\ref{introeoma}) by
$\Delta_{0,\phi}=(\partial^{2}-m^{2}_{\phi})^{-1}\delta^{\scriptscriptstyle{D}}$
and equation (\ref{introeomb}) by
$\Delta_{0,\chi}=(\partial^{2}-m^{2}_{\chi})^{-1}\delta^{\scriptscriptstyle{D}}$
and integrating, one gets the following Schwinger-Dyson equations:
\begin{subequations}
\label{eomfeynmandiagrams}
\begin{eqnarray}
&&
\begin{tabular}{ccc}
\includegraphics[width=3.0cm]{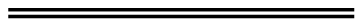}\, & \,=\, & \,\includegraphics[width=3.0cm]{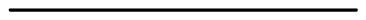} \,+\, \\
\phantom{1} & \phantom{1} & \phantom{1}\\
\phantom{1} & \phantom{1} & \phantom{1}\\
\phantom{1} & \phantom{1} & \phantom{1}\\
\end{tabular}
\includegraphics[width=3.0cm]{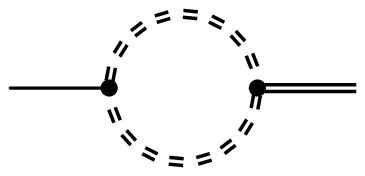}
\label{eomfeynmandiagramsa}\\
&&
\begin{tabular}{cccc}
\includegraphics[width=3.0cm]{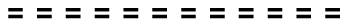}\, &\, =\, & \,\includegraphics[width=3.0cm]{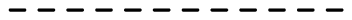} \,+\, & \includegraphics[width=3.0cm]{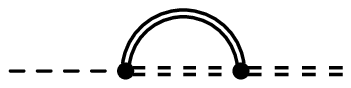} \,+\,\\
\phantom{1} & \phantom{1} & \phantom{1}& \phantom{1}\\
\phantom{1} & \phantom{1} & \phantom{1}& \phantom{1}\\
\phantom{1} & \phantom{1} & \phantom{1}& \phantom{1}\\
\phantom{1} & \phantom{1} & \phantom{1}& \phantom{1}\\
\end{tabular}
\includegraphics[width=3.0cm]{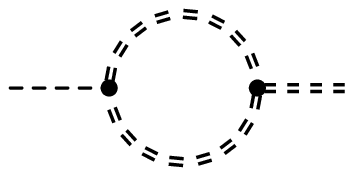}
\label{eomfeynmandiagramsb} \,.
\end{eqnarray}
\end{subequations}
The dressed $\phi$-propagators (double solid lines) can be written
as the sum of the tree level propagators (solid lines) and the
self-mass corrections due to interaction with the dressed
$\chi$-field (double dashed lines), and vice versa for the
environment field.

Let us at this point explicitly state the two main assumptions of
our work. Firstly, we assume that our observer is only sensitive
to Gaussian correlation functions. This implies that we use only
these correlators to calculate the entropy \cite{JFKTPMGS}.
Secondly, we neglect the backreaction from the system field on the
environment field, i.e.: we assume that we can neglect the
self-mass corrections due to the $\phi$-field in equation
(\ref{eomfeynmandiagramsb}). This assumption thus implies that the
environment remains at temperature $T=0$. In particular, the
self-mass corrections to the system propagators in equation
(\ref{eomfeynmandiagramsa}) are now given by:
\begin{equation}\label{eomfeynmandiagrams2}
\includegraphics[width=3.0cm]{Feynmandiagrams3.eps}
\begin{tabular}{ccc}
\phantom{1} & = & \phantom{1}\\
\phantom{1} & \phantom{1} & \phantom{1}\\
\phantom{1} & \phantom{1} & \phantom{1}\\
\phantom{1} & \phantom{1} & \phantom{1}\\
\end{tabular}
\includegraphics[width=3.0cm]{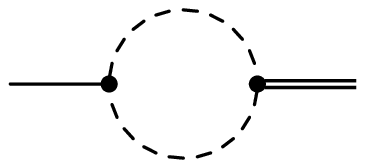}
\begin{tabular}{ccc}
\phantom{1} & + & \phantom{1}\\
\phantom{1} & \phantom{1} & \phantom{1}\\
\phantom{1} & \phantom{1} & \phantom{1}\\
\phantom{1} & \phantom{1} & \phantom{1}\\
\end{tabular}
\includegraphics[width=3.0cm]{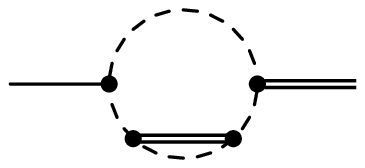}
\begin{tabular}{ccc}
\phantom{1} & + \ldots & \phantom{1}\\
\phantom{1} & \phantom{1} & \,.\\
\phantom{1} & \phantom{1} & \phantom{1} \\
\phantom{1} & \phantom{1} & \phantom{1}\\
\end{tabular}
\end{equation}
The first diagram contains the leading order self-mass correction,
at order $\mathcal{O}(h^{2}/\omega^{2}_{\phi})$, where
$\omega^{2}_{\phi}=m_{\phi}^{2}+k^{2}$. It effectively is a
Gaussian correction to the $\phi$-propagators as it acts just like
a known source to $\phi$. Note that in equation
(\ref{eomfeynmandiagrams2}) we dropped a diagram at order
$\mathcal{O}(h^{2}\lambda^{2}/\omega^{4}_{\phi})$ since it is
irrelevant for the argument presented below.

Looking at equation (\ref{eomfeynmandiagrams2}) we see that these
assumptions are well justified by perturbative arguments provided
there are no secular effects: the backreaction of the system field
on the environment field and the first intrinsically non-Gaussian
correction occurs only at order
$\mathcal{O}(h^{4}/\omega^{4}_{\phi})$, which can be appreciated
by examining the second Feynman diagram on the righthand side of
equation (\ref{eomfeynmandiagrams2}). The 3-point function of the
system, an intrinsically non-Gaussian correlator, is at one-loop
level also perturbatively suppressed at order
$\mathcal{O}(h^{3}/\omega^{3}_{\phi})$.

Finally, let us just mention that the concept of the pointer basis is
frequently discussed in the decoherence literature. The pointer
basis of our theory in the highly squeezed limit is the field
amplitude basis \cite{JFKTPMGS}, occurring for example in
cosmological perturbation theory. Intuitively, this can be
appreciated as follows: the Hamiltonian of a squeezed state is
dominated by the potential term. A system in interaction with an
environment at temperature $T$ minimises its free energy $F=H-TS$.
The system will realise this by increasing its entropy $S$, mainly
due to increasing the spread in momentum $\langle \hat{\pi}^2
\rangle$ since that hardly affects the Hamiltonian, whereas it
does significantly affect the entropy. In other words: the field
amplitude basis is robust under the process of decoherence,
qualifying it as a proper pointer basis. Note that $\phi$ is a
pointer basis only in the statistical sense, such that there is a
well defined probability distribution function from which a
measurement is drawn.

Having discussed the setup of our theory, the assumptions used and
their justification, let us direct our attention to discussing
potential applications. Our results are relevant for several
research areas: the study of decoherence of cosmological
perturbations, of out-of-equilibrium quantum field theory and of
baryogenesis.

\subsection{Decoherence of Cosmological Perturbations}
\label{Decoherence of Cosmological Perturbations}

Although applications of the framework of decoherence are mainly
directed towards experimental efforts (for example related to the
increasingly relevant field of quantum computing), it also
concerns quantum field theory (see e.g.: \cite{Elze:1994qa}).
Research efforts are primarily focused on addressing the
fundamental question of the decoherence of cosmological
perturbations, see e.g.: \cite{Brandenberger:1990bx,
Polarski:1995jg, Lesgourgues:1996jc, Kiefer:1998pb, Kiefer:1998jk,
Kiefer:1998qe, Kiefer:1999es, Campo:2004sz, Campo:2005sy,
Burgess:2006jn, Martineau:2006ki, Prokopec:2006fc, Kiefer:2006je,
Lyth:2006qz, Sharman:2007gi,Kiefer:2007zza, Kiefer:2008ku,
Campo:2008ju, Campo:2008ij, Sudarsky:2009za}. One of the most
important consequences of the inflationary paradigm is that it
provides us with a causal explanation of how initial density
inhomogeneities can be laid out on super Hubble scales that seed
the large scale structure we observe in the Universe today in for
example clusters of galaxies. The decoherence formalism applied to
cosmological perturbations aims at describing the transition
between the quantum nature of the initial density inhomogeneities
as a consequence of inflation and the classical stochastic
behaviour as assumed by large scale structure theory.

In the literature, specific models for example assume that during
inflation the UV (or sub-Hubble) modes of a field, once integrated
out, decohere the IR (or super-Hubble) modes because the former
modes are inaccessible observationally (\cite{Lombardo:1995fg,
Lombardo:2002wt, Lombardo:2005iz}, however also see
\cite{Habib:1992ci}). A similar split of UV and IR modes has been
made in the context of stochastic inflation, see e.g.:
\cite{Starobinsky:1994bd, Buryak:1995tx, Casini:1998wr,
Calzetta:1999zr, Martin:2004ba, Tsamis:2005hd}. In
\cite{Calzetta:1995ys} vacuum fluctuations decohere the mean
field, turning it into a classical stochastic field. In
\cite{Campo:2008ij} it is argued that self-mass corrections to the
equation of motion for the statistical propagator can be rewritten
in terms of a stochastic noise term that in turn decoheres the
system. In \cite{Prokopec:2006fc} it was shown that isocurvature
modes decohere the adiabatic mode.

\subsection{Non-equilibrium Quantum Field Theory}
\label{Non-equilibrium Quantum Field Theory}

In recent years, the study of non-equilibrium quantum field theory
has become more and more tractable (for review articles see:
\cite{Berges:2004yj,Berges:2006rw}). A central ingredient in
performing these studies is the two particle irreducible action,
from which quantum corrections to propagators can be investigated.
Out-of-equilibrium $\lambda\phi^{4}(x)$-theory has extensively
been studied in for example \cite{Aarts:2001yn, Aarts:2001qa,
Aarts:2002dj, Berges:2002cz, Juchem:2003bi,
Juchem:2004cs,Arrizabalaga:2004iw,
Arrizabalaga:2005tf,Berges:2001fi}. The dynamics of
non-equilibrium fermions has been addressed in e.g.:
\cite{Berges:2002wr}.

An interesting study has been performed in \cite{Anisimov:2008dz},
where one also studies, under certain assumptions, the dynamics of
a system field that interacts via a cubic coupling with a thermal
bath, which we also consider. Their thermal bath consists of two
scalar fields with different masses. Very recently, another
interesting calculation for $\lambda\phi^{4}(x)$ self-interaction
has been performed in \cite{Giraud:2009tn} where one calculates a
decoherence parameter and thermalisation of an initial pure state.

Calzetta and Hu consider in \cite{Calzetta:2002ub,
Calzetta:2003dk} also an out-of-equilibrium $\lambda\phi^{4}(x)$
theory. What we would refer to as ``Gaussian von Neumann entropy''
is referred to as ``correlation entropy'' in
\cite{Calzetta:2003dk}. They prove an $H$-theorem for a quantum
mechanical $O(N)$ model.

Renormalising the Kadanoff-Baym equations is a subtle business. In
$\lambda\phi^{4}(x)$-theory it has been examined in different
contexts in \cite{vanHees:2001ik, VanHees:2001pf, vanHees:2002bv,
Berges:2005hc,Borsanyi:2008ar, Borsanyi:2009zza,Blaizot:2003an,
Blaizot:2003br}. We will also come to address the question of
renormalising our cubically interacting field theory. Our main
finding is that the structure of the renormalised equations of
motion differs from the unrenormalised equations, which has in
general to our knowledge not previously been considered in the
literature.

Furthermore, imposing initial conditions at some finite time
$t_{0}$ results in additional infinities that have to be
renormalised separately according to the authors of
\cite{Collins:2003mj, Collins:2005nu, Collins:2006bg,
Collins:2006uy}. Another interesting study has been performed by
\cite{Garny:2009ni} in which the renormalised Kadanoff-Baym
equations, again in $\lambda\phi^{4}(x)$, are numerically
integrated by imposing non-Gaussian initial conditions. We differ
in our approach as we consider the memory effects from the
interacting theory at times before $t_{0}$. We can then impose
appropriate Gaussian initial conditions at $t_{0}$ without
encountering initial time divergences.

\subsection{Baryogenesis}
\label{Baryogenesis}

This work is in part inspired by fundamental questions concerning
the problem of entropy in field theory, and in part by electroweak
scale baryogenesis. The problem is to calculate axial vector
currents generated by a CP violating advancing phase interface of
a true vacuum bubble at the electroweak phase transition. These
currents then feed in hot sphalerons, thus biasing baryon
production. The axial currents are difficult to calculate
reliably, since a controlled calculation would include
non-equilibrium dynamics in a finite temperature plasma in the
presence of a non-adiabatically changing mass parameter. In this
paper we neither include a plasma at finite temperature (this will
be done in a future companion paper), nor do we consider
scattering of fermions on a non-adiabatically changing phase
interface. Yet there are important similarities between the
problem we address here and baryogenesis: our interacting scalar
field model~(\ref{action:tree2}) mimics the Yukawa part of the
lagrangian of the standard model, whereby one scalar field plays
the role of the Higgs field, while the other is a heavy fermion
(top quark or a chargino of a supersymmetric theory). The role of
the axial current is taken by the entropy which are both sensitive
to quantum coherence and the phase interface is a time dependent
mass parameter $m_{\phi}^{2}(t)$. The importance of quantum
coherence in baryogenesis is also treated in
\cite{Garbrecht:2003mn, Garbrecht:2004gv, Garbrecht:2005rr}, where
a coherent mixture of fermions has been used to generate baryons
in grand unified theories during preheating after inflation.
However, the authors of \cite{Garbrecht:2003mn, Garbrecht:2004gv,
Garbrecht:2005rr} treat the interactions phenomenologically in the
relaxation time approximation.

Quantum mechanical scattering on bubble walls in a thermal bath
may become the dominant mechanism for baryon production when the
walls are thin, and has been addressed in several papers in the
mid 1990s \cite{Farrar:1993hn, Farrar:1993sp, Gavela:1993ts,
Huet:1994jb, Gavela:1994ds, Gavela:1994dt} mostly in the context
of baryogenesis within the standard model. Currently, the
consensus is that so far no satisfactory solution to the problem
has been found. Recently Herranen, Kainulainen and
Rahkila~\cite{Herranen:2008hu,Herranen:2008hi,Herranen:2008di}
have reinvigorated interest in the problem, which has gained on
timeliness by the upcoming LHC experiments. Their approach is
based on the observation that the constraint equations for
fermions and scalars admit a third particle shell at a vanishing
energy, $k_0=0$. The authors show that this third shell can be
used to correctly reproduce the Klein paradox both for fermions
and bosons in a step potential, and hope that their intrinsically
off-shell formulation can be used to include interactions in a
field theoretical settings, for which off-shell physics is
essential. The authors have studied both
fermionic~\cite{Herranen:2008hu,Herranen:2008hi} and
bosonic~\cite{Herranen:2008di} quantum mechanical reflection in
the presence of scatterings. However, not all scatterings implied
by the Kadanoff-Baym equations are taken into account. One
important motivation for the present paper is to work within a set
of approximations where all relevant terms in the Kadanoff-Baym
equations are kept.

\section{Entropy and Propagators}
\label{Entropy and Propagators}

\subsection{The Statistical Propagator and Entropy} \label{The
Statistical Propagator and Entropy}

There is a connection between the statistical propagator and the
Gaussian entropy of a system (\cite{JFKTPMGS}, also see:
\cite{Prokopec:1992ia, Campo:2008ij}). In quantum field theory,
one can calculate many propagators, with different properties
associated with each, but not all of them are independent. In this
work, we will be primarily interested in solving for the
statistical propagator of the system. Let us mention that the
information contained in the statistical propagator is also
encoded in the two Wightman functions. Generically, in the
presence of quantum fluctuations, one needs complete knowledge of
the causal propagator in order to solve for the statistical
propagator. In the simple free theory example we consider in
appendix \ref{Quantum Reflection: Free Case}, we can directly
solve for the statistical propagator however and no prior
knowledge of the causal propagator is required.

The statistical propagator describes how states are populated and
is in the Heisenberg picture defined by:
\begin{equation}\label{statisticalpropagator}
F_{\phi}(x;x') = \frac{1}{2} \mathrm{Tr} \left[
\hat{\rho}(t_{0})\{ \hat{\phi}(x'), \hat{\phi}(x) \} \right]=
\frac{1}{2} \mathrm{Tr} \left[ \hat{\rho}(t_{0}) (
\hat{\phi}(x')\hat{\phi}(x) + \hat{\phi}(x)\hat{\phi}(x') )
\right] \,,
\end{equation}
given some density matrix operator $\hat{\rho}(t_{0})$. The causal
propagator roughly describes the number of accessible states and
is given by the commutator of the two fields:
\begin{equation}\label{causalpropagator}
\imath\Delta^{c}_{\phi} (x;x^{\prime}) = \mathrm{Tr} \left(
\hat{\rho}(t_{0})
 [\hat{\phi}(x),\hat{\phi}(x^{\prime})] \right) = \mathrm{Tr}
 \left[ \hat{\rho}(t_{0}) (\hat{\phi}(x)\hat{\phi}(x') -
\hat{\phi}(x')\hat{\phi}(x)) \right]\,.
\end{equation}
In spatially homogeneous backgrounds, we can Fourier transform
e.g. the statistical propagator as follows:
\begin{equation}\label{statpropagatorFourier}
F_{\phi}(k,t,t')=\int \mathrm{d}(\vec{x}-\vec{x}') F_{\phi}(x;x')
e^{-\imath \vec{k}\cdot(\vec{x}-\vec{x}')} \,,
\end{equation}
which in the case we will consider in this paper only depends on
$k=\|\vec{k}\|$. It is only the statistical propagator and its
various time derivatives that determine the entropy. In short, the
entropy is fixed by the area in phase space $\Delta$ the state of
the system occupies and is given by:
\begin{equation}\label{deltaareainphasespace}
\Delta_{k}^{2}(t) = 4\left. \left[
F(k,t,t')\partial_{t}\partial_{t'}F(k,t,t') -
\left\{\partial_{t}F(k,t,t')\right\}^{2} \right] \right|_{t=t'}
\,.
\end{equation}
Throughout the paper, and in particular in this equation we set
$\hbar=1$. We also set $c=1$. The entropy per mode then follows
as:
\begin{equation}\label{entropy}
S_{k}(t) = \frac{ \Delta_{k}(t)+1}{2}
\log\left(\frac{\Delta_{k}(t)+1}{2}\right) - \frac{
\Delta_{k}(t)-1}{2} \log\left(\frac{\Delta_{k}(t)-1}{2}\right) \,.
\end{equation}
Finally, it is interesting to note that the phase space area can
be related to an effective phase space particle number density per
mode or the statistical particle number density per mode as:
\begin{equation}\label{particlenumber}
n_{k}(t) = \frac{ \Delta_{k}(t)-1}{2}\,.
\end{equation}
In appendix \ref{Quantum Reflection: Free Case} we illustrate our
ideas by studying a non-trivial exact case: quantum scattering due
to a changing mass in the free case, i.e.: the interaction
coefficients $h$ and $\lambda$ in equation (\ref{action:tree2c})
are switched off. For a free scalar field with a smoothly changing
mass term, we show that $\Delta_{k}(t)=1$ and hence no entropy has
been generated by the mass change. Secondly, we point out that the
reader should not confuse the statistical particle number density
in equation (\ref{particlenumber}) with the parameter
$|\beta_{k}|^{2}$ characterising non-adiabaticity of the mass
change in equation (\ref{modesolution7b}), which in the literature
is often referred to as a particle number as well
\cite{Birrell:1982ix}. This parameter is non-zero, and possibly
large, simply because the asymptotic in and out vacua differ.

\subsection{Propagators in the Schwinger-Keldysh Formalism}
\label{Propagators in the Schwinger-Keldysh Formalism}

The material included in this subsection may well be familiar to
the experienced reader, but we include it nevertheless for
pedagogical reasons and in order to clearly establish our
notation. Let us consider the expectation value of an operator
$\hat{Q}(t)$ in the Heisenberg picture, given a density matrix
operator $\hat{\rho}(t_{0})$:
\begin{equation} \label{expectationvalues}
\langle\hat{Q}(t) \rangle =
\mathrm{Tr}\left[\hat{\rho}(t_{0})\hat{Q}(t)\right] =
\mathrm{Tr}\left[ \hat{\rho}(t_{0}) \left\{ \overline{T}
\exp\left(\imath \int_{t_0}^t \mathrm{d}t^\prime \hat
H(t^\prime)\right) \right\} \hat Q(t_{0}) \left\{ T \exp
\left(-\imath \int_{t_0}^t \mathrm{d} t^\prime \hat H(t^\prime)
\right) \right\}\right] \,,
\end{equation}
where $t_0<t$ denotes an initial time, $\overline{T}$ and $T$
denote the anti-time ordering and time ordering operations,
respectively, and $\hat H(t)$ denotes the Hamiltonian. If
$\hat{Q}$ in the Scr\"odinger picture depends explicitly on time,
we should replace $\hat{Q}(t_{0})$ by $\hat{Q}_{\mathrm{S}}(t)$.

\begin{figure}[t!]
    \begin{minipage}[t]{.45\textwidth}
        \begin{center}
\includegraphics[width=\textwidth]{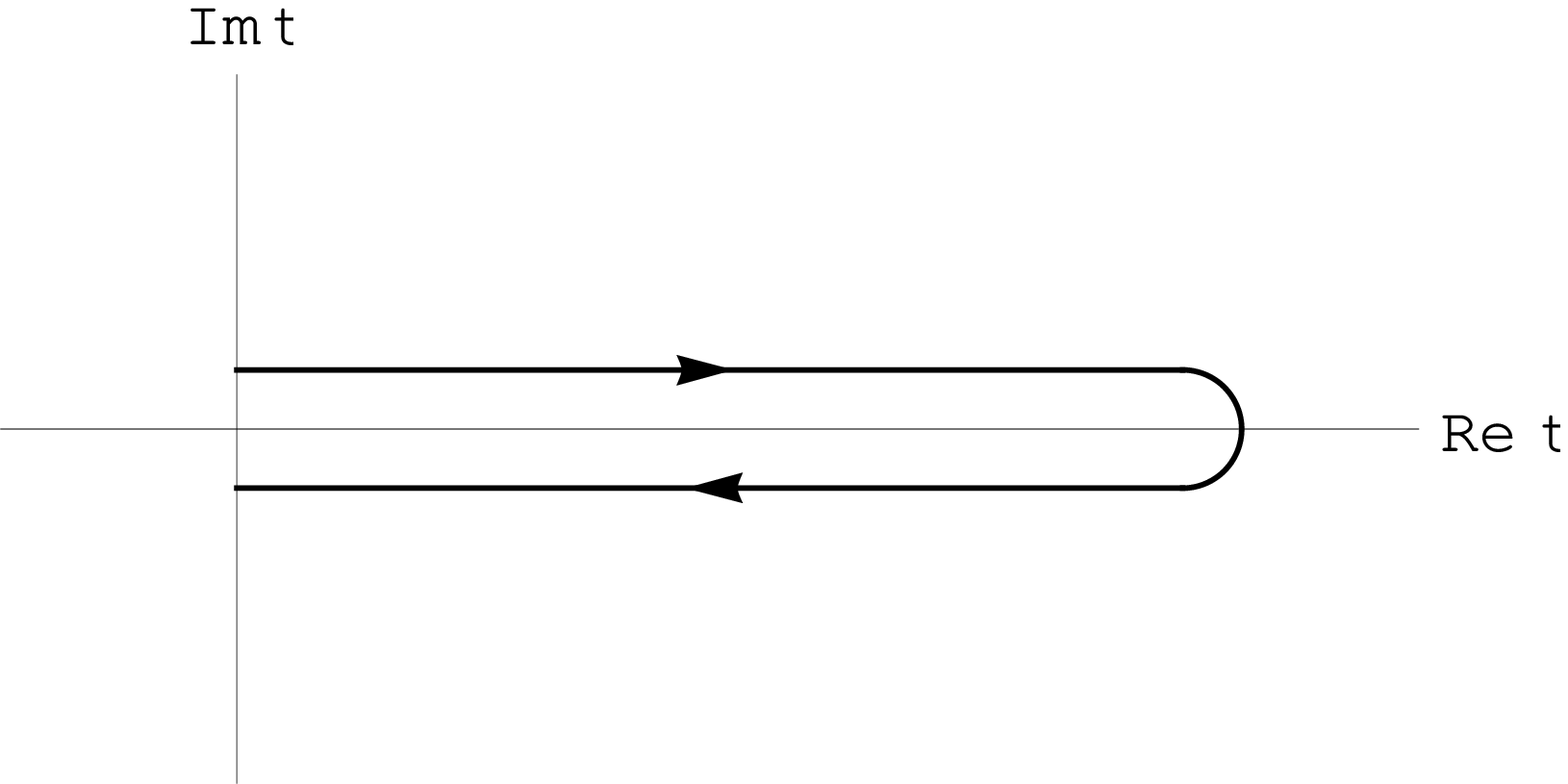}
   {\em \caption{Schwinger-Keldysh contour with finite initial time $t_{0}$ and final time
   $t_{f}$.
   \label{fig:schwingercontour1} }}
        \end{center}
    \end{minipage}
\hfill
    \begin{minipage}[t]{.45\textwidth}
        \begin{center}
\includegraphics[width=\textwidth]{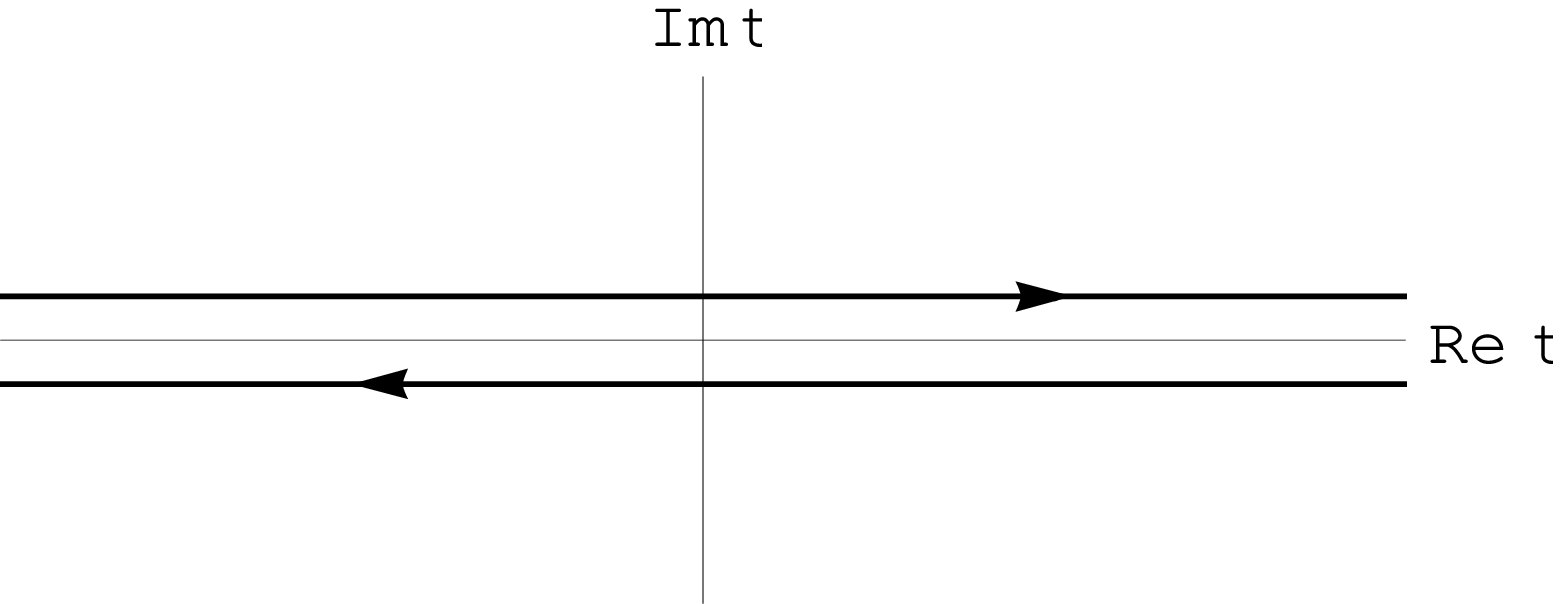}
   {\em \caption{Schwinger-Keldysh contour where the initial and
   final times in figure \ref{fig:schwingercontour1} have been
   extended to negative and positive infinity, respectively.
   \label{fig:schwingercontour2} }}
        \end{center}
    \end{minipage}
\end{figure}

The Schwinger-Keldysh formalism, or closed time path (CTP)
formalism, or in-in formalism, is based on the original papers by
Schwinger \cite{Schwinger:1960qe} and Keldysh
\cite{Keldysh:1964ud} and particularly useful for non-equilibrium
quantum field theory (also see: \cite{Chou:1984es, Jordan:1986ug,
Calzetta:1986cq,Weinberg:2005vy, Koksma:2007uq}). According to the
CTP formalism, the expectation value above can be calculated from
the in-in generating functional in the path integral formulation:
\begin{eqnarray}\label{Z:inin}
&& {\cal Z}[J_{+}^{\phi}, J_{-}^{\phi}, J_{+}^{\chi},
J_{-}^{\chi}, \rho(t_0)] \\
&& \quad  \!=\! \int \! {\cal D}\phi^{+}_{0}{\cal D}\phi^{-}_{0}
{\cal D}\chi^{+}_{0} {\cal D}\chi^{-}_{0} \langle\phi^{+}_{0} ,
\chi^{+}_{0}| \hat{ \rho}(t_{0})|\phi^{-}_{0} , \chi^{-}_{0}
\rangle \!\int_{\phi_{0}^{+}}^{\phi_{0}^{-}}\! {\cal
D}\phi^{+}{\cal D}\phi^{-}
\delta[\phi^{+}(t_{f}\!)-\phi^{-}(t_{f}\!)]\!
\int_{\chi_{0}^{+}}^{\chi_{0}^{-}} \!{\cal D}\chi^{+}{\cal
D}\chi^{-} \delta[\chi^{+}(t_{f}\!)-\chi^{-}(t_{f}\!)]
\nonumber \\
&& \qquad\qquad \times {\rm exp}\left[\imath \int
\mathrm{d}^{D-1}x\int_{t_{0}}^{t_{f}} \mathrm{d}t^\prime
\left({\cal L}[\phi^{+},\chi^{+},t']-{\cal
L}[\phi^{-},\chi^{-},t'] +J_{+}^{\phi}\phi^{+} +
J_{-}^{\phi}\phi^{-} + J_{+}^{\chi} \chi^{+} +
J_{-}^{\chi}\chi^{-} \right)\right]  \nonumber \,,
\end{eqnarray}
where the Lagrangian is given in equation (\ref{action:tree1}). We
can use the well-known Schwinger-Keldysh contour depicted in
figures \ref{fig:schwingercontour1} and
\ref{fig:schwingercontour2}.
It runs from $t_{0}$ up to $t_{f}$, where both times can in
principle be extended to negative and positive infinity,
respectively (as depicted in figure \ref{fig:schwingercontour2}).
As we will come to discuss, the two paths are not equivalent in an
interacting quantum field theory, where memory effects play an
important role. In this paper, we will extend $t_{0}$ to negative
infinity at some point, but let us for the moment keep it finite.
Clearly, these contours are closely related to the two evolution
operators in equation (\ref{expectationvalues}). The fields $\phi$
and $\chi$ and their corresponding sources $J^{\phi}$ and
$J^{\chi}$ split up on the upper (+) and lower (-) parts of the
contour, where necessarily the conditions
$\phi^{+}(t_{f})=\phi^{-}(t_{f})$ and $\chi^{+}(t_{f}) =
\chi^{-}(t_{f})$ apply. These conditions are indeed enforced by
the two functional $\delta$-distributions. The first functional
integrals in (\ref{Z:inin}) are over the initial configuration
space at $t_{0}$, where the system is specified by the density
operator (density matrix) $\hat\rho(t_0)$. The path integrals in
(\ref{Z:inin}) run over the Schwinger-Keldysh contour in figure
\ref{fig:schwingercontour1} or \ref{fig:schwingercontour2}.

Expectation values of $n$-point functions are obtained by varying
the generating functional~(\ref{Z:inin}) as follows:
\begin{eqnarray} \label{npointfunctions}
\left. \mathrm{Tr}\left[ \hat{\rho}(t_{0})
\overline{T}[\hat\phi(x_1)\dots \hat\phi(x_n)]
T[\hat\phi(y_1)\dots \hat\phi(y_k)] \right] =
\frac{\delta^{n+k}{\cal Z}[J, \rho(t_0)] } {\imath\delta \!
J_{-}^{\phi}(x_{1})\cdots \imath\delta\! J_{-}^{\phi}(x_{n})
\imath\delta\! J_{+}^{\phi}(y_{1})\cdots \imath\delta\!
J_{+}^{\phi}(y_{k})} \right|_{J = 0} \,, \nonumber
\end{eqnarray}
provided that $x^{0}_{j}\leq t_{f}$ and $y^{0}_{j}\leq t_{f}$ for
all $j$, and where $J=(J^{\phi}_{\pm},J^{\chi}_{\pm})$. We can now
define the following propagators:
\begin{subequations}
\label{propagators}
\begin{eqnarray}
\imath\Delta^{++}_{\phi}(x;x^\prime) &=&
\mathrm{Tr}\left[\hat{\rho}(t_{0})
T[\hat\phi(x^\prime)\hat\phi(x)] \right] =
 \mathrm{Tr}\left[\hat{\rho}(t_{0}) \hat\phi^+(x^\prime)\hat\phi^+(x)\right] =
\left.\frac{\delta^2{\cal Z}[J, \rho(t_0)]} {\imath\delta\!
J_{+}^{\phi}(x) \imath\delta\! J_{+}^{\phi}(x^\prime)}
\right|_{J=0}
\label{propagatorsa} \\
\imath\Delta^{--}_{\phi}(x;x^\prime) &=&
 \mathrm{Tr}\left[\hat{\rho}(t_{0}) \overline{T} [ \hat\phi(x^\prime)\hat\phi(x)]
\right] =
 \mathrm{Tr}\left[\hat{\rho}(t_{0}) \hat\phi^-(x^\prime)\hat\phi^-(x)\right] =
\left. \frac{\delta^2{\cal Z} [J, \rho(t_0)]} {\imath\delta\!
J_{-}^{\phi}(x)\imath\delta\! J_{-}^{\phi}(x^\prime)}
\right|_{J=0}
\label{propagatorsb} \\
\imath\Delta^{-+}_{\phi}(x;x^\prime) &=&
 \mathrm{Tr}\left[\hat{\rho}(t_{0})  \hat\phi(x)\hat\phi(x^\prime)\right] =
 \mathrm{Tr}\left[\hat{\rho}(t_{0}) \hat\phi^-(x)\hat\phi^+(x^\prime)\right] =
\left.\frac{\delta^2{\cal Z} [J, \rho(t_0)] } {\imath\delta\!
J_{-}^{\phi}(x)\imath\delta\! J_{+}^{\phi}(x^\prime)}
\right|_{J=0} \label{propagatorsc}
\\
\imath\Delta^{+-}_{\phi}(x;x^\prime) &=&
 \mathrm{Tr}\left[\hat{\rho}(t_{0})\hat\phi(x^\prime)\hat\phi(x)\right] =
 \mathrm{Tr}\left[\hat{\rho}(t_{0})\hat\phi^-(x^\prime)\hat\phi^+(x)\right]=
\left.\frac{\delta^2{\cal Z} [J, \rho(t_0)] } {\imath\delta\!
J_{+}^{\phi}(x)\imath\delta\! J_{-}^{\phi}(x^\prime)}
\right|_{J=0} \,,\label{propagatorsd}
\end{eqnarray}
\end{subequations}
We define the various propagators for the $\chi$-field
analogously. In the absence of a condensate for $\chi$ all mixed
two point functions, such as:
\begin{equation}\label{propagatorexclude}
\langle \Omega |\hat{\phi}(x^\prime)\hat{\chi}(x)|\Omega\rangle
\,,
\end{equation}
vanish by virtue of the interaction term (\ref{action:tree2c}). In
equation (\ref{propagators}),
$\imath\Delta^{++}_{\phi}(x;x^\prime) \equiv
\imath\Delta^{F}_{\phi}(x;x^\prime)$ denotes the Feynman or time
ordered propagator and $\imath\Delta^{--}_{\phi}(x;x^\prime)$
represents the anti-time ordered propagator. The two Wightman
functions are given by $\imath\Delta^{-+}_{\phi}(x;x^\prime)$ and
$\imath\Delta^{+-}_{\phi}(x;x^\prime)$. From
equation~(\ref{action:tree1}) we infer that the free Feynman
propagator obeys:
\begin{equation} \label{Feynman propagator}
{\cal D}_x\imath \Delta^{++}_{\phi, 0} (x;x^\prime) \equiv
(\partial_x^2 - m^2)\imath \Delta^{++}_{\phi, 0}(x;x^\prime) =
\imath \delta^{\scriptscriptstyle{D}}\!(x-x^\prime)\,,
\end{equation}
where $\partial_x^2 = \eta^{\mu\nu}\partial_\mu\partial_\nu $ and
where the same identity holds for $\imath \Delta^{++}_{\chi, 0}
(x;x^\prime)$. In the presence of interactions the equation of
motion for the Feynman propagator becomes much more involved and
we will discus it shortly. One can easily show that at tree level
the Wightman functions obey the homogeneous equation:
\begin{equation} \label{Wightman:eom:free}
{\cal D}_x \imath \Delta_{\phi, 0}^{+-}(x;x^\prime) = 0 = {\cal
D}_x \imath \Delta_{\phi, 0}^{-+}(x;x^\prime) \,.
\end{equation}
Identical relations hold for the free $\chi$-propagators. The four
propagators defined above are not independent. The Wightman
functions for example constitute the time ordered and anti-time
ordered propagators:
\begin{subequations}
\label{propagatoridentities}
\begin{eqnarray}
\imath\Delta^{++}_{\phi}(x;x^\prime) &=& \theta(t-t^\prime)\imath
\Delta^{-+}_{\phi}(x;x^\prime) + \theta(t^\prime-t)\imath
\Delta^{+-}_{\phi}(x;x^\prime) \label{propagatoridentitiesa}
\\
\imath\Delta^{--}_{\phi}(x;x^\prime) &=& \theta(t^\prime-t)\imath
\Delta^{-+}_{\phi}(x;x^\prime) + \theta(t-t^\prime)\imath
\Delta^{+-}_{\phi}(x;x^\prime) \label{propagatoridentitiesb}
\end{eqnarray}
where $t=x^0$, $t^\prime = {x^0}^{\,\prime}$, and where this
identity holds for the $\chi$-propagators as well. Appreciate
that:
\begin{eqnarray}
\imath\Delta^{++}_{\phi}(x;x^\prime) +
\imath\Delta^{--}_{\phi}(x;x^\prime) &=& \imath
\Delta^{-+}_{\phi}(x;x^\prime) +\imath
\Delta^{+-}_{\phi}(x;x^\prime) \label{propagatoridentitiesc}
\\
\imath \Delta^{-+}_{\phi}(x;x^\prime)&=&\imath
\Delta^{+-}_{\phi}(x^\prime;x)\,, \label{propagatoridentitiesd}
\end{eqnarray}
\end{subequations}
are exact identities and they are also satisfied by the
$\chi$-propagators. We write the four Green's functions in the
$2\times 2$ Keldysh propagator matrix form:
\begin{equation}\label{Keldysh propagator}
\imath{\cal G}_{\phi}(x;x^\prime) = \left(
\begin{array}{cc}
\imath \Delta^{++}_{\phi} & \imath \Delta^{+-}_{\phi} \cr \imath
\Delta^{-+}_{\phi} & \imath \Delta^{--}_{\phi} \cr
\end{array}\right)
\,,
\end{equation}
which at tree level obeys,
\begin{equation}
 {\cal D}_x \imath{\cal G}_{\phi, 0}(x;x^\prime) = \imath \sigma^3 \delta^{\scriptscriptstyle{D}}\!(x-x^\prime)
\,,
\end{equation}
where $\sigma^3={\rm diag}(1,-1)$ is the third Pauli matrix, which
we can also write as:
\begin{equation}
(\sigma^3)^{ab} = a \delta^{ab} \,,
\end{equation}
where $a,b=\pm$.

Let us define some more Green's functions. In subsection \ref{The
Statistical Propagator and Entropy} we already defined the causal
and statistical propagator, but let us for completeness list them
again. The causal Green's function, also known as the Pauli-Jordan
or Schwinger or spectral two-point function,
$\imath\Delta^{c}_{\phi} \equiv \imath\Delta^{PJ}_{\phi} \equiv
{\cal A}_{\phi} \equiv \rho_{\phi}$, is given by:
\begin{equation} \label{Delta:causal}
\imath\Delta^{c}_{\phi} (x;x^\prime) = \mathrm{Tr}\left(
\hat{\rho}(t_{0})  [\hat\phi(x),\hat\phi(x^\prime)]\right)= \imath
\Delta^{-+}_{\phi}(x;x^\prime) - \imath
\Delta^{+-}_{\phi}(x;x^\prime)\,,
\end{equation}
and the statistical or Hadamard two-point function, $F_{\phi}
\equiv \Delta^{H}_{\phi}$, is given by:
\begin{equation} \label{Hadamard}
F_{\phi}(x;x') = \frac{1}{2} \mathrm{Tr}\left[ \hat{\rho}(t_{0})
\{\hat \phi(x'),\hat\phi(x)\} \right]=
\frac{1}{2}\Big(\imath\Delta^{-+}_{\phi}(x;x') +
\imath\Delta^{+-}_{\phi}(x;x')\Big) \,.
\end{equation}
The retarded ($\imath\Delta^{\mathrm{r}}$) and advanced
($\imath\Delta^{\mathrm{a}}$) propagators are defined as:
\begin{subequations}
\label{propagators2}
\begin{eqnarray}
\imath\Delta^{\mathrm{r}}_{\phi}(x;x^\prime) &=& \imath
\Delta^{++}_{\phi}(x;x^\prime) - \imath
\Delta^{+-}_{\phi}(x;x^\prime)
\nonumber\\
&=& -[\imath \Delta^{--}_{\phi}(x;x^\prime) - \imath
\Delta^{-+}_{\phi}(x;x^\prime)] = \theta(t-t^\prime)\imath
\Delta^c_{\phi}(x;x^\prime) \label{Delta:ret}
\\
\imath\Delta^{\mathrm{a}}_{\phi}(x;x^\prime)&=& \imath
\Delta^{++}_{\phi}(x;x^\prime) - \imath
\Delta^{-+}_{\phi}(x;x^\prime)
\nonumber\\
&=& -[\imath \Delta^{--}_{\phi}(x;x^\prime) - \imath
\Delta^{+-}_{\phi}(x;x^\prime)] = -\theta(t^\prime-t)\imath
\Delta^{c}_{\phi}(x;x^\prime) \,. \label{Delta:adv}
\end{eqnarray}
\end{subequations}
Moreover, we can express all propagators $\imath
\Delta^{ab}_{\phi}$ solely in terms of the causal and statistical
propagators:
\begin{subequations}
\label{reduction:F+Deltac}
\begin{eqnarray}
\imath \Delta^{+-}_{\phi}(x;x^\prime) &=& F_{\phi}(x;x^\prime)-
\frac{1}{2}\imath\Delta^{c}_{\phi}(x;x^\prime)
\label{reduction:F+Deltaca}\\
\imath \Delta^{-+}_{\phi}(x;x^\prime) &=& F_{\phi}(x;x^\prime) +
\frac{1}{2}\imath\Delta^{c}_{\phi}(x;x^\prime)
\label{reduction:F+Deltacb}\\
\imath \Delta^{++}_{\phi}(x;x^\prime) &=& F_{\phi}(x;x^\prime) +
\frac{1}{2}\mathrm{sgn}(t- t^\prime)
\imath\Delta^{c}_{\phi}(x;x^\prime)
\label{reduction:F+Deltacc}\\
\imath \Delta^{--}_{\phi}(x;x^\prime) &=& F_{\phi}(x;x^\prime)-
\frac{1}{2}\mathrm{sgn}(t-t^\prime)\imath\Delta^{c}_{\phi}(x;x^\prime)
\,. \label{reduction:F+Deltacd}
\end{eqnarray}
\end{subequations}
Since $F_\phi^\dagger=F_\phi$ and $(\imath\Delta^c_\phi)^\dagger=
-\imath\Delta^c_\phi$, the relations above correspond to splitting
the various Green's functions into their hermitian and
anti-hermitian parts (for that reason we do not put an $\imath$ in
front of $F_\phi$). The definitions of the retarded, advanced,
causal and the statistical propagators and the relations between
them easily extend to the $\chi$-field.

\subsection{The Kadanoff-Baym Equations}
\label{The Kadanoff-Baym
equations}

In order to study the effect of perturbative loop corrections on
classical expectation values, one often considers the effective
action. In this subsection we will calculate the 2PI effective
action, using the Schwinger-Keldysh formalism outlined above. The
2PI effective action is the relevant functional to consider
because it captures the interaction of the $\phi$- and
$\chi$-fields in the right way. Varying the 2PI effective action
with respect to the propagators yields the so-called Kadanoff-Baym
equations that govern their dynamics. These equations of motion
contain the non-local scalar self-energy corrections or self-mass
corrections to the propagator.

In the present subsection, we shall mainly follow
\cite{Prokopec:2003pj, Berges:2004yj, Weinberg:2005vy,
vanderMeulen:2007ah}. We can extract the Feynman rules from the
interaction part of the tree level action~(\ref{action:tree2c}):
\begin{equation}\label{lagrangian:interaction}
{\cal L}_{\mathrm{int}}[\phi,\chi] = -  \sum_{a=\pm} a \left(
\frac{\lambda}{3!}(\chi^{a}(x))^{3} + \frac{1}{2} h
(\chi^{a}(x))^{2} \phi^{a}(x) \right) \,.
\end{equation}
The Feynman propagator is promoted to $\imath {\cal G}_{\phi}$ and
each vertex has two polarities: plus ($+$) and minus ($-$), such
that the minus vertex gains an extra minus sign $+\imath \lambda$
as compared to the standard perturbation theory plus vertex
$-\imath \lambda$.

\begin{figure}[t!]
 \centering
  \begin{minipage}[t]{.70\textwidth}
   \includegraphics[width=\textwidth]{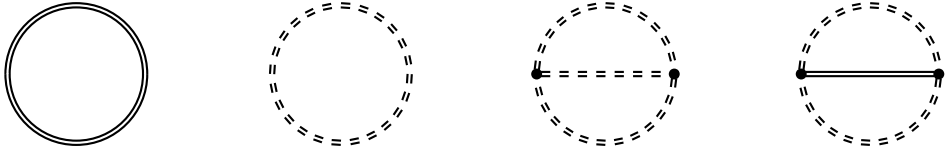}
   {\em \caption{Contributions to the 2PI effective action up to two
    loop order. The double solid lines denote $\phi$-propagators, whereas the double dashed lines correspond
    to $\chi$-propagators. \label{fig:2PIEfAction}}}
   \end{minipage}
\end{figure}

\begin{figure}[t!]
 \centering
  \begin{minipage}[t]{.75\textwidth}
   \includegraphics[width=\textwidth]{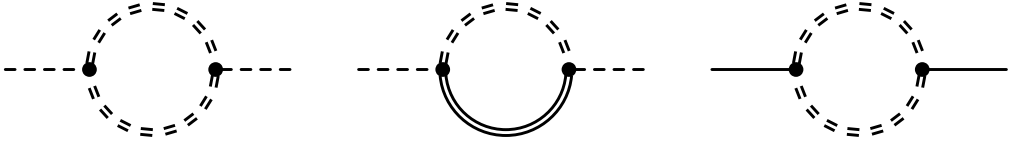}
   {\em \caption{Contributions to the self-masses up to one-loop order.
    Again, the double solid lines denote $\phi$-propagators, whereas the
    double dashed lines correspond to $\chi$-propagators. Hence, the first
    two Feynman diagrams contribute to the self-mass of $\chi(x)$, and
    only the third diagram contributes to the self-mass of $\phi(x)$.
    \label{fig:SelfMass}}}
   \end{minipage}
\end{figure}

The 2PI effective action can be obtained as a double Legendre
transform from the generating functional $W$ for connected Green's
functions with respect to the linear source $J$ and another
quadratic source (see e.g.: \cite{Berges:2004yj}). In the absence
of field condensates the background fields vanish:
\begin{subequations}
\label{nobackground}
\begin{eqnarray}
\bar{\phi}^{a}\equiv \langle\Omega|\hat{\phi}^{a}|\Omega\rangle
&=& 0
\label{nobackgrounda}\\
\bar{\chi}^{a}\equiv \langle\Omega|\hat{\chi}^{a}|\Omega\rangle
&=& 0 \label{nobackgroundb}\,,
\end{eqnarray}
\end{subequations}
in which case the variation with respect to the linear or
quadratic sources can easily be related. In particular, the
definitions of the four propagators in equation
(\ref{propagators}) remain valid. The effective action formally
reads \cite{Cornwall:1974vz, Jackiw:1974cv, Calzetta:1986cq,
Berges:2004yj}:
\begin{eqnarray}\label{effectiveaction}
\Gamma[\bar{\phi}^{a},\bar{\chi}^{a},\imath\Delta_{\phi}^{ab},\imath\Delta_{\chi}^{ab}]
&=& S[\bar{\phi}^{a},\bar{\chi}^{a}] + \frac{\imath}{2}
\mathrm{Tr} \ln [ (\imath\Delta_{\phi}^{ab})^{-1}] +
\frac{\imath}{2} \mathrm{Tr} \ln [
(\imath\Delta_{\chi}^{ab})^{-1}] \\
&& +  \frac{\imath}{2} \mathrm{Tr} \frac{\delta^{2}\!
S[\bar{\phi}^{a},\bar{\chi}^{a}]}{\delta\!\bar{\phi}^{a} \delta\!
\bar{\phi}^{b}} \imath\Delta_{\phi}^{ab} + \frac{\imath}{2}
\mathrm{Tr} \frac{\delta^{2}
S[\bar{\phi}^{a},\bar{\chi}^{a}]}{\delta\! \bar{\chi}^{a} \delta\!
\bar{\chi}^{b}} \imath\Delta_{\chi}^{ab} +
\Gamma^{(2)}[\bar{\phi}^{a},\bar{\chi}^{a},\imath\Delta_{\phi}^{ab},\imath\Delta_{\chi}^{ab}]
\nonumber \,.
\end{eqnarray}
Here, $\Gamma^{(2)}$ denotes the 2PI contribution to the effective
action. Moreover, we omitted the dependence on all variables for
notational convenience. Several Feynman diagrams contribute to the
effective action of which the one and two loop order contributions
are given in figure \ref{fig:2PIEfAction}.
We can now write the effective action up to two loops as:
\begin{equation}\label{effectiveaction2}
\Gamma[\imath\Delta_{\phi}^{ab},\imath\Delta_{\chi}^{ab} ] =
\Gamma_0[\imath\Delta_{\phi}^{ab},\imath\Delta_{\chi}^{ab}] +
\Gamma_1 [\imath\Delta_{\phi}^{ab},\imath\Delta_{\chi}^{ab}] +
\Gamma_2[\imath\Delta_{\phi}^{ab},\imath\Delta_{\chi}^{ab}]
\end{equation}
where the subscript denotes the number of loops and where:
\begin{subequations}
\label{Gamma}
\begin{eqnarray}
\Gamma_{0}[\imath\Delta_{\phi}^{ab},\imath\Delta_{\chi}^{ab} ] &=&
\phantom{+} \int \mathrm{d}^{\scriptscriptstyle{D}}\!x
\mathrm{d}^{\scriptscriptstyle{D}}\!x' \sum_{a,b=\pm}
\frac{a}{2}(\partial_{x}^{2} -
m^{2}_{\phi})\delta^{\scriptscriptstyle{D}}\!(x-x') \delta^{ab}
\imath\Delta^{ba}_{\phi}(x';x)  \label{Gamma0} \\
&& + \int \mathrm{d}^{\scriptscriptstyle{D}}\! x
\mathrm{d}^{\scriptscriptstyle{D}}\! x' \sum_{a,b=\pm}
\frac{a}{2}(\partial_{x}^{2} -
m^{2}_{\chi})\delta^{\scriptscriptstyle{D}}\! (x-x') \delta^{ab}
\imath \Delta^{ba}_{\chi} (x';x) \nonumber
\\
\Gamma_{1} [\imath\Delta_{\phi}^{ab},\imath\Delta_{\chi}^{ab} ]
&=& -\frac{\imath}{2}{\rm Tr} \ln\left[\imath \Delta^{aa}_{\phi}
(x;x)\right]  -\frac{\imath}{2}{\rm Tr} \ln\left[\imath
\Delta^{aa}_{\chi} (x;x)\right] \label{Gamma1}
\\
\Gamma_{2}[\imath\Delta_{\phi}^{ab},\imath\Delta_{\chi}^{ab} ] &=&
\frac{\imath\lambda^{2}}{12} \int
\mathrm{d}^{\scriptscriptstyle{D}}\! x
\mathrm{d}^{\scriptscriptstyle{D}}\! x' \sum_{a,b=\pm} ab
\left(\imath\Delta^{ab}_{\chi}(x';x)\right)^{3} + \frac{\imath
h^{2}}{4} \int \mathrm{d}^{\scriptscriptstyle{D}}\! x
\mathrm{d}^{\scriptscriptstyle{D}}\! x' \sum_{a,b=\pm} ab
\left(\imath\Delta^{ab}_{\chi}(x;x')\right)^{2}
\imath\Delta^{ab}_{\phi}(x';x) \label{Gamma2} ,
\end{eqnarray}
\end{subequations}
where ${\rm Tr}$ denotes a trace over both spacetime variables and
the Keldysh indices $\pm$. The equations of motion for the
propagators result as usual from the variational principle:
\begin{subequations}
\label{EOM1}
\begin{eqnarray}
\frac{ \delta \Gamma[\imath\Delta_{\phi}^{ab},
\imath\Delta_{\chi}^{ab}
]}{\delta \imath\Delta_{\phi}^{ab}} &=& 0 \label{EOM1a} \\
\frac{\delta
\Gamma[\imath\Delta_{\phi}^{ab},\imath\Delta_{\chi}^{ab} ]}{\delta
\imath\Delta_{\chi}^{ab}} &=& 0 \label{EOM1b} \,.
\end{eqnarray}
\end{subequations}
Explicitly, they yield:
\begin{subequations}
\label{EOM2}
\begin{eqnarray}
\frac{a}{2}(\partial_{x}^{2} -
m^{2}_{\phi})\delta^{\scriptscriptstyle{D}}\!(x-x^\prime)\delta^{ab}
-\frac{\imath}{2}\left[\imath\Delta^{ab}_{\phi}(x;x^\prime)\right]^{-1}
+ \frac{\imath h^{2}}{4}ab
\left(\imath\Delta^{ab}_{\chi}(x;x')\right)^2 &=& 0
\label{EOM2a} \\
\frac{a}{2}(\partial_{x}^{2} -
m^{2}_{\chi})\delta^{\scriptscriptstyle{D}}\!(x-x^\prime)\delta^{ab}
-\frac{\imath}{2}\left[\imath\Delta^{ab}_{\chi}(x;x^\prime)\right]^{-1}
+ \frac{\imath \lambda^{2}}{4}ab
\left(\imath\Delta^{ab}_{\chi}(x;x')\right)^2 + \frac{\imath
h^{2}}{2}ab \, \imath\Delta^{ab}_{\chi}(x;x')
\imath\Delta^{ab}_{\phi}(x;x') &=& 0 \label{EOM2b} \,.
\end{eqnarray}
\end{subequations}
We will bring these equations into a more familiar form by
multiplying by $2a
\imath\Delta^{bc}_{\phi}(x^\prime;x^{\prime\prime})$ and $2a
\imath\Delta^{bc}_{\chi}(x^\prime;x^{\prime\prime})$,
respectively, and then integrating over $x^\prime$ and summing
over $b=\pm$. This results in the following one-loop
Kadanoff-Baym~\cite{KadanoffBaym:1962} equations for the elements
of the Keldysh propagator $\imath{\cal G}(x;x^\prime)$:
\begin{subequations}
\label{EOM3}
\begin{eqnarray}
(\partial_{x}^{2}-m^{2}_{\phi})\imath
\Delta^{ab}_{\phi}(x;x^\prime) -\sum_{c=\pm}c\int
\mathrm{d}^{\scriptscriptstyle{D}}\! x_1
M^{ac}_{\phi}(x;x_1)\imath \Delta^{cb}_{\phi}(x_1;x^\prime) &=&
a\delta^{ab}\imath\delta^{\scriptscriptstyle{D}}\!(x-x^\prime)
\label{EOM3a} \\
(\partial_{x}^{2}-m^{2}_{\chi})\imath
\Delta^{ab}_{\chi}(x;x^\prime) -\sum_{c=\pm}c\int
\mathrm{d}^{\scriptscriptstyle{D}}\! x_1
M^{ac}_{\chi}(x;x_1)\imath \Delta^{cb}_{\chi}(x_1;x^\prime) &=&
a\delta^{ab}\imath\delta^{\scriptscriptstyle{D}}\!(x-x^\prime)
\label{EOM3b} \,.
\end{eqnarray}
\end{subequations}
where the self-masses at one loop have the form:
\begin{subequations}
\label{selfMass}
\begin{eqnarray}
\imath M^{ac}_{\phi}(x;x_1) &=&
-2ac\frac{\delta\Gamma_{2}[\imath\Delta_{\phi}^{ab},
\imath\Delta_{\chi}^{ab} ]}{\delta\imath\Delta^{ca}_{\phi}(x_1;x)}
= -\frac{\imath h^{2}}{2} \left(\imath
\Delta^{ac}_{\chi}(x;x_1)\right)^{2}
\label{selfMassa} \\
 \imath M^{ac}_{\chi}(x;x_1) &=&
-2ac\frac{\delta\Gamma_{2}[\imath\Delta_{\phi}^{ab},
\imath\Delta_{\chi}^{ab} ]}{\delta\imath\Delta^{ca}_{\chi}(x_1;x)}
= -\frac{\imath \lambda^{2}}{2} \left(\imath
\Delta^{ac}_{\chi}(x;x_1)\right)^{2} - \imath h^{2} \imath
\Delta^{ac}_{\chi}(x;x_1) \imath \Delta^{ac}_{\phi}(x;x_1)
\label{selfMassb} \,,
\end{eqnarray}
\end{subequations}
where in the last step we used the hermiticity symmetry of the
operator $\imath {\cal G}$, according to which, $\imath
\Delta^{ac}(x;x^\prime) = \imath \Delta^{ca}(x^\prime;x)$. The
Feynman diagrams contributing to the one-loop self-mass are given
in figure \ref{fig:SelfMass}.
We have chosen the definition of (\ref{selfMass}) such that the
structure of the self-mass resembles that of the propagators. The
factor 1/2 in~(\ref{selfMass}) originates from the symmetry factor
of the one-loop self-mass diagram.

Equation~(\ref{EOM3a}) consists of the following four equations:
\begin{subequations}
\label{EOM3aExtended}
\begin{eqnarray}
(\partial_{x}^{2}-m^{2}_{\phi}) \imath
\Delta^{++}_{\phi}(x;x^\prime) - \int
\mathrm{d}^{\scriptscriptstyle{D}}\! y \left[\imath
M^{++}_{\phi}(x;y)\imath \Delta^{++}_{\phi}(y;x^\prime) - \imath
M^{+-}_{\phi}(x;y)\imath \Delta^{-+}_{\phi}(y;x^\prime)\right] &=&
\, \imath\delta^{\scriptscriptstyle{D}}\!(x-x^\prime)
\label{EOM3a++} \\
(\partial_{x}^{2} - m^{2}_{\phi})\imath
\Delta^{+-}_{\phi}(x;x^\prime)- \int
\mathrm{d}^{\scriptscriptstyle{D}}\!y \left[\imath
M^{++}_{\phi}(x;y)\imath \Delta^{+-}_{\phi}(y;x^\prime) - \imath
M^{+-}_{\phi}(x;y)\imath \Delta^{--}_{\phi}(y;x^\prime)\right] &=&
0
\label{EOM3a+-} \\
(\partial_{x}^{2} - m^{2}_{\phi})\imath
\Delta^{-+}_{\phi}(x;x^\prime) - \int
\mathrm{d}^{\scriptscriptstyle{D}}\!y \left[\imath
M^{-+}_{\phi}(x;y)\imath \Delta^{++}_{\phi}(y;x^\prime) - \imath
M^{--}_{\phi}(x;y)\imath \Delta^{-+}_{\phi}(y;x^\prime)\right] &=&
0
\label{EOM3a-+} \\
(\partial_{x}^{2}-m^{2}_{\phi})\imath
\Delta^{--}_{\phi}(x;x^\prime) - \int
\mathrm{d}^{\scriptscriptstyle{D}}\!y \left[\imath
M^{-+}_{\phi}(x;y)\imath \Delta^{+-}_{\phi}(y;x^\prime) - \imath
M^{--}_{\phi}(x;y)\imath \Delta^{--}_{\phi} (y;x^\prime)\right]
&=& -\imath \delta^{\scriptscriptstyle{D}}\!(x-x^\prime)  \,,
\label{EOM3a--}
\end{eqnarray}
\end{subequations}
but in the light of equation (\ref{propagatoridentities}), only
two of them are independent. Note that we have another set of four
equations of motion for the $\chi$-field. In the end, we will be
interested in solving this equation of motion in Fourier space,
e.g.:
\begin{subequations}
\label{Fouriertransformdef}
\begin{eqnarray}
\imath \Delta_{\phi}^{ab}(x;x')= \int
\frac{\mathrm{d}^{\scriptscriptstyle{D}-1}\vec{k}}{(2\pi)^{\scriptscriptstyle{D}-1}}
\imath \Delta_{\phi}^{ab}(\vec{k},t,t') e^{\imath
\vec{k}(\vec{x}-\vec{x}')}
\label{Fouriertransformdefa} \\
\imath \Delta_{\phi}^{ab}(\vec{k},t,t')= \int
\mathrm{d}^{\scriptscriptstyle{D}-1}(\vec{x}-\vec{x}') \imath
\Delta_{\phi}^{ab}(x ; x') e^{-\imath \vec{k}(\vec{x}-\vec{x}')}
\label{Fouriertransformdefb} \,.
\end{eqnarray}
\end{subequations}
Such that equation (\ref{EOM3aExtended}) transforms into:
\begin{subequations}
\label{EOM4aExtended}
\begin{eqnarray}
(\partial_{t}^{2}+k^{2}+m^{2}_{\phi}) \imath
\Delta^{++}_{\phi}(k,t,t') + \int_{-\infty}^{\infty}
\mathrm{d}t_{1} \left[\imath M^{++}_{\phi}(k,t,t_{1})\imath
\Delta^{++}_{\phi}(k,t_{1},t') - \imath
M^{+-}_{\phi}(k,t,t_{1})\imath
\Delta^{-+}_{\phi}(k,t_{1},t')\right] &=&
\label{EOM4a++} \\
&& \, \imath\delta(t-t')\nonumber \\
(\partial_{t}^{2}+k^{2}+m^{2}_{\phi})\imath
\Delta^{+-}_{\phi}(k,t,t')+ \int_{-\infty}^{\infty}
\mathrm{d}t_{1} \left[\imath M^{++}_{\phi}(k,t,t_{1})\imath
\Delta^{+-}_{\phi}(k,t_{1},t') - \imath
M^{+-}_{\phi}(k,t,t_{1})\imath
\Delta^{--}_{\phi}(k,t_{1},t')\right] &=& 0
\label{EOM4a+-} \\
(\partial_{t}^{2}+k^{2}+m^{2}_{\phi})\imath
\Delta^{-+}_{\phi}(k,t,t') + \int_{-\infty}^{\infty}
\mathrm{d}t_{1} \left[\imath M^{-+}_{\phi}(k,t,t_{1})\imath
\Delta^{++}_{\phi}(k,t_{1},t') - \imath
M^{--}_{\phi}(k,t,t')\imath \Delta^{-+}_{\phi}(k,t_{1},t')\right]
&=& 0
\label{EOM4a-+} \\
(\partial_{t}^{2}+k^{2}+m^{2}_{\phi})\imath
\Delta^{--}_{\phi}(k,t,t') + \int_{-\infty}^{\infty}
\mathrm{d}t_{1} \left[\imath M^{-+}_{\phi}(k,t,t_{1})\imath
\Delta^{+-}_{\phi}(k,t_{1},t') - \imath
M^{--}_{\phi}(k,t,t_{1})\imath \Delta^{--}_{\phi}
(k,t_{1},t')\right] &=& \label{EOM4a--} \\
&& -\imath \delta(t-t')  \nonumber \,.
\end{eqnarray}
\end{subequations}
Note that we have extended $t_{0} \rightarrow -\infty$ in the
equation above. Again, we have an analogous set of equations of
motion for the $\chi$-field. In principle we can solve these
coupled equations of motion only numerically in full generality.
Our strategy is to push the analytical calculation forward as far
as possible, before relying on numerical methods. Before we make
an important simplifying assumption, let us first consider the
renormalisation of our theory.

\section{Renormalising the Kadanoff-Baym Equations}
\label{Renormalising the Kadanoff-Baym equations}

In order to renormalise equation of motion (\ref{EOM3aExtended})
or (\ref{EOM4aExtended}) above, we need to Fourier transform also
with respect to the difference of the time variables:
\begin{subequations}
\label{Fouriertransformdef2}
\begin{eqnarray}
\imath \Delta_{\phi}^{ab}(x;x') &=& \int
\frac{\mathrm{d}^{\scriptscriptstyle{D}}k}{(2\pi)^{\scriptscriptstyle{D}}}
 \imath \Delta_{\phi}^{ab}(k^{\mu}){\rm e}^{\imath k \cdot (x-x')}
\label{Fouriertransformdef2a} \\
\imath \Delta_{\phi}^{ab}(k^{\mu}) &=& \int
\mathrm{d}^{\scriptscriptstyle{D}}(x-x')\imath
\Delta_{\phi}^{ab}(x;x')  {\rm e}^{-\imath k \cdot (x-x')} \,,
\label{Fouriertransformdef2b}
\end{eqnarray}
\end{subequations}
There is a subtlety: for the moment we neglect the time dependence
in the mass term. We only use this assumption to renormalise. In
the end it turns out that we need a mass independent counterterm
to cancel all divergences in our theory, which allows us to
consider a time varying mass term again. In fact, as we assume
there is no residual dependence on the average time coordinate
$(t+t')/2$ in $\imath \Delta_{\phi}^{ab}(k^{\mu})$, equation
(\ref{Fouriertransformdef2}) coincides with a Wigner transform.
Fourier transforming equation of motion (\ref{EOM3aExtended})
yields:
\begin{subequations}
\label{EOM4Fourier}
\begin{eqnarray}
(-k_{\mu}k^{\mu}-m^{2}_{\phi}- \imath M^{++}_{\phi}(k^{\mu}))
\imath \Delta^{++}_{\phi}(k^{\mu}) + \imath M^{+-}_{\phi}(k^{\mu})
\imath \Delta^{-+}_{\phi}(k^{\mu}) &=& \, \imath
\label{EOM4Fourier++} \\
(-k_{\mu}k^{\mu} - m^{2}_{\phi} -\imath M^{++}_{\phi}(k^{\mu})
)\imath \Delta^{+-}_{\phi}(k^{\mu}) + \imath
M^{+-}_{\phi}(k^{\mu})\imath \Delta^{--}_{\phi}(k^{\mu}) &=& 0
\label{EOM4Fourier+-} \\
(-k_{\mu}k^{\mu} - m^{2}_{\phi} + \imath M^{--}_{\phi}(k^{\mu})
)\imath \Delta^{-+}_{\phi}(k^{\mu}) - \imath
M^{-+}_{\phi}(k^{\mu})\imath \Delta^{++}_{\phi}(k^{\mu}) &=& 0
\label{EOM4Fourier-+} \\
(-k_{\mu}k^{\mu}-m^{2}_{\phi} + \imath
M^{--}_{\phi}(k^{\mu}))\imath \Delta^{--}_{\phi}(k^{\mu}) - \imath
M^{-+}_{\phi}(k^{\mu})\imath \Delta^{+-}_{\phi}(k^{\mu}) &=&
-\imath \,, \label{EOM4Fourier--}
\end{eqnarray}
\end{subequations}
Here and henceforth, we use the notation $k_{\mu}k^{\mu}=
-k_{0}^{2}+k^{2}$ to distinguish the four-vector length from the
spatial three-vector length $k=\|\vec{k}\|$. Because of the
convolution, the equations of motion above are local in Fourier
space. Let us remind the reader again that analogous equations
hold for the $\chi$-propagators. As already announced in the
introduction, we shall not solve the dynamical equations for both
$\phi$- and $\chi$-propagators. Instead, we shall assume the
following hierarchy of couplings:
\begin{equation}\label{h<<lambda}
h\ll\lambda
\end{equation}
and expand the solution in powers of $h/\lambda\ll 1$. In fact, we
shall solve the system only at order $(h/\lambda)^0$. This does
not imply that the $h \chi^{2} \phi$ interaction is unimportant:
we will only assume that $\lambda$ is large such that the
$\chi$-field is thermalised by its strong self-interaction. This
allows us to approximate the solutions of the dynamical equations
for $\chi$ as thermal propagators which we derived in the appendix
in equation (\ref{ThermalPropagatorApp}), see
\cite{LeBellac:1996}:
\begin{subequations}
\label{ThermalPropagator}
\begin{eqnarray}
\imath\Delta_{\chi}^{++}(k^{\mu}) &=&
\frac{-\imath}{k_{\mu}k^{\mu}+m_{\chi}^{2} -\imath\epsilon} + 2\pi
\delta(k_{\mu}k^{\mu}+m_{\chi}^{2})
n_{\chi}^{\mathrm{eq}}(|k_{0}|) \label{ThermalPropagator++}
\\
\imath\Delta_{\chi}^{--}(k^{\mu}) &=&
\frac{\imath}{k_{\mu}k^{\mu}+m_{\chi}^{2}+\imath\epsilon} +  2\pi
\delta(k_{\mu}k^{\mu}+m_{\chi}^{2}) n_{\chi}^{\mathrm{eq}}(|k^{0}|)\label{ThermalPropagator--}\\
\imath\Delta_{\chi}^{+-}(k^{\mu}) &=& 2\pi
\delta(k_{\mu}k^{\mu}+m_{\chi}^{2}) \left[ \theta(-k^{0}) +
n_{\chi}^{\mathrm{eq}}(|k^{0}|)\right] \label{ThermalPropagator+-}
\\
\imath\Delta_{\chi}^{-+}(k^{\mu}) &=& 2\pi
\delta(k_{\mu}k^{\mu}+m_{\chi}^{2}) \left[ \theta(k^{0}) +
n_{\chi}^{\mathrm{eq}}(|k^{0}|)\right] \,,
\label{ThermalPropagator-+}
\end{eqnarray}
\end{subequations}
where the Bose-Einstein distribution is given by:
\begin{equation}
n_\chi^{\rm eq}(k^0) = \frac{1}{{\rm e}^{\beta k^0} - 1} \,,\qquad
\beta = \frac{1}{k_BT} \,, \label{BoseEinstein}
\end{equation}
with $k_B$ denoting the Stefan-Boltzmann constant and $T$ the
temperature. Let us remark that assumption (\ref{h<<lambda})
allows us to compute the quantum corrections to the
$\phi$-propagators as it depends solely on $\chi$-propagators
running in the loop. We neglect the backreaction of the system
field on the environment field, such that the latter remains in
thermal equilibrium at temperature $T$. This assumption is
perturbatively well justified as already discussed in the
introduction. Furthermore, we neglected for simplicity the
$\mathcal{O}(\lambda^{2})$ correction to the propagators above
that will slightly change the equilibrium of the environment
field. Note finally that, in our approximation scheme, the
dynamics of the system-propagators is effectively influenced only
by the usual 1PI self-mass correction.

In this paper, we consider only an environment field $\chi$ in its
vacuum state at $T=0$ and we postpone the finite temperature
corrections to a future publication. Any divergences, if present,
originate from the vacuum contributions to the self-masses, i.e.:
the vacuum propagators at $T=0$ are a useful case to consider anyway:
\begin{subequations}
\label{VacuumPropagator}
\begin{eqnarray}
\imath\Delta_{\chi}^{++}(k^{\mu}) &=&
\frac{-\imath}{k_{\mu}k^{\mu}+m_{\chi}^{2} -\imath\epsilon}
\label{VacuumPropagator++}
\\
\imath\Delta_{\chi}^{--}(k^{\mu}) &=&
\frac{\imath}{k_{\mu}k^{\mu}+m_{\chi}^{2}+\imath\epsilon} \label{VacuumPropagator--}\\
\imath\Delta_{\chi}^{+-}(k^{\mu}) &=& 2\pi
\delta(k_{\mu}k^{\mu}+m_{\chi}^{2}) \theta(-k^{0})
\label{VacuumPropagator+-}
\\
\imath\Delta_{\chi}^{-+}(k^{\mu}) &=& 2\pi
\delta(k_{\mu}k^{\mu}+m_{\chi}^{2}) \theta(k^{0}) \,.
\label{VacuumPropagator-+}
\end{eqnarray}
\end{subequations}
We evaluate the Feynman self-mass $\imath M_{\phi}^{++}(x;x')$
following from equation (\ref{selfMassa}) where we make the
simplifying assumption $m_{\chi}\rightarrow 0$. Let us briefly
expatiate justifying this assumption as at first sight it seems
that $m_{\chi}\rightarrow 0$ makes our approximation scheme more
susceptible to undesired backreaction effects\footnote{We thank
Julien Serreau for this useful comment.}. It is a priori not at
all clear that the backreaction is negligible: if we examine the
second Feynman diagram on the right-hand side of equation
(\ref{eomfeynmandiagramsb}) we see that the leading order
backreaction occurs at order
$\mathcal{O}(h^{2}/\omega_{\chi}^{2})$. Since in our setup
$\omega_{\chi}^{2}=k^{2}+m_{\chi}^{2}=k^{2}$, it is clear that the
backreaction on deep IR (infrared) Fourier modes of the
environment field is perturbatively unsuppressed. Despite that, it
does not spoil the perturbative arguments employed in the
introduction: the influence of the environment field on the system
field is still perturbatively under control. In order to see this,
let us consider the first non-Gaussian contribution on the
right-hand side of equation (\ref{eomfeynmandiagrams2}). Indeed,
one can show that the IR part of the inner loop is phase space
suppressed: the IR part of this integral is given by
$\int_{0}^{\bar{\mu}} \mathrm{d}^{4}q
[(k^{\nu}k_{\nu}-m^{2}_{\phi})(q_{\sigma}+k_{\sigma})(q^{\sigma}+k^{\sigma})]^{-1}
\sim \bar{\mu}^{2}/m_{\phi}^{2}$ when $m_{\phi}>h \simeq
\bar{\mu}$. Nevertheless, we admit it would be worthwhile to
examine these integrals for $m_{\chi}\neq 0$ and see whether the
results presented in this paper are robust under this change.

We thus need to evaluate:
\begin{equation}\label{Feynmanpropposition}
\imath \Delta_{\chi}^{++}(x;x')= \int
\frac{\mathrm{d}^{\scriptscriptstyle{D}}k}{(2\pi)^{\scriptscriptstyle{D}}}
\, \imath \Delta_{\chi}^{++}(k^{\mu})e^{\imath k(x-x')} \,.
\end{equation}
This integral can be performed in arbitrary dimensions by making
use of two straightforward contour integrations and
\cite{GradshteynRyzhik,Janssen:2008px}:
\begin{equation}\label{spericalsymmetry}
\int\frac{\mathrm{d}^{\scriptscriptstyle{D-1}}\vec{k}}{(2\pi)^{\scriptscriptstyle{D-1}}}
e^{i\vec{k}\cdot \vec{x}} f(k)  =
\frac{2}{(4\pi)^{\frac{\scriptscriptstyle{D-1}}{2}}}
\int_{0}^{\infty} \mathrm{d}k \, k^{\scriptscriptstyle{D-2}}
\frac{J_{\frac{\scriptscriptstyle{D-3}}{2}}(k x)}{(\frac{1}{2}k x
)^{\frac{\scriptscriptstyle{D-3}}{2}}} f(k) \,,
\end{equation}
which is valid for any function $f(k)$ that depends solely on
$k=\|\vec{k}\|$. $ J_{\mu}(k x)$ is a Bessel function of the first
kind. This yields:
\begin{equation}\label{Feynmanpropposition2}
\imath \Delta_{\chi}^{++}(x;x')= \frac{\Gamma(\frac{D}{2}-1)}{4
\pi^{\frac{D}{2}}} \frac{1}{ \Delta
x_{++}^{\scriptscriptstyle{D}-2}(x;x')} \,.
\end{equation}
Here, $\Delta x_{++}^{2}(x,x')$ is one of the distance functions
between two spacetime points $x$ and $x'$ frequently used in the
Schwinger-Keldysh formalism and given by:
\begin{subequations}
\label{x}
\begin{eqnarray}
\Delta x_{++}^{2}(x,x') &=& - \left(\left|t - t' \right| - i
\varepsilon \right)^{2} + \| \vec{x} -
\vec{x}'\|^{2} \label{x++}  \\
\Delta x_{+-}^{2}(x,x') &=& - \left( \phantom{|}t - t'\phantom{|}
+ i \varepsilon \right)^{2} + \| \vec{x} - \vec{x}'\|^{2}
\label{x+-} \\
\Delta x_{-+}^{2}(x,x') &=& - \left( \phantom{|} t - t'\phantom{|}
 - i \varepsilon \right)^{2} + \| \vec{x} - \vec{x}'\|^{2}
\label{x-+} \\
\Delta x_{--}^{2}(x,x') &=& - \left(\left|t - t' \right| + i
\varepsilon \right)^{2} + \| \vec{x} - \vec{x}'\|^{2} \label{x--}
\,.
\end{eqnarray}
\end{subequations}
We thus immediately find from equations (\ref{selfMassa}) and
(\ref{Feynmanpropposition2}):
\begin{equation}\label{SelfMassPosspace}
\imath M_{\phi}^{++}(x;x')= - \frac{\imath h^{2}}{2}
\frac{\Gamma^{2}(\frac{D}{2}-1)}{16 \pi^{\scriptscriptstyle{D}}}
\frac{1}{ \Delta x_{++}^{2\scriptscriptstyle{D}-4}(x;x')} \,.
\end{equation}
The other self-masses can be obtained from this expression using
the appropriate $\varepsilon$ pole prescription as indicated in
equation (\ref{x}). We will now rewrite this expression slightly
in order to extract the divergence. For an arbitrary exponent
$\beta \neq D$, $\beta \neq 2$, we can easily derive:
\begin{equation}\label{SelfMassPosspace2}
\frac{1}{\Delta x_{++}^{\beta}(x;x')} = \frac{1}{(\beta-2)(\beta -
D)}
\partial^{2} \frac{1}{\Delta x_{++}^{\beta-2}(x;x')}\,.
\end{equation}
Furthermore, recall \cite{Janssen:2007ht, Janssen:2008dp}:
\begin{subequations}
\label{SelfMassPosspace3}
\begin{equation}
\partial^{2} \frac{1}{\Delta x_{++}^{\scriptscriptstyle{D}-2}(x;x')} = \frac{4
\pi^{\frac{D}{2}}}{\Gamma(\frac{D-2}{2})} \imath
\delta^{\scriptscriptstyle{D}} (x-x') \label{SelfMassPosspace3a}
\,.
\end{equation}
Let us also recall the similar identities for the other distance
functions:
\begin{eqnarray}
\partial^{2} \frac{1}{\Delta x_{--}^{\scriptscriptstyle{D}-2}(x;x')} &=& - \frac{4
\pi^{\frac{D}{2}}}{\Gamma(\frac{D-2}{2})} \imath
\delta^{\scriptscriptstyle{D}} (x-x')  \label{SelfMassPosspace3b} \\
\partial^{2} \frac{1}{\Delta x_{+-}^{\scriptscriptstyle{D}-2}(x;x')} &=& 0 \label{SelfMassPosspace3c}\\
\partial^{2} \frac{1}{\Delta x_{-+}^{\scriptscriptstyle{D}-2}(x;x')} &=& 0 \label{SelfMassPosspace3d}\,.
\end{eqnarray}
\end{subequations}
We now arrange equation (\ref{SelfMassPosspace}), using
(\ref{SelfMassPosspace2}) and (\ref{SelfMassPosspace3a}):
\begin{equation}\label{SelfMassPosspace4}
\imath M_{\phi}^{++}(x;x')= - \frac{\imath h^{2}
\Gamma^{2}(\frac{D}{2}-1) }{64 \pi^{\scriptscriptstyle{D}}}
\frac{1}{(D-3)(D-4)} \left[ \partial^{2}\left\{ \frac{1}{ \Delta
x_{++}^{2\scriptscriptstyle{D}-6}(x;x')} -
\frac{\mu^{\scriptscriptstyle{D}-4}}{ \Delta
x_{++}^{\scriptscriptstyle{D}-2}(x;x')} \right\} + \frac{ 4
\pi^{\frac{D}{2}} \mu^{\scriptscriptstyle{D}-4}
}{\Gamma(\frac{D-2}{2})} \imath \delta^{\scriptscriptstyle{D}}
(x-x') \right]\,.
\end{equation}
Here, the scale $\mu$ has been introduced on dimensional grounds.
If we Taylor expand the term in curly brackets\footnote{Note that
in the minimal subtraction scheme, one would also expand the term
multiplying the Dirac delta function around $D=4$, which gives
rise, once integrated at the level of the equation of motion, to a
finite local contribution to the mass of $\phi$.} around $D=4$, we
find:
\begin{equation}\label{SelfMassPosspace4Taylor}
\imath M_{\phi}^{++}(x;x')= - \frac{\imath h^{2}
\Gamma(\frac{D}{2}-1) \mu^{\scriptscriptstyle{D}-4} }{16
\pi^{\frac{D}{2}} (D-3)(D-4)}\imath \delta^{\scriptscriptstyle{D}}
(x-x') + \frac{\imath h^{2}}{128 \pi^{4}} \partial^{2}\left[
\frac{\log(\mu^{2}\Delta x_{++}^{2}(x;x'))}{ \Delta
x_{++}^{2}(x;x')}\right] + \mathcal{O}(D-4) \,.
\end{equation}
We have been able to separate a local $(D-4)^{-1}$ divergence and
a non-local finite term to the self-mass. In order to precisely
cancel the divergence, we can thus add a local counterterm, i.e.:
an ordinary mass term of the form:
\begin{equation}\label{SelfMasscounterterm}
\imath M_{\phi,\mathrm{ct}}^{\pm \pm}(x;x')= \mp \frac{\imath
h^{2} \Gamma(\frac{D}{2}-1) \mu^{\scriptscriptstyle{D}-4} }{16
\pi^{\frac{D}{2}} (D-3)(D-4)}\imath \delta^{\scriptscriptstyle{D}}
(x-x') \,.
\end{equation}
The relative sign difference of $\imath
M_{\phi,\mathrm{ct}}^{--}(x;x')$ is due to equation
(\ref{SelfMassPosspace3b}). We are left with the following
renormalised self-mass:
\begin{equation}\label{SelfMassPosspace5}
\imath M_{\phi,\mathrm{ren}}^{++}(x;x')= - \frac{\imath h^{2}
\Gamma^{2}(\frac{D}{2}-1) }{64 \pi^{\scriptscriptstyle{D}}}
\frac{1}{(D-3)(D-4)} \partial^{2}\left\{ \frac{1}{ \Delta
x_{++}^{2\scriptscriptstyle{D}-6}(x;x')} -
\frac{\mu^{\scriptscriptstyle{D}-4}}{ \Delta
x_{++}^{\scriptscriptstyle{D}-2}(x;x')} \right\} \,.
\end{equation}
We will now perform a spatial Fourier transform in order to solve
for the dynamics this term generates:
\begin{equation}\label{SelfMassFourierSp1}
\imath M_{\phi,\mathrm{ren}}^{++}(\vec{k},t,t')= \int
\mathrm{d}^{\scriptscriptstyle{D}-1}(\vec{x}-\vec{x}') \imath
M_{\phi,\mathrm{ren}}^{++}(x ; x') e^{-\imath
\vec{k}(\vec{x}-\vec{x}')} \,.
\end{equation}
By introducing a regulator in order to dispose of the overall
surface terms arising from two partial integrations, we can easily
convert the partial derivatives. Using several analytic
extensions, we obtain:
\begin{eqnarray}\label{SelfMassFourierSp2}
\imath M_{\phi,\mathrm{ren}}^{++}(k,t,t') &=& \frac{\imath h^{2}
\Gamma^{2}(\frac{D}{2}-1)2^{\frac{D-13}{2}} \pi^{-\frac{D+1}{2}}
}{ k^{\frac{D-3}{2}} (D-3)(D-4)} \left(
\partial^{2}_{t} + k^{2}\right) \Bigg[
-\frac{\mu^{\scriptscriptstyle{D}-4} 2^{\frac{3-D}{2}}
\pi^{\frac{1}{2}} k^{\frac{D-5}{2}} }{\Gamma(\frac{D-2}{2})}
e^{-\imath k (|\Delta t| -\imath\epsilon)} \\
&& \qquad\qquad\qquad\qquad\qquad\qquad\qquad\qquad +
\frac{k^{\scriptscriptstyle{D}-4}(\imath |\Delta
t|+\epsilon)^{\frac{5-D}{2}}}{2^{\scriptscriptstyle{D}-4}\Gamma(D-3)}K_{\frac{D-5}{2}}\left(k(i|\Delta
t|+\epsilon)\right) \Bigg] \nonumber \,,
\end{eqnarray}
where $k=\|\vec{k}\|$, $\Delta t = t-t'$ and where $K_{\nu}(z)$ is
the modified Bessel function of the second kind. We expand this
result around $D=4$:
\begin{eqnarray}\label{SelfMassFourierSp3}
\imath M_{\phi,\mathrm{ren}}^{++}(k,t,t') &=& \frac{\imath
h^{2}}{32 \pi^{2}\sqrt{2 k \pi}} \left(
\partial^{2}_{t} + k^{2}\right) \Bigg[ \sqrt{\frac{\pi}{2k}} e^{-\imath k (|\Delta t|
-\imath\epsilon)}\left( \gamma_{\mathrm{E}} + \log\left[
\frac{k}{2 \imath \mu^{2} (|\Delta t| - \imath \epsilon)}\right]\right) \\
&& \qquad\qquad\qquad\qquad\quad - \sqrt{\imath |\Delta t|+
\epsilon}\,
\partial_{\nu}K_{\nu}\left(\imath k (|\Delta t|- \imath
\epsilon)\right)\Big|_{\nu=1/2} \Bigg]  + \mathcal{O}(D-4)
\nonumber \,.
\end{eqnarray}
Here, $\gamma_{\mathrm{E}}$ is the Euler–-Mascheroni constant.
Moreover, the scale $\mu$ introduced earlier combines nicely with
the other terms to make the argument of the logarithm
dimensionless as it should. Indeed, we need to find an expression
for the derivative with respect to the order $\nu$ of
$K_{\nu}(z)$. Starting from the general expansion:
\begin{equation}\label{ModBesselFunction}
K_{\nu}(z) = \frac{\pi \csc (\pi\nu)}{2} \sum_{k=0}^{\infty}
\left\{
\frac{1}{\Gamma(k-\nu+1)k!}\left(\frac{z}{2}\right)^{2k-\nu} -
\frac{1}{\Gamma(k+\nu+1)k!}\left(\frac{z}{2}\right)^{2k+\nu}
\right\}\,,
\end{equation}
we immediately derive:
\begin{eqnarray}\label{ModBesselFunctionderivative}
\partial_{\nu}K_{\nu}(z)\Big|_{\nu=1/2} &=&
-\sqrt{\frac{\pi}{2z}}\,e^{z}\left[\mathrm{Chi}(2z)-\mathrm{Shi}(2z)\right]
\\
&=& -\sqrt{\frac{\pi}{2z}}\,e^{z}\left[ \gamma_{\mathrm{E}} +
\log(2z) + \int_{0}^{2z} \mathrm{d}t \frac{\cosh{t}-1}{t} -
\int_{0}^{2z} \mathrm{d}t \frac{\sinh{t}}{t}\right] \nonumber \,,
\end{eqnarray}
where $\mathrm{Chi}(2z)$ and $\mathrm{Shi}(2z)$ are the hyperbolic
cosine and hyperbolic sine integral functions, respectively,
defined by the expressions on the second line. In our case, the
variable $z$ is imaginary, so it proves useful to extract an
$\imath$ and convert this expression to the somewhat more familiar
sine and cosine integral functions, defined by:
\begin{subequations}
\begin{eqnarray}\label{SineCosineIntFuncs}
\mathrm{si}(z) &=& - \int_{z}^{\infty} \mathrm{d}t
\frac{\sin{t}}{t} \label{SineCosineIntFuncsa} \\
\mathrm{ci}(z) &=& - \int_{z}^{\infty} \mathrm{d}t
\frac{\cos{t}}{t} \label{SineCosineIntFuncsb} \,.
\end{eqnarray}
\end{subequations}
We finally arrive at:
\begin{eqnarray}\label{SelfMassFourierSp4}
\imath M_{\phi,\mathrm{ren}}^{++}(k,t,t') &=& \frac{\imath
h^{2}}{64 k \pi^{2}} \left(
\partial^{2}_{t} + k^{2}\right) \Bigg[ e^{-\imath k |\Delta t|} \left( \gamma_{\mathrm{E}} + \log\left[
\frac{k}{2 \imath \mu^{2} (|\Delta t| - \imath \epsilon)}\right]\right) \\
&& \qquad\qquad\qquad\qquad + e^{\imath k |\Delta t|} \Big(
\mathrm{ci}(2k(|\Delta t| -\imath\epsilon)) - \imath
\mathrm{si}(2k(|\Delta t| -\imath\epsilon))\Big)\Bigg]  +
\mathcal{O}(D-4) \nonumber \,,
\end{eqnarray}
where we have set the $\epsilon$ regulators in the exponents to
zero as the expression is well defined. Rather than going several
times through the calculation above to determine the other
self-masses, we make use of a few analytic extensions. Observe for
example that if $\Delta t >0$, $\Delta x_{++}(x,x')$ and $\Delta
x_{-+}(x,x')$ coincide, hence the expressions for self-masses
$\imath M_{\phi,\mathrm{ren}}^{++}(k,t,t')$ and $\imath
M_{\phi}^{-+}(k,t,t')$ should also coincide in that region. All we
need to do is to sensibly analytically extend to $\Delta t < 0$.
We will thus need:
\begin{equation}\label{SineIntFuncNegativeArgument}
\mathrm{si}( - z) = - \mathrm{si}(z) - \pi \,.
\end{equation}
If $\Delta t < 0$, we have to carefully make use of the $\epsilon$
pole prescription in the cosine integral function:
\begin{eqnarray}\label{CosineIntFuncNegativeArgument}
\mathrm{ci}(- 2k(-\Delta t +\imath\epsilon)) &=& - \int_{-
2k(-\Delta t +\imath\epsilon)}^{\infty} \mathrm{d}t\frac{\cos
t}{t} = - \left[ \int_{- 2k(-\Delta t)}^{-\imath\epsilon}
\mathrm{d}t + \int_{-\imath\epsilon}^{2 k (-\Delta t)} \mathrm{d}t
+ \int_{2 k (-\Delta t)}^{\infty} \mathrm{d}t\right] \frac{\cos
t}{t} \\
&=& -\log(\imath \epsilon) + \log(-\imath \epsilon) +
\mathrm{ci}(2k(-\Delta t)) \nonumber \\
&=& -\imath \pi + \mathrm{ci}(2k(-\Delta t)) \nonumber\,.
\end{eqnarray}
We thus find the following expressions for the renormalised
self-masses:
\begin{subequations}
\label{SelfMassFourierSpFinalResult}
\begin{equation}
\imath M_{\phi,\mathrm{ren}}^{ab}(k,t,t') = \left(
\partial^{2}_{t} + k^{2}\right) \imath Z^{ab}_{\phi}(k,t,t')
\,,
\end{equation}
where:
\begin{eqnarray}
Z_{\phi}^{\pm \pm }(k,t,t') &=& \frac{h^{2}}{64 k \pi^{2}} \left[
e^{\mp \imath k |\Delta t|} \left( \gamma_{\mathrm{E}} +
\log\left[ \frac{k}{2 \mu^{2} |\Delta t|}\right] \mp \imath
\frac{\pi}{2}\right) + e^{\pm \imath k |\Delta t|} \Big(
\mathrm{ci}(2k|\Delta t|) \mp \imath \mathrm{si}(2k|\Delta
t|)\Big)\right] \label{SelfMassFourierSpFinalResulta}
\\
Z_{\phi}^{\mp\pm }(k,t,t') &=& \frac{h^{2}}{64 k \pi^{2}} \left[
e^{\mp \imath k \Delta t}\! \left( \gamma_{\mathrm{E}} +
\log\left[ \frac{k}{2 \mu^{2}|\Delta t|} \right]\! \mp \imath
\frac{\pi}{2} \mathrm{sgn}(\Delta t) \right) \!+ e^{\pm \imath k
\Delta t } \Big( \mathrm{ci}(2k|\Delta t|) \mp
\imath\mathrm{sgn}(\Delta t) \mathrm{si}(2k|\Delta t|)\!\Big)\!
\right] \! . \qquad\phantom{1}
\label{SelfMassFourierSpFinalResultb}
\end{eqnarray}
\end{subequations}
Firstly, appreciate that $\imath M_{\phi}^{-+}(k,t,t') $ and
$\imath M_{\phi}^{+-}(k,t,t') $ need not be renormalised. The
reason is that these expressions do not contain a divergence in
$D=4$, which can be seen from equations
(\ref{SelfMassPosspace3c}--\ref{SelfMassPosspace3d}). Moreover,
the local counterterm which we add to renormalise $\imath
M_{\phi}^{--}(k,t,t')$ contains the opposite sign as compared to
$\imath M_{\phi}^{++}(k,t,t')$ because of equation
(\ref{SelfMassPosspace3b}), which we already stated in equation
(\ref{SelfMasscounterterm}). Finally, we have sent all $\epsilon$
regulators to zero as the expression above is well defined in the
limit $\Delta t \rightarrow 0$.

We performed two independent checks of the calculation above.
Firstly, one can renormalise via a calculation in Fourier space
(rather than position space). We show that the two results agree
in appendix \ref{Alternative Method of Renormalising the
Self-masses}. Secondly, one can calculate the retarded self-mass
directly from the position space result using
(\ref{SelfMassPosspace4Taylor}) and compare with the result
obtained from (\ref{SelfMassFourierSpFinalResult}). We show that
the two results agree in appendix \ref{Retarded Self-mass}.

If one were to evaluate the two time derivatives in the
expressions above, one would find a divergent answer in the limit
when $\Delta t \rightarrow 0$. We also show this in appendix
\ref{Alternative Method of Renormalising the Self-masses}. This
does not reflect an incorrect renormalisation procedure. It is
crucial to extract the two time derivatives as presented above in
order to properly take the effect of the self-masses into account
as only now $Z_{\phi}^{ab}(k,t,t')$ is finite at coincidence
$\Delta t \rightarrow 0$. Indeed, this is most easily seen in
position space\footnote{One can easily recognise that the
structure of the renormalised self-masses in equation
(\ref{SelfMassFourierSpFinalResult}) is identical to the
d'Alembertian in Fourier space. The presence of
$Z_{\phi}^{ab}(k,t,t')$ induces time dependence in the propagator.
A similar phenomenon has been observed in \cite{Miao:2006gj},
where this phenomenon is referred to as a ``finite wave function
renormalisation'', in which the effect of gravitons on fermions in
an expanding Universe is investigated.}.

Let us compare these expressions with existing literature. In e.g.
\cite{Berges:2004yj,Berges:2005hc} it is derived that the
renormalised equations for $\lambda\phi^{4}(x)$ theory have an
identical structure as the unrenormalised equations. In our
theory, clearly, the structure of the two equations changes as we
need to extract an operator of the form
$(\partial^{2}_{t}+k^{2})$, as derived in equation
(\ref{SelfMassFourierSpFinalResult}).

\section{Decoupling the Kadanoff-Baym Equations}
\label{Decoupling the Kadanoff-Baym equations}

Having renormalised our theory, we are ready to massage the
Kadanoff-Baym equations (\ref{EOM4aExtended}) in two different
ways. Firstly, we will write Kadanoff-Baym equations in terms of
the causal and statistical propagator such that they decouple.
This is of course a vital step required to solve the Kadanoff-Baym
equations in the next section. Secondly, we show that when we
write the equations in terms of the advanced and retarded
propagators, the one-loop contributions preserve causality as they
should.

Note that the structure of the self-mass (\ref{selfMass}) is such
that we can construct relations analogous to equation
(\ref{reduction:F+Deltac}), which of course hold identically for
$\chi$:
\begin{subequations}
\label{reduction:selfMass:MF+Mc}
\begin{eqnarray}
Z^{+-}_{\phi}(k,t,t') &=& Z^{F}_{\phi}(k,t,t') - \frac{1}{2}\imath
Z^{c}_{\phi}(k,t,t')
\label{reduction:selfMass:MF+Mca}\\
Z^{-+}_{\phi}(k,t,t') &=& Z^{F}_{\phi}(k,t,t') + \frac{1}{2}\imath
Z^{c}_{\phi}(k,t,t')
\label{reduction:selfMass:MF+Mcb}\\
Z^{++}_{\phi}(k,t,t') &=& Z^{F}_{\phi}(k,t,t') + \frac{1}{2}{\rm
sgn}(t-t^\prime)\imath Z^{c}_{\phi}(k,t,t')
\label{reduction:selfMass:MF+Mcc}\\
Z^{--}_{\phi}(k,t,t') &=& Z^{F}_{\phi}(k,t,t')- \frac{1}{2}{\rm
sgn}(t-t^\prime)\imath Z^{c}_{\phi}(k,t,t')
\label{reduction:selfMass:MF+Mcd}\,,
\end{eqnarray}
\end{subequations}
such that we find from (\ref{SelfMassFourierSpFinalResult}):
\begin{subequations}
\label{selfmasses}
\begin{eqnarray}
Z^{F}_{\phi}(k,t,t') &=& \frac{1}{2} \left[Z^{-+}_{\phi}(k,t,t') + Z^{+-}_{\phi}(k,t,t')\right] \label{selfMass:MFphi}\\
&=& \frac{h^{2}}{64 k \pi^{2}} \left[ \cos(k\Delta
t)\left(\gamma_{\mathrm{E}} + \log\left[ \frac{k}{2 \mu^{2}|\Delta
t|} \right]+ \mathrm{ci}(2k|\Delta t|) \right)+\sin(k|\Delta
t|)\left(\mathrm{si}(2k|\Delta t|) - \frac{\pi}{2}\right)\right]
\nonumber \\
Z^{c}_{\phi}(k,t,t') &=& \imath \left[Z^{+-}_{\phi}(k,t,t')
-  Z^{-+}_{\phi}(k,t,t')\right]\label{selfMass:Mcphi}\\
&=& \frac{h^{2}}{64 k \pi^{2}} \left[ -2 \cos(k\Delta
t)\mathrm{sgn}(\Delta t) \left( \mathrm{si}(2k|\Delta t|) +
\frac{\pi}{2} \right)+2\sin(k\Delta t)\left( \mathrm{ci}(2k|\Delta
t|) - \gamma_{\mathrm{E}} - \log\left[ \frac{k}{2 \mu^{2}|\Delta
t|} \right]  \right)\right] \nonumber \,.
\end{eqnarray}
\end{subequations}
The expressions for $\chi$ differ due to (\ref{selfMass}). We can
derive a system of two closed equations for the causal and
statistical propagator by adding and subtracting equations
(\ref{EOM4a-+}) and (\ref{EOM4a+-}). In order to obtain the
equation of motion for the causal propagator
(\ref{causalpropagator}), we subtract (\ref{EOM4a+-}) from
(\ref{EOM4a-+}) to find:
\begin{eqnarray}
(\partial_{t}^{2}+k^{2}+m^{2}_{\phi}) \imath
\Delta^{c}_{\phi}(k,t,t') &+& \frac{1}{2} \int_{-\infty}^{\infty}
\mathrm{d}t_{1} \Bigg[\left\{\imath M^{-+}_{\phi}(k,t,t_{1})
-\imath M^{+-}_{\phi}(k,t,t_{1})\right\} \mathrm{sgn}(t_{1}-t')
\nonumber \\
&& \qquad\qquad\qquad +\imath
M^{++}_{\phi,\mathrm{ren}}(k,t,t_{1}) - \imath
M^{--}_{\phi,\mathrm{ren}}(k,t,t_{1})\Bigg]
\imath\Delta^{c}_{\phi}(k,t_{1},t') = 0 \label{eomcausalprop1} \,.
\end{eqnarray}
Using equations (\ref{SelfMassFourierSpFinalResult}) and
(\ref{reduction:selfMass:MF+Mc}) we find:
\begin{equation}
(\partial_{t}^{2}+k^{2}+m^{2}_{\phi}) \Delta^{c}_{\phi}(k,t,t') -
\left(\partial_{t}^{2}+k^{2}\right) \int_{t'}^{t} \mathrm{d}t_{1}
Z^{c}_{\phi}(k,t,t_{1}) \Delta^{c}_{\phi}(k,t_{1},t')=0
\label{eomcausalprop2} \,.
\end{equation}
Note that equation~(\ref{eomcausalprop2}) is causal, in the sense
that no knowledge in the future of the maximum of $t,t'$ is needed
to specify $\imath \Delta^{c}_{\phi} (k,t,t')$. Moreover, at one
loop the evolution of $\imath \Delta^{c}_{\phi}$ requires only
knowledge of the Green's functions in the time interval between
$t^\prime$ and $t$, and is thus independent of the initial
conditions at $t_{0}=-\infty$. Finally note that we deleted the
$\imath$ in front of $\imath \Delta^{c}_{\phi}$ in the equation of
motion above to stress that $\Delta^{c}_{\phi}$ is real to prepare
this equation for numerical integration.

In order to get an equation for the statistical Hadamard function
(\ref{statisticalpropagator}), we add equation (\ref{EOM3a+-}) to
(\ref{EOM3a-+}) to get:
\begin{eqnarray}
&& (\partial_{t}^{2}+k^{2}+m^{2}_{\phi}) F_{\phi}(k,t,t')
\label{eomstatprop1}\\
&& + \frac{1}{2} \int_{-\infty}^{\infty} \mathrm{d}t_{1}
\left[\imath M^{++}_{\phi,\mathrm{ren}}(k,t,t_{1}) -\imath
M^{+-}_{\phi}(k,t,t_{1}) + \imath M^{-+}_{\phi}(k,t,t_{1}) -
\imath M^{--}_{\phi,\mathrm{ren}}(k,t,t_{1})\right]
F_{\phi}(k,t_{1},t') \nonumber \\
&& +\frac{1}{4} \int_{-\infty}^{\infty}  \mathrm{d}t_{1}
\left[\left\{ \imath M^{+-}_{\phi}(k,t,t_{1}) + \imath
M^{-+}_{\phi}(k,t,t_{1}) \right\}\mathrm{sgn}(t_{1}-t') - \imath
M^{++}_{\phi,\mathrm{ren}}(k,t,t_{1}) - \imath
M^{--}_{\phi,\mathrm{ren}}(k,t,t_{1})\right]  \imath
\Delta^{c}_{\phi}(k,t_{1},t') = 0 \nonumber .
\end{eqnarray}
Again using (\ref{SelfMassFourierSpFinalResult}) and
(\ref{reduction:selfMass:MF+Mc}) we find the relevant differential
equation for the statistical propagator:
\begin{equation}
(\partial_{t}^{2}+k^{2}+m^{2}_{\phi}) F_{\phi}(k,t,t') -
\left(\partial_{t}^{2}+k^{2}\right)\left[ \int_{-\infty}^{t}
\mathrm{d}t_{1} Z^{c}_{\phi}(k,t,t_{1}) F_{\phi}(k,t_{1},t') -
\int_{-\infty}^{t'} \mathrm{d}t_{1} Z^{F}_{\phi}(k,t,t_{1})
\Delta^{c}_{\phi}(k,t_{1},t') \right]=0 \label{eomstatprop2} \,.
\end{equation}
We can thus say that the equations of motion for the causal and
statistical propagator have decoupled in the following sense: the
differential equations haven been brought in triangular form. Note
that equations (\ref{eomcausalprop2}) and (\ref{eomstatprop2})
together with the causal and statistical self-masses in equation
(\ref{selfmasses}), represent a closed causal system of equations
suitable for integration in terms of an initial value problem.
Given the knowledge of $F$ and $\imath \Delta^c$ for both $\chi$
and $\phi$, all other Green's functions can be reconstructed from
equation (\ref{reduction:F+Deltac}). This strategy was used (see
\cite{Berges:2004yj} and references therein) to study the dynamics
of out-of-equilibrium quantum statistical (scalar and fermionic)
field theories. Indeed, we will solve equations
(\ref{eomcausalprop2}) and (\ref{eomstatprop2}) numerically in the
next section. We emphasise however that the form of equations
(\ref{eomstatprop2}) and (\ref{eomcausalprop2}) differs from the
ones found in \cite{Berges:2004yj}. The renormalised equations of
motion have a different structure than the unrenormalised ones,
which is not taken into account in e.g.
\cite{Berges:2004yj,Berges:2005hc}.

Before doing so, let us show that the one-loop self-masses do not
spoil causality in another way: the retarded and advanced Green's
functions only receive information from the past and future light
cone, respectively. Now subtracting equation (\ref{EOM4a+-}) from
(\ref{EOM4a++}) one obtains:
\begin{equation}
(\partial_{t}^{2}+k^{2}+m^{2}_{\phi})
\imath\Delta^{\mathrm{r}}_{\phi}(k,t,t') +
\int_{-\infty}^{\infty}\mathrm{d}t_{1} \imath
M^{\mathrm{r}}_{\phi,\mathrm{ren}}(k,t,t_{1})
\imath\Delta^{\mathrm{r}}_{\phi}(k,t_{1},t') =
\imath\delta(t-t^\prime) \label{eom:Delta:ret0} \,.
\end{equation}
Making use of equation (\ref{RetardedSelfMassFinalResult}) we
find:
\begin{equation}
\imath M^{\mathrm{r}}_{\phi,\mathrm{ren}}(k,t,t_{1}) = - \left(
\partial^{2}_{t} + k^{2}\right)
\theta(t-t_{1}) Z_{\phi}^{c}(k,t,t_{1}) \label{selfmasscausalret}
\,.
\end{equation}
Equation of motion (\ref{eom:Delta:ret0}) transforms into:
\begin{equation}
(\partial_{t}^{2}+k^{2}+m^{2}_{\phi})
\imath\Delta^{\mathrm{r}}_{\phi}(k,t,t') -
(\partial^{2}_{t}+k^{2}) \int_{-\infty}^{t}\mathrm{d}t_{1}
Z_{\phi}^{c}(k,t,t_{1})
\imath\Delta^{\mathrm{r}}_{\phi}(k,t_{1},t') =
\imath\delta(t-t^\prime) \label{eom:Delta:ret} \,.
\end{equation}
The retarded self-mass gets contributions only from within the
past light cone, i.e.: when $t_{1} < t$.

Similar to equation (\ref{eom:Delta:ret}), we can subtract
equation (\ref{EOM4a-+}) from (\ref{EOM4a++}) to obtain the
equation of motion for the advanced propagator:
\begin{eqnarray}
(\partial_{t}^{2}+k^{2}+m^{2}_{\phi})
\imath\Delta^{\mathrm{a}}_{\phi}(k,t,t') +
\int_{-\infty}^{\infty}\mathrm{d}t_{1} \Bigg[ \left(\imath
M^{++}_{\phi,\mathrm{ren}}(k,t,t_{1}) - \imath
M^{-+}_{\phi}(k,t,t_{1}) \right)
\imath\Delta^{++}_{\phi}(k,t_{1},t') \qquad \phantom{1} && \nonumber \\
- \left(\imath M^{+-}_{\phi}(k,t,t_{1}) - \imath
M^{--}_{\phi,\mathrm{ren}}(k,t,t_{1}) \right)
\imath\Delta^{-+}_{\phi}(k,t_{1},t')\Bigg] &=&
\imath\delta(t-t^\prime)  \label{eom:Delta:adv0} \,.
\end{eqnarray}
This yields:
\begin{equation}
(\partial_{t}^{2}+k^{2}+m^{2}_{\phi})
\imath\Delta^{\mathrm{a}}_{\phi}(k,t,t') -
(\partial^{2}_{t}+k^{2}) \int_{t}^{\infty}\mathrm{d}t_{1}
Z_{\phi}^{c}(k,t,t_{1})
\imath\Delta^{\mathrm{a}}_{\phi}(k,t_{1},t') =
\imath\delta(t-t^\prime) \label{eom:Delta:adv} \,,
\end{equation}
where we find an analogous relation for the advanced self-mass:
\begin{equation}
\imath M^{\mathrm{a}}_{\phi,\mathrm{ren}}(k,t,t_{1}) = -
 \left(
\partial^{2}_{t} + k^{2}\right)
\theta(t_{1}-t)Z_{\phi}^{c}(k,t,t_{1}) \label{selfmasscausaladv}
\,.
\end{equation}
As expected, $\imath
M^{\mathrm{a}}_{\phi,\mathrm{ren}}(k,t,t_{1})$ acquires
contributions from the future only, i.e.: when $t_{1}>t$. Rather
than solving for the causal propagator, we could alternatively
solve for the retarded propagator or the advanced propagator. We
will however not pursue this in the present work.

\section{Numerically Solving the Kadanoff-Baym Equations}
\label{Numerically Solving the Kadanoff-Baym equations}

Let us once more explicitly write down the equations of motion of
the causal and statistical propagators (\ref{eomcausalprop2}) and
(\ref{eomstatprop2}) we will numerically tackle in this section:
\begin{subequations}
\label{eomnumerical}
\begin{eqnarray}
(\partial_{t}^{2}+k^{2}+m^{2}_{\phi}) \Delta^{c}_{\phi}(k,t,t') -
\left(\partial_{t}^{2}+k^{2}\right) \int_{t'}^{t} \mathrm{d}t_{1}
Z^{c}_{\phi}(k,t,t_{1})
\Delta^{c}_{\phi}(k,t_{1},t')\!&=&\!0 \label{eomnumericala}\\
(\partial_{t}^{2}+k^{2}+m^{2}_{\phi}) F_{\phi}(k,t,t') -
\left(\partial_{t}^{2}+k^{2}\right) \left[ \int_{-\infty}^{t}
\mathrm{d}t_{1} Z^{c}_{\phi}(k,t,t_{1}) F_{\phi}(k,t_{1},t') -
\int_{-\infty}^{t'} \mathrm{d}t_{1} Z^{F}_{\phi}(k,t,t_{1})
\Delta^{c}_{\phi}(k,t_{1},t') \right]\!&=& \! 0 .\qquad
\label{eomnumericalb}
\end{eqnarray}
\end{subequations}
The causal and statistical self-masses are given in equation
(\ref{selfmasses}). In particular, we will be interested in two
cases:
\begin{subequations}
\label{massbehaviour}
\begin{eqnarray}
m_{\phi}(t) &=& m_{0} = \mathrm{const} \label{massbehavioura} \\
m^{2}_{\phi}(t) &=& A+B\tanh(\rho t-50) \label{massbehaviourb} \,,
\end{eqnarray}
\end{subequations}
where we let $A$ and $B$ take different values. Let us take a
closer look at the two equations of motion above. Clearly, we
first need to determine the causal propagator. Note that equations
of motion (\ref{eomnumerical}) depend on two variables, i.e.: for
each $t'$, we have to solve this equation of
motion\footnote{Alternatively, we could have written down the
equations of motion of the causal and statistical propagator where
the operator acts on the other leg of the propagator, on $t'$.
Then we would have to solve these four equations of motion
simultaneously. Needless to say the two methods are completely
equivalent.}. The self-mass corrections contribute only through a
``memory kernel'' (memory integral over time) between $t'$ and
$t$. The boundary conditions for determining the causal propagator
are as follows:
\begin{subequations}
\label{boundaryconditionscausual}
\begin{eqnarray}
\Delta^{c}_{\phi}(t,t) &=& 0 \label{boundaryconditionscausuala} \\
\partial_{t} \Delta^{c}_{\phi}(t,t')|_{t=t'} &=& -1
\label{boundaryconditionscausualb} \,,
\end{eqnarray}
\end{subequations}
Condition (\ref{boundaryconditionscausuala}) has to be satisfied
by definition and condition (\ref{boundaryconditionscausualb})
follows from the Wronskian normalisation condition due to the
commutation relations.

Once we have solved for the causal propagator, we turn our
attention to the second equation (\ref{eomnumericalb}). Suppose we
would not have sent $t_{0} \rightarrow - \infty$. The equation for
the statistical propagator then would have been of the following
form:
\begin{equation}
(\partial_{t}^{2}+k^{2}+m^{2}_{\phi}) F_{\phi}(k,t,t') -
\left(\partial_{t}^{2}+k^{2}\right) \left[ \int_{t_{0}}^{t}
\mathrm{d}t_{1} Z^{c}_{\phi}(k,t,t_{1}) F_{\phi}(k,t_{1},t') -
\int_{t_{0}}^{t'} \mathrm{d}t_{1} Z^{F}_{\phi}(k,t,t_{1})
\Delta^{c}_{\phi}(k,t_{1},t') \right]\!= \! 0 .
\label{eomnumericalbALT}
\end{equation}
Clearly, equations (\ref{eomnumericalbALT}) and
(\ref{eomnumericalb}) are not equivalent. Equation
(\ref{eomnumericalb}) contains a memory kernel from the infinite
past up to $t$ and $t'$, whereas equation (\ref{eomnumericalbALT})
only contains memory kernels from $t_{0}$ onwards. This
corresponds to an interaction that is switched on
non-adiabatically at time $t_{0}$. To understand this, consider
replacing the coupling constant $h$ hidden in the self-masses
(\ref{selfmasses}) with\footnote{Note that $t>t_{1}$ by
construction.}:
\begin{equation}
h \rightarrow h\theta(t_{1}-t_{0}) \label{timedeph}\,.
\end{equation}
The step-function would then have transformed equation
(\ref{eomnumericalb}) to (\ref{eomnumericalbALT}) which mimics
switching on the interaction between the two fields at some finite
time $t_{0}$. The two standard Schwinger-Keldysh contours
presented in figures \ref{fig:schwingercontour1} and
\ref{fig:schwingercontour2} are thus not equivalent in interacting
quantum field theories where memory effects play an important
role. Alternatively, we could say that non-locality, generic for
any interacting quantum field theory, enforces the memory kernel
to start at the infinite past. This effect has, in the context of
electromagnetic radiation, been recognised and investigated by
Serreau \cite{Serreau:2003wr}. In the work of Borsanyi and Reinosa
\cite{Borsanyi:2008ar, Borsanyi:2009zza} the memory integral,
extended to negative infinity, plays an important role too. They
suggest to use that in connection with a generalised
dissipation-fluctuation theorem.

Needless to say, we have to start at some finite time in our
numerical analysis. We therefore make the assumption to
approximate the propagators in the memory kernels from the
negative past to $t_{0}$ with the free propagators:
\begin{eqnarray}
(\partial_{t}^{2}+k^{2}+m^{2}_{\phi}) F_{\phi}(k,t,t') -
\left(\partial_{t}^{2}+k^{2}\right) &\Bigg[&
\int_{-\infty}^{t_{0}} \mathrm{d}t_{1} Z^{c}_{\phi}(k,t,t_{1})
F_{\phi}^{\mathrm{free}}(k,t_{1},t') + \int_{t_{0}}^{t}
\mathrm{d}t_{1} Z^{c}_{\phi}(k,t,t_{1}) F_{\phi}(k,t_{1},t')
\label{eomnumericalc}\\
&& - \int_{-\infty}^{t_{0}} \mathrm{d}t_{1}
Z^{F}_{\phi}(k,t,t_{1})
\Delta^{c,\mathrm{free}}_{\phi}(k,t_{1},t') - \int_{t_{0}}^{t'}
\mathrm{d}t_{1} Z^{F}_{\phi}(k,t,t_{1})
\Delta^{c}_{\phi}(k,t_{1},t') \Bigg]\! = \! 0 \nonumber,
\end{eqnarray}
where $F_{\phi}^{\mathrm{free}}(k,t_{1},t')$ and
$\Delta^{c,\mathrm{free}}_{\phi}(k,t_{1},t')$ are the free
propagators obtained in equation (\ref{propagatorconstantmassF}).
This approximation induces an error of the order
$\mathcal{O}(h^{4}/\omega_{\phi}^{4})$. An alternative approach
has been outlined in \cite{Garny:2009ni} where, for
$\lambda\phi^{4}(x)$ theory, non-Gaussian initial conditions at
$t_{0}$ are imposed. We can explicitly evaluate the infinite past
memory kernel:
\begin{eqnarray}
M_{F}^{\mathrm{free}}(k,t,t',t_{0})&=&
\left(\partial_{t}^{2}+k^{2}\right) \int_{-\infty}^{t_{0}}
\mathrm{d}t_{1} \left[ Z^{c}_{\phi}(k,t,t_{1})
F_{\phi}^{\mathrm{free}}(k,t_{1},t') - Z^{F}_{\phi}(k,t,t_{1})
\Delta^{c,\mathrm{free}}_{\phi}(k,t_{1},t') \right] \nonumber\\
&=& \frac{h^{2}}{32 \pi^{2} \omega_{\mathrm{in}}}
\int_{-\infty}^{t_{0}} \mathrm{d}t_{1}
\frac{\cos(kt+\omega_{\mathrm{in}}t'-(k+\omega_{\mathrm{in}})t_{1})}{t-t_{1}}\label{freememoryintegral1}\,.
\end{eqnarray}
We change variables to $\tau=t-t_{1}$ to find:
\begin{eqnarray}
M_{F}^{\mathrm{free}}(k,t,t',t_{0})&=& \frac{h^{2}}{32 \pi^{2}
\omega_{\mathrm{in}}} \int_{t-t_{0}}^{\infty} \mathrm{d}\tau
\frac{\cos((k+\omega_{\mathrm{in}})\tau -\omega_{\mathrm{in}}(
t-t'))}{\tau}\label{freememoryintegral2}\\
&=& - \frac{h^{2}}{32 \pi^{2}
\omega_{\mathrm{in}}}\left[\cos(\omega_{\mathrm{in}}(t-t'))
\mathrm{ci}((k+\omega_{\mathrm{in}})(t-t_{0})) +
\sin(\omega_{\mathrm{in}}(t-t'))
\mathrm{si}((k+\omega_{\mathrm{in}})(t-t_{0}))\right] \nonumber\,.
\end{eqnarray}
We postpone the discussion of imposing boundary conditions for
$F_{\phi}$ at $t_{0}$ to subsection \ref{Constant Mass Solutions}.
Equation (\ref{eomnumerical}) transforms into:
\begin{subequations}
\label{EOMSYSTEM}
\begin{eqnarray}
(\partial_{t}^{2}+k^{2}+m^{2}_{\phi}) \Delta^{c}_{\phi}(k,t,t') -
\left(\partial_{t}^{2}+k^{2}\right) \int_{t'}^{t} \mathrm{d}t_{1}
Z^{c}_{\phi}(k,t,t_{1})
\Delta^{c}_{\phi}(k,t_{1},t')\!&=&\!0 \label{EOMSYSTEMa}\\
(\partial_{t}^{2}+k^{2}+m^{2}_{\phi}) F_{\phi}(k,t,t') -
M_{F}^{\mathrm{free}}(k,t,t',t_{0}) \qquad\qquad\qquad\qquad\qquad\qquad\qquad\qquad && \label{EOMSYSTEMb} \\
- \left(\partial_{t}^{2}+k^{2}\right) \left[ \int_{t_{0}}^{t}
\mathrm{d}t_{1} Z^{c}_{\phi}(k,t,t_{1}) F_{\phi}(k,t_{1},t') -
\int_{t_{0}}^{t'} \mathrm{d}t_{1} Z^{F}_{\phi}(k,t,t_{1})
\Delta^{c}_{\phi}(k,t_{1},t') \right]\!&=& \! 0 \nonumber \,,
\end{eqnarray}
\end{subequations}
We have now prepared the problem for numerical integration. In the
numerical code, we take $t_{0}=0$ and we let $\rho t$ and $\rho
t'$ run between 0 and 100. In order to solve differential equation
(\ref{EOMSYSTEMb}), we thus need to evaluate two more memory
kernels. The one involving the causal propagator can be computed
immediately. Once we have solved for the statistical propagator,
our life becomes much easier as we can immediately find the phase
space area via relation (\ref{deltaareainphasespace}). The phase
space area fixes the entropy.

Differential equation (\ref{EOMSYSTEMb}) merits another remark. If
we let $t\rightarrow t_{0}$, we encounter a logarithmic divergence
in $M_{F}^{\mathrm{free}}(k,t,t',t_{0})$ as $\mathrm{ci}(x)
\propto \log(x)$ as $x\rightarrow 0$. This divergence is only
apparent. Intuitively, this should of course be the case as we
introduced the boundary time $t_{0}$ by hand and no divergences
should arise consequently. If $h=\mathrm{const}$, the time $t_{0}$
is introduced as a fictitious time, hence observables cannot
depend on $t_{0}$. Of course, neglecting the memory integral from
negative past infinity to $t_{0}$ introduces a dependence on
$t_{0}$. Thus, removing the distant memory integrals completely is
equivalent to setting $h \rightarrow h\theta(t_{1}-t_{0})$ as in
equation (\ref{timedeph}). We will prove that this logarithmic
divergence is only apparent rigorously by rewriting equation
(\ref{EOMSYSTEMa}) for the causal propagator and
(\ref{EOMSYSTEMb}) for the statistical propagator in a different
form, and by using the symmetry properties of the propagators.
Focussing first on the equation of motion for the causal
propagator (\ref{EOMSYSTEMa}), note that we can transfer the $t$
derivative to a $t_{1}$ derivative by using the fact that the
causal self-mass (\ref{selfMass:Mcphi}) is a function of $\Delta
t=t-t_{1}$ only:
\begin{eqnarray}
\partial_{t}^{2} \int_{t'}^{t} \mathrm{d}t_{1}
Z^{c}_{\phi}(k,t,t_{1}) \Delta^{c}_{\phi}(k,t_{1},t') &=& -
\partial_{t}\left[ \int_{t'}^{t} \mathrm{d}t_{1}
\partial_{t_{1}} Z^{c}_{\phi}(k,t,t_{1})
\Delta^{c}_{\phi}(k,t_{1},t')\right]
\label{eomnumerical2} \\
&=&\partial_{t}\left[ \int_{t'}^{t} \mathrm{d}t_{1}
 Z^{c}_{\phi}(k,t,t_{1}) \partial_{t_{1}}
\Delta^{c}_{\phi}(k,t_{1},t')\right]\nonumber \\
&=& - Z^{c}_{\phi}(k,t,t') + \int_{t'}^{t} \mathrm{d}t_{1}
 Z^{c}_{\phi}(k,t,t_{1}) \partial_{t_{1}}^{2}
\Delta^{c}_{\phi}(k,t_{1},t') \nonumber \,,
\end{eqnarray}
where we partially integrated in the second line (the boundary
terms vanish by virtue of equation
(\ref{boundaryconditionscausuala}) and $Z^{c}_{\phi}(k,t,t) = 0$),
and we used the commutation relations in the third. We transform
the equation of motion of the statistical propagator
(\ref{EOMSYSTEMb}) analogously to find:
\begin{subequations}
\label{eomnumerica3}
\begin{eqnarray}
(\partial_{t}^{2}+k^{2}+m^{2}_{\phi}) \Delta^{c}_{\phi}(k,t,t') +
Z^{c}_{\phi}(k,t,t') - \int_{t'}^{t} \mathrm{d}t_{1}
Z^{c}_{\phi}(k,t,t_{1}) \left(\partial_{t_{1}}^{2} +k^{2}\right)
\Delta^{c}_{\phi}(k,t_{1},t')   && = 0 \label{eomnumerica3a}  \\
(\partial_{t}^{2} + k^{2} + m^{2}_{\phi}) F_{\phi}(k,t,t') -
M_{F}^{\mathrm{free}}(k,t,t',t_{0}) -
\Big[\partial_{t}Z^{c}_{\phi}(k,t,t_{0}) F_{\phi}(k,t_{0},t') -
\partial_{t}Z^{F}_{\phi}(k,t,t_{0})
\Delta^{c}_{\phi}(k,t_{0},t')\,\,\phantom{1}
 && \nonumber \\
- Z^{F}_{\phi}(k,t,t')  + Z^{c}_{\phi}(k,t,t_{0}) \partial_{t_{0}}
F_{\phi}(k,t_{0},t') - Z^{F}_{\phi}(k,t,t_{0})
\partial_{t_{0}} \Delta^{c}_{\phi}(k,t_{0},t')\Big] && \nonumber \\
- \int_{t_{0}}^{t}  \mathrm{d}t_{1} Z^{c}_{\phi}(k,t,t_{1})
 \left(\partial_{t_{1}}^{2} + k^{2}\right) F_{\phi}(k,t_{1},t')+\int_{t_{0}}^{t'}  \mathrm{d}t_{1} Z^{F}_{\phi}(k,t,t_{1})
\left(\partial_{t_{1}}^{2} + k^{2}\right)
\Delta^{c}_{\phi}(k,t_{1},t') && = 0 \,. \phantom{1}
\label{eomnumerica3b} \quad\phantom{1}
\end{eqnarray}
\end{subequations}
These two differential equations should be completely equivalent
to equation (\ref{EOMSYSTEM}). In fact, we have found a
non-trivial test of our numerical code: the results of equation
(\ref{EOMSYSTEM}) and of the two equations above should agree. We
will show this in due course.

Now, we can see another logarithmic divergence appearing in
$\partial_{t}Z^{c}_{\phi}(k,t,t_{0})$ in equation
(\ref{eomnumerica3b}) when we send $t~\rightarrow~t_{0}$. The
reader can easily verify that the logarithmic divergences in
$M_{F}^{\mathrm{free}}(k,t,t',t_{0})$ and
$\partial_{t}Z^{c}_{\phi}(k,t,t_{0}) F_{\phi}(k,t_{0},t')$ in
equation (\ref{eomnumerica3b}) above cancel to leave a finite
result when $t\rightarrow t_{0}$ if we set
$2F_{\phi}(k,t_{0},t')=\cos(\omega_{\mathrm{in}}(t_{0}-t'))/\omega_{\mathrm{in}}$.
We thus find that at order $\mathcal{O}(h^{2}/\omega^{2}_{\phi})$
no divergences at $t_{0}$ remain and we expect that a similar
treatment would cure these type of apparent divergences at higher
order: clearly, $t_{0}$ has been introduced by hand so this should
not lead to any irregularities.

Let us finally make some remarks about the literature. The authors
of \cite{Collins:2003mj, Collins:2005nu, Collins:2006bg,
Collins:2006uy} study out-of-equilibrium $\lambda\phi^{4}(x)$.
They encounter, after renormalisation, a residual divergence in
their theory at the surface of initial boundary conditions at
$t_{0}$ which they choose to renormalise separately. We differ in
their approach as we do not find these residual divergences. The
infinite past memory kernel precisely takes care of these as can
be appreciated from the previous discussion. This is also the case
in the approach of \cite{Borsanyi:2008ar, Borsanyi:2009zza}
mentioned before.

\subsection{Constant Mass Solutions}
\label{Constant Mass Solutions}

The constant mass case is interesting as we can make non-trivial
statements based on some analytical calculations. The bottom line
is that the generated entropy is constant. The argument is rather
simple. When $m_{\phi}=\mathrm{const}$, we have:
\begin{equation}
F_{\phi}(k,t,t') = F_{\phi}(k,t-t') \label{Fconstantmass} \,.
\end{equation}
Using a few Fourier transforms (\ref{Fouriertransformdef2}), we
have:
\begin{subequations}
\label{Fconstantmass2}
\begin{eqnarray}
F_{\phi}(k,0) &=& \int_{-\infty}^{\infty} \frac{\mathrm{d}k^{0}}{2\pi} F_{\phi}(k^{\mu}) \label{Fconstantmass2a} \\
\left.\partial_{t} F_{\phi}(k, \Delta t) \right|_{\Delta t =0} &=&
- \imath \int_{-\infty}^{\infty} \frac{\mathrm{d}k^{0}}{2\pi}
k^{0}
F_{\phi}(k^{\mu}) \label{Fconstantmass2b} \\
\left.\partial_{t'}\partial_{t} F_{\phi}(k, \Delta t)
\right|_{\Delta t =0} &=& \int_{-\infty}^{\infty}
\frac{\mathrm{d}k^{0}}{2\pi} k_{0}^{2} F_{\phi}(k^{\mu})
\label{Fconstantmass2c} \,,
\end{eqnarray}
\end{subequations}
The right-hand sides do no longer contain any time dependence.
Hence, the left-hand sides are also time independent. This implies
that the phase space area $\Delta_{k}$ is constant, and so is the
generated entropy. As we insert some finite $t_{0}$, we expect to
observe some transient dependence of the entropy on time because
we approximated the propagators in the infinite past memory kernel
with free propagators. When this behaviour has died out, the
entropy should settle to its constant value derived from equations
(\ref{Fconstantmass2}), (\ref{deltaareainphasespace}) and
(\ref{entropy}). Indeed, this constant entropy does not
necessarily equal 0. We interpret this non-zero entropy as the
entropy generated by the coupling to the second field, which in
the effective action acts as a source for $F_{\phi}$. Effectively,
this opens up phase space for the system field that previously was
inaccessible for it. More accessible phase space for the system
field in turn, implies that less information about the system
field is accessible to us and hence we observe an increase in
entropy.

\begin{figure}[t!]
    \begin{minipage}[t]{.43\textwidth}
        \begin{center}
\includegraphics[width=\textwidth]{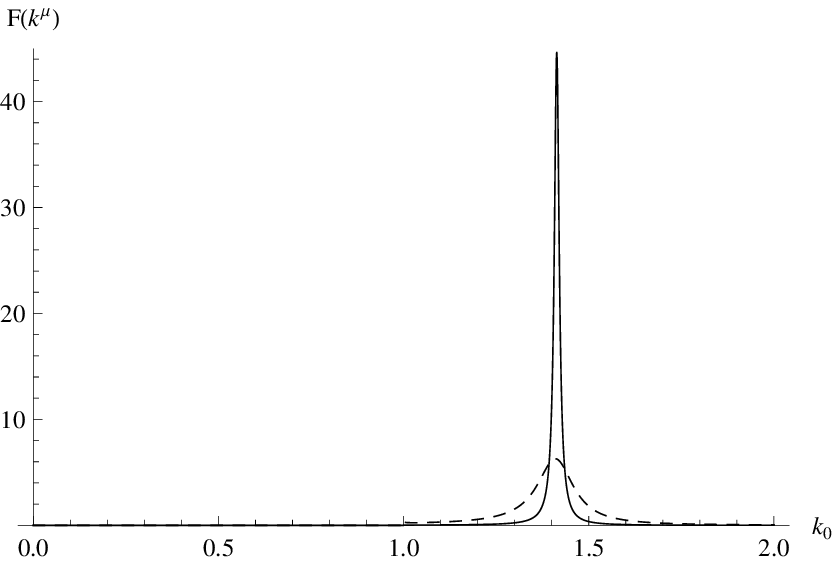}
   {\em \caption{Statistical propagator in Fourier space for a small
   coupling $h/\rho=3/2$ (black) and a larger one $h/\rho=4$ (dashed). Because of a non-zero coupling, we
   observe that the $\delta$-function, present in the original dispersion relation, has
   broadened to a ``quasi-particle peak'', roughly of a Breit-Wigner form. If the coupling increases, the ``quasi particle peak''
   broadens further. Clearly, when $h \gg \omega_{\phi}$ in the strongly
   coupled regime, we have a ``collection of quasi particles''. We used $k/\rho=1$,
   $m_{\phi}/\rho=1$ and $\mu/\rho=1$.
    \label{fig:FWigner}}}
        \end{center}
    \end{minipage}
\hfill
    \begin{minipage}[t]{.43\textwidth}
        \begin{center}
\includegraphics[width=\textwidth]{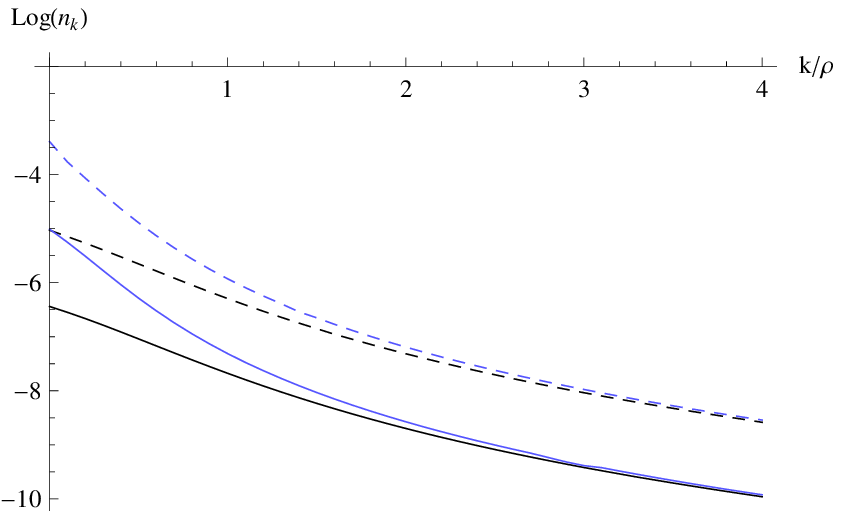}
   {\em \caption{Statistical particle number density $n_{k}$ as a function
   of $k/\rho$. We
   used: $h/\rho=1$ and $m_{\phi}/\rho=1$ (black), $h/\rho=2$ and $m_{\phi}/\rho=1$ (black
   dashed), $h/\rho=1$ and $m_{\phi}/\rho=0.5$ (blue) and $h/\rho=2$ and $m_{\phi}/\rho=0.5$
   (blue
   dashed). In the UV $n_{k}$ vanishes
   irrespective of the value of $h/\rho$ or $m_{\phi}/\rho$. Particles are only produced by the interaction in the statistical sense in the IR.
   The mass only influences the IR behaviour.
   Moreover, one can show by appropriate rescalings of the
   functions above that $n_{k}$ is given by equation
   (\ref{particlenumberfunction}) in the UV.
   \label{fig:particlenumber}}}
        \end{center}
    \end{minipage}
\end{figure}
In order to evaluate the integrals above, we have derived the
statistical propagator in Fourier space in appendix \ref{The
Statistical Propagator in Fourier Space}. The result is:
\begin{eqnarray}
F_{\phi}(k^{\mu}) &=& - \frac{\imath}{2} \mathrm{sgn}(k^{0})
\theta(k_{0}^{2}-k^{2})
\Bigg[\frac{1}{k_{\mu}k^{\mu}+m_{\phi}^{2}+\frac{h^{2}}{32
\pi^{2}}\left(\log\left(\frac{|k_{\mu}k^{\mu}|}{4\mu^{2}}\right)+2\gamma_{\mathrm{E}}
\right)- \frac{\imath
h^{2}}{32\pi}\mathrm{sgn}(k^{0})\theta(k_{0}^{2}-k^{2})} \nonumber \\
&& \qquad\qquad\qquad\qquad\qquad -
\frac{1}{k_{\mu}k^{\mu}+m_{\phi}^{2}+\frac{h^{2}}{32
\pi^{2}}\left(\log\left(\frac{|k_{\mu}k^{\mu}|}{4\mu^{2}}\right)+2\gamma_{\mathrm{E}}
\right)+ \frac{\imath
h^{2}}{32\pi}\mathrm{sgn}(k^{0})\theta(k_{0}^{2}-k^{2})} \Bigg]
\label{FourierStatisticalA} \,.
\end{eqnarray}
To gain some intuitive understanding, we depicted the statistical
propagator for $h/\rho=3/2$ and for $h/\rho=4$ in figure
\ref{fig:FWigner}. When $h/\rho=0$, we have a $\delta$-function
dispersion relation as usual but in the presence of a non-zero
coupling, the $\delta$-function broadens to a so-called ``quasi
particle peak'' of a Breit-Wigner form. For $h/\rho=4$, this peak
is still well pronounced, but when we enter the strongly coupled
regime, this simple picture breaks down when the resonance becomes
broad and we can no longer sensibly talk about a ``quasi
particle'', but rather we should think of a ``collection of quasi
particles''.

The integrals in equation (\ref{Fconstantmass2}) can now be
evaluated numerically to yield the appropriate initial conditions.
For example when $k/\rho=1$, $m_{\phi}/\rho=1$ and $h/\rho=4$, we
find:
\begin{subequations}
\label{Fconstantmass3}
\begin{eqnarray}
\left. F_{\phi}(k/\rho=1,\Delta t)\right|_{\Delta t =0} &=&
0.35196
\label{Fconstantmass3a} \\
\left.\partial_{t} F_{\phi}(k/\rho=1, \Delta t) \right|_{\Delta t
=0} &=& 0 \label{Fconstantmass3b} \\
\left.\partial_{t'}\partial_{t} F_{\phi}(k/\rho=1, \Delta t)
\right|_{\Delta t =0} &=& 0.73120  \label{Fconstantmass3c} \,,
\end{eqnarray}
\end{subequations}
Clearly, equation (\ref{Fconstantmass3b}) holds for all values of
$k$, $m_{\phi}$ and $h$ as the integrand is an odd function of
$k^{0}$, which can be appreciated from equations
(\ref{Fconstantmass2b}) and (\ref{FourierStatisticalA}). The
numerical value of the phase space area in this case follows from
(\ref{Fconstantmass3}) and (\ref{deltaareainphasespace}) as:
\begin{equation}
\Delta_{\mathrm{ms}} = 1.01461 > 1 \label{DeltaconstantMass} \,.
\end{equation}
Hence also:
\begin{equation}
S_{\mathrm{ms}} = 0.04327 > 0 \label{SconstantMass} \,,
\end{equation}
where the subscript ms is an abbreviation for ``mixed state''.

From the phase space area $\Delta_{\mathrm{ms}}$ we can easily
obtain the statistical particle number density
(\ref{particlenumber}). It is interesting to study its behaviour
as a function of $k$. Figure \ref{fig:particlenumber} clearly
shows that in the deep UV (ultraviolet) the particle number
density vanishes: the interaction between the two fields only
produces particles (in the statistical sense) in the IR. Moreover,
using figure \ref{fig:particlenumber}, we can show:
\begin{equation}
n_{k}(h,m_{\phi}) \rightarrow n_{\mathrm{UV}}\left(\frac{h}{k}
\right)= \zeta \frac{h^{2}}{k^{2}} \label{particlenumberfunction}
\,,
\end{equation}
in the deep UV. In fact, we can estimate the constant of
proportionality $\zeta$ appearing in equation
(\ref{particlenumberfunction}) as $\zeta \simeq 0.0008$ which
turns out to be insensitive to the value of the mass of the system
field $m_{\phi}$ and the coupling $h$. The mass only influences
the IR behaviour, as expected, which can also be appreciated from
figure \ref{fig:particlenumber}. Note finally that the formal
divergence of derived quantities, such as the total particle
number per volume $N/V=\int \mathrm{d}^{3}k/(2\pi)^{3} \, n_{k}$
or the total entropy per volume $S/V=\int
\mathrm{d}^{3}k/(2\pi)^{3} \, S_{k}$ does not pose any problems
for the dynamics we are about to solve since these quantities do
not enter the equations of motion.

So far, we postponed the discussion of imposing boundary
conditions for numerically determining the statistical propagator.
We just proved that, independently of how one imposes initial
conditions, the phase space area should settle to a constant value
and for a specific choice of parameters, we have been able to
calculate this constant in equation (\ref{DeltaconstantMass}). One
could think of at least two separate ways of imposing boundary
conditions: ``pure state initial conditions'' and what we will
henceforth refer to as ``mixed state initial conditions''. If we
constrain the statistical propagator to occupy the minimal allowed
phase space area initially, we set:
\begin{subequations}
\label{PSboundaryconditionsstat}
\begin{eqnarray}
F_{\phi}(t_{0},t_{0}) &=& \frac{1}{2 \omega_{\mathrm{in}}} \label{PSboundaryconditionsstata} \\
\partial_{t} F_{\phi}(t,t_{0})|_{t=t_{0}} &=& 0
\label{PSboundaryconditionsstatb} \\
\partial_{t'}\partial_{t} F_{\phi}(t,t')|_{t=t'=t_{0}} &=&
\frac{\omega_{\mathrm{in}}}{2} \label{PSboundaryconditionsstatc}
\,,
\end{eqnarray}
\end{subequations}
and $\omega_{\mathrm{in}}$ is determined from equation
(\ref{solomegain}). This yields:
\begin{equation}
\Delta_{k}(t_{0}) = 1 \label{PSboundaryconditionsstatd} \,,
\end{equation}
such that $S_{k}(t_{0})=0$. Physically, this means that despite
the fact that interactions enlarge the accessible phase space of
the system field, we force it to occupy a minimal area initially
and let it evolve\footnote{If we would not include the infinite
past memory kernel $M_{F}^{\mathrm{free}}(k,t,t',t_{0})$ and
indeed consider a coupling between two fields that is switched on
non-adiabatically at some finite time $t_{0}$ as previously
discussed, the pure state initial condition would be the natural
choice for this problem.}. Alternatively, we can impose mixed
state initial conditions, i.e.: the values calculated from
equations (\ref{Fconstantmass2}) and (\ref{FourierStatisticalA}):
\begin{subequations}
\label{MSboundaryconditionsstat}
\begin{eqnarray}
F_{\phi}(k,t_{0},t_{0}) &=& \int_{-\infty}^{\infty} \frac{\mathrm{d}k^{0}}{2\pi} F_{\phi}(k^{\mu}) \label{MSboundaryconditionsstata} \\
\left.\partial_{t} F_{\phi}(k, t,t') \right|_{t=t'=t_{0}} &=& 0 \label{MSboundaryconditionsstatb} \\
\left.\partial_{t'}\partial_{t} F_{\phi}(k, t,t') \right|_{t
=t'=t_{0}} &=& \int_{-\infty}^{\infty}
\frac{\mathrm{d}k^{0}}{2\pi} k_{0}^{2} F_{\phi}(k^{\mu})
\label{MSboundaryconditionsstatc} \,,
\end{eqnarray}
\end{subequations}
such that initially:
\begin{equation}
\Delta_{k}(t_{0}) = \Delta_{\mathrm{ms}}=\mathrm{const}
\label{MSboundaryconditionsstatd} \,,
\end{equation}
and also $S_{\mathrm{ms}}>0$. Clearly, boundary conditions
(\ref{MSboundaryconditionsstat}) can only be evaluated numerically
for each choice of parameters. Needless to say we are completely
free to impose any other type of initial conditions as well, but
we consider the two cases above to be physically well motivated if
the system is close to its minimum energy state.

\begin{figure}[t!]
    \begin{minipage}[t]{.43\textwidth}
        \begin{center}
\includegraphics[width=\textwidth]{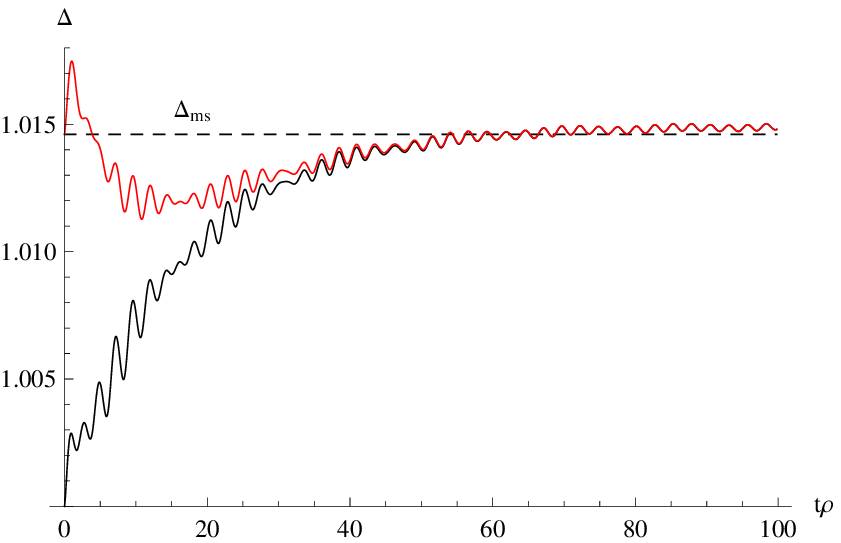}
   {\em \caption{Phase space evolution for constant $m_{\phi}$ for
   pure state initial conditions $\Delta_{k}(t_{0})=1$ (black line) and mixed
   state initial conditions (red line)
   $\Delta_{k}(t_{0})=\Delta_{\mathrm{ms}}$. In both cases the phase space area settles to the
   constant value $\Delta_{\mathrm{ms}}$, indicated by the
   dashed line and calculated in equation (\ref{DeltaconstantMass}). We use $k/\rho=1$,
   $m_{\phi}/\rho=1$, $h/\rho=4$ and $N=2000$.
    \label{fig:AreaConstMass}}}
%
%
\includegraphics[width=\textwidth]{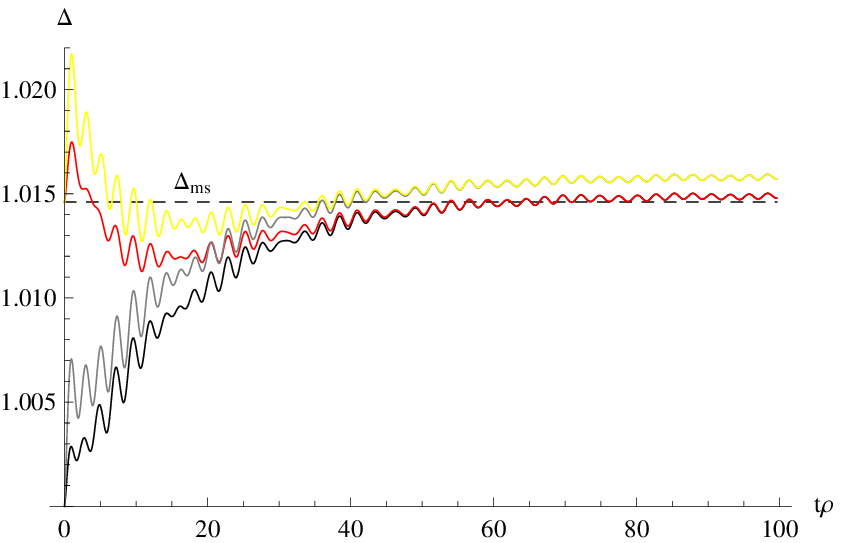}
   {\em \caption{Convergence for the phase space evolution
   presented in figure \ref{fig:AreaConstMass}. The black and red lines
   are identical to the ones in figure \ref{fig:AreaConstMass} and most accurate ($N=2000$).
   The gray and yellow lines are calculated with $N=1000$. The other parameters are kept fixed.
   Clearly, the difference between $\Delta_{\mathrm{ms}}$ and the
   numerical asymptotes decreases as the accuracy increases.
   \label{fig:AreaConstMassAccuracy} }}
        \end{center}
    \end{minipage}
\hfill
    \begin{minipage}[t]{.43\textwidth}
        \begin{center}
\includegraphics[width=\textwidth]{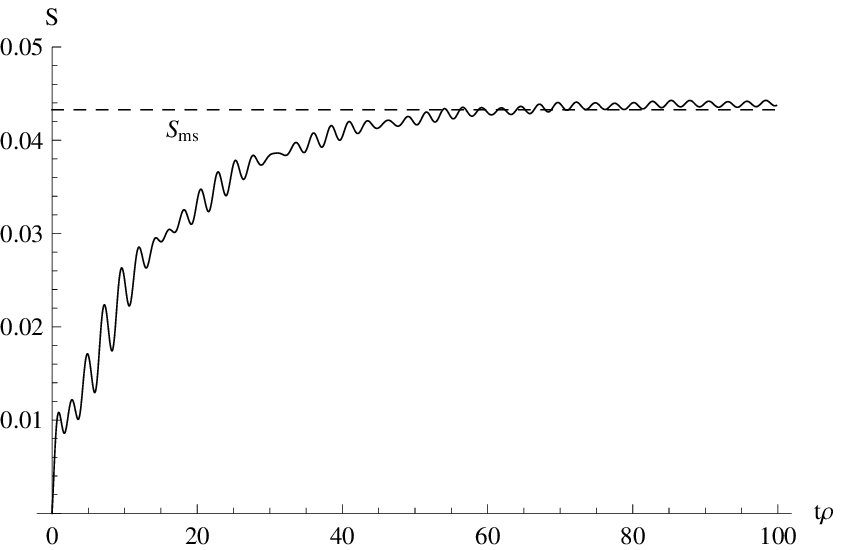}
   {\em \caption{Entropy generation for the system field $\phi$ through interaction with
   the environment $\chi$ in the vacuum state. As both fields are in a vacuum state, the entropy generation is
   relatively small. We used pure state
   initial conditions $S_{k}(t_{0})=0$ and $k/\rho=1$,
   $m_{\phi}/\rho=1$, $h/\rho=4$ and $N=2000$. The entropy settles
   to a constant value $S_{\mathrm{ms}}$ calculated from the value of
   $\Delta_{\mathrm{ms}}$ of figure \ref{fig:AreaConstMass} and
   equation (\ref{entropy}).
   \label{fig:EntropyConstMass} }}
%
%
\includegraphics[width=\textwidth]{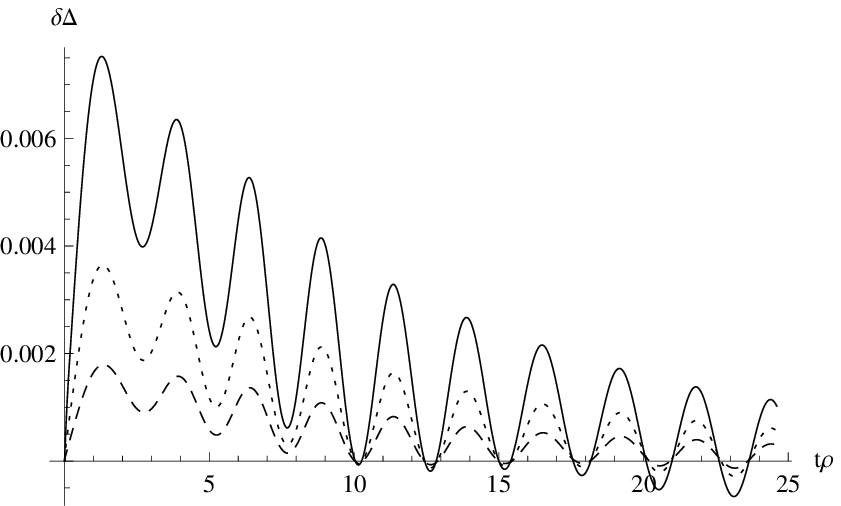}
   {\em \caption{Test of our numerical code. This plot shows the
   difference of the phase space area calculated from equation
   (\ref{EOMSYSTEM}) and (\ref{eomnumerica3}) for different values
   of $N$ but the same values for the other parameters $k/\rho=1$,
   $m_{\phi}/\rho=1$ and $h/\rho=4$ for pure state initial conditions. We used: $N=1000$ at
   $t=100$ (solid line), $N=1000$ at $t=50$ (dotted line) and $N=1000$ at $t=25$ (dashed line).
   The difference between the two methods disappears
   as the accuracy of the numerical evolution increases.
   \label{fig:AreaConstMassDifference} }}
        \end{center}
    \end{minipage}
\end{figure}

In figure \ref{fig:AreaConstMass}, we show the phase space
evolution for both pure state initial conditions (black line) and
mixed state initial conditions (red line). For pure state initial
conditions, the evolution is precisely as anticipated. The phase
space area increases from its minimal area $\Delta_{k}(t_{0})=1$,
to the asymptotic value $\Delta_{\mathrm{ms}}$ calculated in
equation (\ref{DeltaconstantMass}). For mixed state initial
conditions, we firstly observe some transient behaviour which
eventually decays. We then smoothly evolve to
$\Delta_{\mathrm{ms}}$. The initial transient is due to our
assumption to approximate the propagators in the memory kernel
from past infinity to $t_{0}$ with free propagators. As is
apparent from (\ref{freememoryintegral2}), its effect becomes less
important as time elapses.

From the evolution of the phase space area (for pure state initial
conditions), we can immediately find the time evolution of the
entropy $S_{k}(t)$ in figure \ref{fig:EntropyConstMass}. This
shows that entropy has been generated by interaction with an
environment that is in the vacuum state assuming that some
observer is only sensitive to Gaussian correlators. The entropy
eventually settles to its asymptotic value $S_{\mathrm{ms}}$
calculated from $\Delta_{\mathrm{ms}}$ and equation
(\ref{entropy}). As can be anticipated, the generated entropy per
mode is small: both system and environment are in a state close to
the minimum energy state ($T=0$).

We conclude that when the mass $m_{\phi}$ is a constant, no
further entropy is generated if we start with mixed state initial
conditions. However, we do observe a generation of entropy if we
start with pure state initial conditions. This increase in entropy
can be understood by the system's tendency to evolve towards the
vacuum state of the interacting theory.

Let us return once more to figure \ref{fig:AreaConstMass}. The
fact that our numerical asymptote is located slightly above the
one calculated from (\ref{DeltaconstantMass}) can be attributed to
the accuracy of the implementation of the infinite past memory
kernel. This can be appreciated from figure
\ref{fig:AreaConstMassAccuracy} where we test the accuracy of our
code. Clearly, the numerically found asymptote decreases towards
$\Delta_{\mathrm{ms}}$ as accuracy increases. Moreover, observe
that the initial violent oscillations in $\Delta_{k}(t)$ decrease
as accuracy improves. Also, we have chosen $\omega\Delta t$, where
$\Delta t$ is the step size of the numerical integration, such
that we resolve all the oscillations. For a $N=2000$ run at
$t\rho=100$, we have $\omega\Delta t \simeq 0.071$ for the
parameters used in figure \ref{fig:AreaConstMass}. Also, we can
observe a ``beating'' phenomenon that persists even if the
accuracy increases (and that can hence not be attributed to
numerical artifacts). It is caused by a frequency mismatch by
approximating the propagators in the infinite past memory kernel
by free propagators. Finally, let us discuss figure
\ref{fig:AreaConstMassDifference}. Here, we show the difference
between the phase space evolution calculated from equation
(\ref{EOMSYSTEM}) and from (\ref{eomnumerica3}). The dashed line
is more accurate than the dotted one, which in turn is more
accurate than the solid line. Clearly, the differences disappear
when the accuracy improves. This confirms our numerical analysis
in a non-trivial way.

\subsection{The Decoherence Time Scale}
\label{The Decoherence Time Scale}

We define the decoherence time scale to be the characteristic time
it takes for the phase space area $\Delta_{k}(t)$ to settle to its
constant mixed state value $\Delta_{\mathrm{ms}}$. One can suppose
that such a process is described by a differential equation of the
following form:
\begin{equation}\label{timescale1}
\delta\dot{\Delta}_{k}(t) + \Gamma_{k}(h,\omega_{\phi})
\delta\Delta_{k}(t) =0 \,,
\end{equation}
where $\delta\Delta_{k}(t)=\Delta_{k}(t)-\Delta_{\mathrm{ms}}$ and
where $\Gamma_{k}(h,\omega_{\phi})$ is the decoherence rate. This
equation is equivalent to $\dot{n}_{k}= - \Gamma_{k}(
n_{k}-n_{\mathrm{ms}})$, where $n_{k}$ is defined in equation
(\ref{particlenumber}) and $n_{\mathrm{ms}}$ is the stationary $n$
corresponding to $\Delta_{\mathrm{ms}}$. We anticipate that the
decoherence rate depends both on the coupling constant and on the
energy of our system field\footnote{Ideally, we should of course
take the $\mathcal{O}(h^2/\omega_{\phi}^{2})$ to $\omega_{\phi}$
through the dispersion relation into account.}. The following
intuitive picture is helpful: the solution of equation
(\ref{timescale1}) results in an exponential decay to the mixed
state value $\delta \Delta_{k}(t)\propto
\exp[-\Gamma_{k}(h,\omega_{\phi}) t]$. Furthermore, a stronger
coupling $h$ should result in a larger value of
$\Gamma_{k}(h,\omega_{\phi})$. However, a larger energy
$\omega^{2}_{\phi}=m^{2}_{\phi}+k^{2}$ should be reflected in a
smaller value of $\Gamma_{k}(h,\omega_{\phi})$. On dimensional
grounds, we thus anticipate:
\begin{equation}\label{timescale2}
\Gamma_{k}(h,\omega_{\phi}) = \frac{h^{2}}{\omega_{\phi}} \gamma
\,,
\end{equation}
where $\gamma=\mathrm{const}$. Let us now test this expected
scaling relation.
\begin{figure}[t!]
    \begin{minipage}[t]{.43\textwidth}
        \begin{center}
\includegraphics[width=\textwidth]{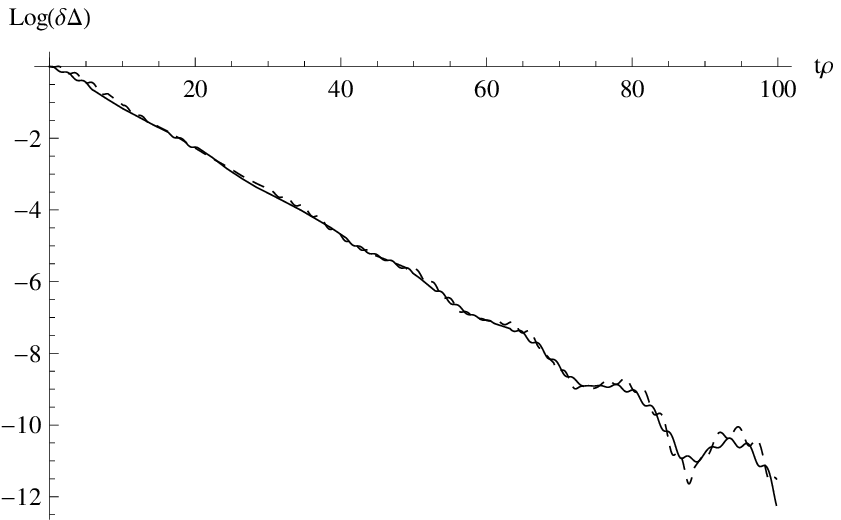}
   {\em \caption{Exponential approach to $\Delta_{\mathrm{ms}}$.
   We study, for different initial conditions, differences of
   $\Delta_{k}(t)$ on a logarithmic scale defined by equation (\ref{timescale3}). Clearly, the
   decoherence rate does not depend on the initial conditions. We
   used $k/\rho=1$, $m_{\phi}/\rho=1$, $h/\rho=4$ and $N=2000$.
    \label{fig:ExpDecay}}}
        \end{center}
    \end{minipage}
\hfill
    \begin{minipage}[t]{.43\textwidth}
        \begin{center}
\includegraphics[width=\textwidth]{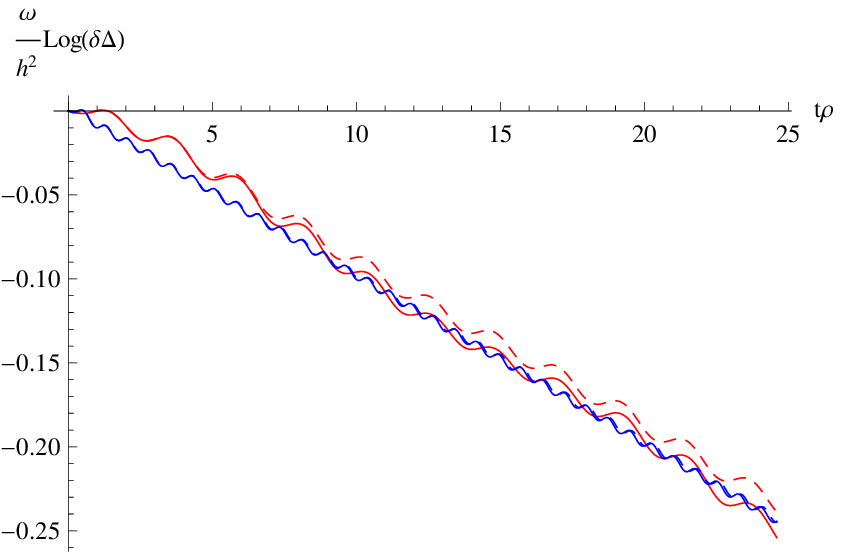}
   {\em \caption{Decoherence time scale. We confirm the scaling
   relation for the decoherence time scale anticipated in equation
   (\ref{timescale2}). For all plots we took $k/\rho=1$ and $N=1000$, and
   furthermore we used $m_{\phi}/\rho=1$, $h/\rho=4$ (solid red
   line), $m_{\phi}/\rho=1$, $h/\rho=1$ (dashed red line), $m_{\phi}/\rho=4$, $h/\rho=4$ (solid blue line) and $m_{\phi}/\rho=4$,
   $h/\rho=1$ (dashed blue line).
   \label{fig:DecoherenceTimeScale}}}
        \end{center}
    \end{minipage}
\end{figure}
Looking back at figure \ref{fig:AreaConstMass}, we see in the
first few time steps that $\Delta_{k}(t)$ oscillates. Clearly, the
time scale of these fluctuations has nothing to do with the
decoherence time scale, but rather can be fully attributed to
numerical accuracy. To capture the decoherence time scale
correctly, we thus consider the difference
$\Delta_{k}^{\mathrm{ms}}(t) - \Delta_{k}^{\mathrm{ps}}(t)$ of the
evolution of the phase space area $\Delta_{k}^{\mathrm{ms}}(t)$
using mixed state initial conditions and
$\Delta_{k}^{\mathrm{ps}}(t)$ using pure state initial conditions.
On a logarithmic scale, we observe in figure \ref{fig:ExpDecay} an
exponential decay towards $\Delta_{\mathrm{ms}}$ (solid line).
Moreover, the slope should not depend on the particular choice of
initial conditions. To this end, we also calculate the difference
of $\Delta_{k}^{\mathrm{ms}}(t) - \Delta_{k}^{\mathrm{nps}}(t)$,
where $\Delta_{k}^{\mathrm{nps}}(t)$ follows from setting
$1<\Delta_{k}^{\mathrm{nps}}(t_{0})<\Delta_{k}^{\mathrm{ms}}(t_{0})$
initially. In order to do this, we kept the value of
$F_{\phi}(k,t_{0},t_{0})$ identical to the value it had for the
mixed state boundary conditions but reduced the value of
$\left.\partial_{t'}\partial_{t} F_{\phi}(k, t,t') \right|_{t
=t'=t_{0}}$ such that the inequality
$1<\Delta_{k}^{\mathrm{nps}}(t_{0})<\Delta_{k}^{\mathrm{ms}}(t_{0})$
is satisfied. The resulting decoherence rates precisely coincide
as seen in figure \ref{fig:ExpDecay}, where we plot:
\begin{equation}\label{timescale3}
\log(\delta \Delta) \equiv
\log\left(\frac{\Delta^{\mathrm{ms}}_{k}(t) -
\Delta_{k}^{\mathrm{ps}}(t)}{\Delta^{\mathrm{ms}}_{k}(t_{0}) -
\Delta_{k}^{\mathrm{ps}}(t_{0})} \right)\,,
\end{equation}
and likewise for $\Delta_{k}^{\mathrm{nps}}(t)$ (dashed line).

We can repeat the steps outlined above for a different choice of
parameters $m$ and $h$. By rescaling the obtained decoherence
rates by a factor of $\omega_{\phi}/h^2$, we can test the scaling
relation (\ref{timescale1}). All decoherence rates now precisely
overlap as we depicted in figure \ref{fig:DecoherenceTimeScale}.
We can thus estimate the value of the constant of proportionality
$\gamma$ appearing in equation (\ref{timescale2}):
\begin{equation}\label{timescale4}
\Gamma_{k}(h,\omega_{\phi}) = (0.0101 \pm 0.0003)
\frac{h^{2}}{\omega_{\phi}}\,.
\end{equation}
This relation gives the decoherence rate for our particular model.
This result is nothing but the single particle decay rate:
\begin{equation}\label{timescale_comparison}
\Gamma_{\phi \rightarrow \chi \chi} = - \frac{\mathrm{Im}(\imath
M^{++})}{\omega_{\phi}} = \frac{1}{32 \pi}
\frac{h^{2}}{\omega_{\phi}} \,,
\end{equation}
where we have used equation (\ref{SelfMassFourier2}) and e.g.
\cite{Weldon:1983jn, Anisimov:2008dz}. Let us compare the result
(\ref{timescale4}) to the literature. Let us remark that most of
the calculations have been performed in an expanding Universe
setting, or with a different model, so it is hard to compare this
result quantitatively. In \cite{Lombardo:2005iz} it was found
that, for a different model during inflation, the decoherence rate
is proportional to the spatial volume, which we do not find.

\subsection{Changing Mass Solutions}
\label{Changing Mass Solutions}

Finally, let us discuss the evolution of the phase space area when
$m_{\phi}^{2}(t)$ is changing according to equation
(\ref{massbehaviourb}). The analytic expression for the
statistical propagator in Fourier space we previously derived in
equation (\ref{FourierStatisticalA}) is no longer valid.
Introducing a time dependent mass $m_{\phi}^{2}(t)$, generated by
a time dependent Higgs-like scalar field, breaks the time
translation invariance of the problem. Consequently, the
statistical propagator $F(k,t,t')$ no longer depends only on
the time difference of its time variables $\Delta t = t-t'$,
because considering $m_{\phi}^{2}(t)$ introduces a proper time
dependence on the average time coordinate $\tau = (t+t')/2$ in the
problem. When the mass is changing rapidly, we can only rely on
numerical methods. However, asymptotically, where the mass settles
again to a constant value, the analysis performed in the previous
subsection should still apply.

We impose mixed state boundary conditions as in equation
(\ref{MSboundaryconditionsstat}) such that
$\Delta_{k}(t_{0})=\Delta_{\mathrm{ms}}$ initially. Of course, the
value of the mass inserted to calculate these initial conditions,
is the value of the initial mass, valid before the mass jump.
\begin{figure}[t!]
    \begin{minipage}[t]{.43\textwidth}
        \begin{center}
\includegraphics[width=\textwidth]{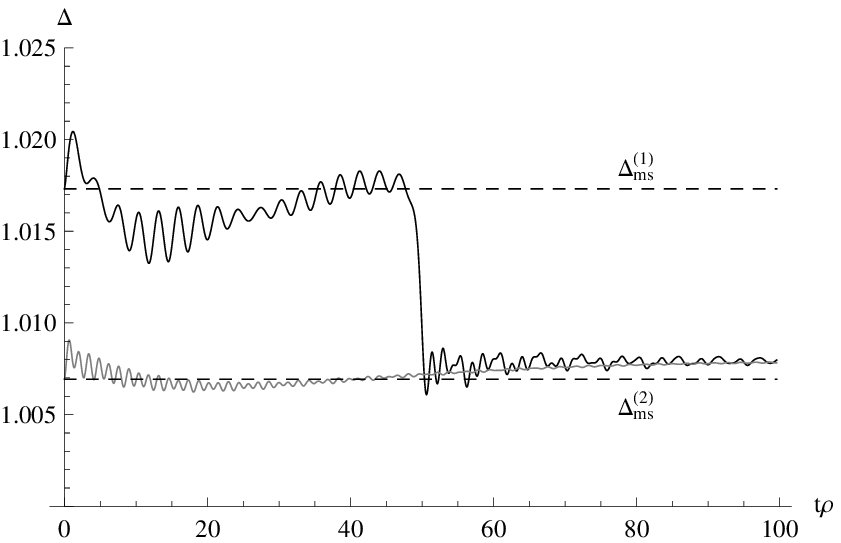}
   {\em \caption{Phase space area decrease due to a mass increase
   from $m_{\phi}/\rho=0.75$ to $m_{\phi}/\rho=2$ (solid black line). We
   used $h/\rho=4$, $k/\rho=1$ and $N=2000$. We thus observe a slight entropy decrease. The solid gray line
   denotes the constant mass phase space evolution for
   $m_{\phi}/\rho=2$. As the two asymptotes coincide, we conclude
   no entropy has been generated at late times by the mass change.
   $\Delta_{\mathrm{ms}}^{(1)}$ and $\Delta_{\mathrm{ms}}^{(2)}$ are
   the constant mixed phase space areas calculated for $m_{\phi}/\rho=0.75$
   and $m_{\phi}/\rho=2$, respectively.
    \label{fig:Area2MassIncrease}}}
\includegraphics[width=\textwidth]{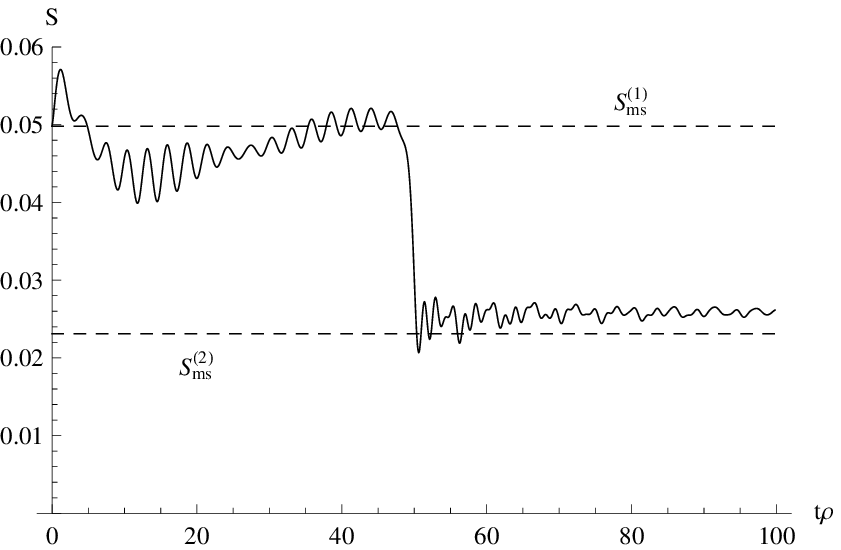}
   {\em \caption{Entropy decrease for the case presented in figure
   \ref{fig:Area2MassIncrease}. Because of the mass increase, the
   phase space area decreases which results consequently in a drop
   in the entropy.
    \label{fig:Entropy2MassIncrease}}}
        \end{center}
    \end{minipage}
\hfill
    \begin{minipage}[t]{.43\textwidth}
        \begin{center}
\includegraphics[width=\textwidth]{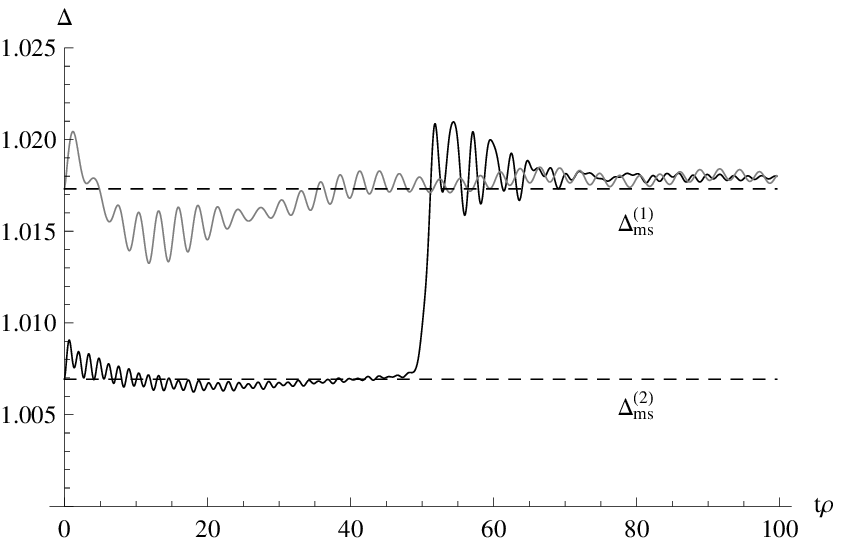}
   {\em \caption{Phase space area increase due to a mass decrease
   from $m_{\phi}/\rho=2$ to $m_{\phi}/\rho=0.75$ (solid black line). The other parameters are the same as in figure \ref{fig:Area2MassIncrease}. The solid gray line
   denotes the constant mass phase space evolution for
   $m_{\phi}/\rho=0.75$. Again the two asymptotes coincide and no entropy has been generated at late times by the mass change.
    \label{fig:Area2MassDecrease}}}
\vskip 0.8cm
\includegraphics[width=\textwidth]{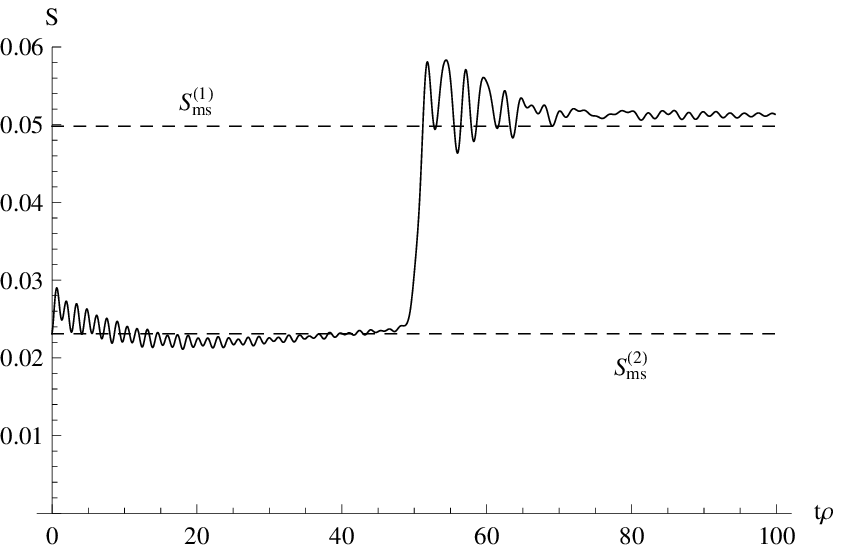}
   {\em \caption{Entropy increase for the case presented in figure
   \ref{fig:Area2MassDecrease}. Clearly, a mass decrease gives
   rise to a slight entropy increase.
    \label{fig:Entropy2MassDecrease}}}
        \end{center}
    \end{minipage}
\end{figure}

If the mass changes non-adiabatically, this results in a
significant particle creation according to the discussion in
appendix \ref{Quantum Reflection: Free Case}. We can thus identify
the following regimes:
\begin{subequations}
\label{adiabatic}
\begin{eqnarray}
|\beta_{k}|^{2} \ll 1 \qquad && \mathrm{adiabatic \,\, regime} \label{adiabatica}\\
|\beta_{k}|^{2} \gg 1 \qquad && \mathrm{non-adiabatic \,\, regime}
\label{adiabatib}\,,
\end{eqnarray}
\end{subequations}
where $\beta_{k}$ is given is equation (\ref{modesolution7b}). For
the parameters we used in figures \ref{fig:Area2MassIncrease} and
\ref{fig:Area2MassDecrease} we are in the adiabatic regime
($|\beta_{k}|^{2}= 3.5 \cdot 10^{-4}$).

In figure \ref{fig:Area2MassIncrease} we study a mass increase
from $m_{\phi}/\rho=0.75$ to $m_{\phi}/\rho=2$ (black line). This
decreases the phase space area and consequently the entropy
decreases, which we depicted in figure
\ref{fig:Entropy2MassIncrease}. Intuitively, a larger mass of the
$\phi$-field reduces the effect of the quantum corrections of the
$\chi$-field. Hence
$\Delta_{\mathrm{ms}}^{(2)}<\Delta_{\mathrm{ms}}^{(1)}$, where
$\Delta_{\mathrm{ms}}^{(2)}$ is the constant phase space area
calculated for $m_{\phi}/\rho=2.0$ from three Fourier integrals as
in equation (\ref{Fconstantmass3}). Likewise,
$\Delta_{\mathrm{ms}}^{(1)}$ corresponds to the phase space area
calculated for $m_{\phi}/\rho=0.75$. The relevant behaviour to
compare with is the constant mass phase space evolution for
$m_{\phi}/\rho=2$ which we also depicted in figure
\ref{fig:Area2MassIncrease} in gray. Clearly, the late time
asymptotes of the two functions coincide and we conclude that,
also at late times, no entropy has been generated. As we are in
the deep adiabatic regime, this is to be expected.

Now let us study the opposite: a mass decrease from
$m_{\phi}/\rho=2$ to $m_{\phi}/\rho=0.75$. This is depicted (black
line) in figure \ref{fig:Area2MassDecrease}. Clearly, the phase
space area increases from $\Delta_{\mathrm{ms}}^{(2)}$ to
$\Delta_{\mathrm{ms}}^{(1)}$ and we plotted the resulting entropy
increase in figure \ref{fig:Entropy2MassDecrease}. Again, we
compared this evolution with the phase space area calculated for a
constant mass $m_{\phi}/\rho=0.75$ (gray line). As the two
asymptotes also coincide in this case, we conclude that no entropy
has been generated at late times by the mass change. Of course, it
would be very interesting to see what happens when we study the
same process in the non-adiabatic regime and we hope to address
this question in a future publication.

If we compare the evolution of the entropy in time in these two
cases with the free case $S_{k}(t)=0$, the interacting case
reveals much more interesting behaviour. Firstly, due to the
presence of an environment field, the constant value to which the
entropy settles asymptotically is different from zero, unlike the
free case. Secondly, a changing mass induces dynamics: the entropy
depends on time and evolves from one value $S_{\mathrm{ms}}^{(1)}$
to another $S_{\mathrm{ms}}^{(2)}$ or vice versa.

\section{Conclusion}
\label{Conclusion}

We apply the decoherence framework to quantum field theory. We
consider two scalar fields, one corresponding to a ``system
field'', the second to an ``environment field'', in interaction
via a cubic coupling. Here, we consider an environment in its
vacuum state ($T=0$) and postpone finite temperature contributions
to a future publication. We neglect the backreaction of the
environment field on the system field, assuming that the former
remains at $T=0$.

We advocate the following point of view regarding a sensible
application of decoherence to quantum field theory: for some
observer inaccessible higher order correlators give rise, once
neglected, to an increase in entropy $S_{k}$ of the system. This
is inspired by realising that correlators are measured in quantum
field theories and that higher order irreducible $n$-point
functions are usually perturbatively suppressed. In this work, we
assume that our observer is only sensitive to Gaussian correlators
and will hence neglect higher order, non-Gaussian correlators.
Neglecting the information stored in these higher order
correlators, gives rise to an increase of the entropy of the
system. If the system initially occupies the minimal area in phase
space (characterised by a pure state with $S_{k}(t_{0})=0$), we
numerically calculate the evolution of the entropy $S_{k}(t)$ in
figure \ref{fig:EntropyConstMass}. Also, we calculate the
asymptotic value of the phase space area $\Delta_{\mathrm{ms}}$
and the entropy $S_{\mathrm{ms}}$ as a function of the coupling
$h$, the mass $m_{\phi}$ and $k$ to which these functions evolve.
This increase in entropy can be understood from the system's
tendency to evolve towards the vacuum state of the interacting
theory. Even though we do not solve the full 2PI equations at one
loop, we have strong numerical evidence that within our
approximation scheme, the system evolves towards its correct
stationary interacting vacuum state. If the system, however,
starts out initially occupying the state characterised by
$S_{\mathrm{ms}}$, we observe no further increase in the entropy.
Furthermore, we calculate the decoherence rate in equation
(\ref{timescale4}) which charaterises the exponential rate at
which the system approaches its stationary state.

Secondly, we study the effect of a time dependent mass
$m_{\phi}^{2}(t)$ of the system field. Now, calculating $S_{k}(t)$
can only be addressed numerically. Starting out at mixed state
initial conditions $S_{k}(t_{0})=S_{\mathrm{ms}}>0$, we observe an
entropy increase (decrease) due to a mass decrease (increase) as
depicted in figures \ref{fig:Entropy2MassDecrease} and
\ref{fig:Entropy2MassIncrease}, respectively. By comparing with
the constant mass evolution for the entropy, we conclude that no
additional entropy has been generated asymptotically by the mass
jump. As we study mass changes in the deep adiabatic regime, it
remains to be investigated whether this statement also holds in
the non-adiabatic regime.

We also would like to draw a few somewhat more technical
conclusions. It is important to stress that in interacting field
theories where memory effects play a crucial role, one cannot just
insert initial conditions at some arbitrary finite time $t_{0}$,
because one then neglects the memory effects existing from the
infinite asymptotic past to $t_{0}$. In numerical computations
however, one has to start at some finite time. We therefore
approximate the propagators in the memory integral from the past
infinity to $t_{0}$ by free propagators, inducing a perturbatively
suppressed error.

Also, it has not been previously appreciated in the literature
that renormalising the vacuum contribution in the Kadanoff-Baym
equations can actually significantly change the structure of these
equations. In order to properly take account of the renormalised
self-mass contribution we had to extract two time derivatives
which can readily be seen from equation
(\ref{SelfMassFourierSpFinalResult}).

\

\noindent \textbf{Acknowledgements}

\noindent JFK thanks Jeroen Diederix for many useful suggestions.
JFK and TP thank Tomas Janssen and Gerben Stavenga and acknowledge
financial support from FOM grant 07PR2522 and from Utrecht
University. The authors thank Julien Serreau for many useful
comments and for thoroughly reading the manuscript. The authors
also gratefully acknowledge the hospitality of the Nordic
Institute for Theoretical Physics (NORDITA) during their stay at
the ``Electroweak Phase Transition'' workshop in June 2009.

\appendix

\section{Quantum Effects of a Changing Mass: Free Case}
\label{Quantum Reflection: Free Case}

It is interesting to compare our results to a non-trivial exact
case: scattering of a changing mass field in the spirit of Birrell
and Davies \cite{Birrell:1982ix}. The solutions presented here
stem from the cosmological particle creation literature (based on
\cite{Parker:1968mv, Parker:1969au}) and are originally due to
Bernard and Duncan \cite{Bernard:1977pq}. Let us consider the
action of a free scalar field:
\begin{equation}\label{action1}
S[\phi] = \int \mathrm{d}^{4}\!x \left\{-\frac{1}{2}
\partial_\mu\phi(x)
\partial_\nu \phi(x) \eta^{\mu\nu} - \frac{1}{2} m^{2}_{\phi}(t)
\phi^{2}(x) \right\}\,,
\end{equation}
where as usual $\eta_{\mu\nu} = {\rm diag}(-1,1,1,1)$ is the
Minkowski metric, and where we consider the following behaviour of
the mass $m_{\phi}(t)$ of the scalar field:
\begin{equation}\label{masstanh}
m^{2}_{\phi}(t) = \left(A + B \tanh(\rho t)\right)\,.
\end{equation}
From (\ref{action1}), it follows that:
\begin{equation} \label{eom1}
\left(\partial_{t}^{2}-\partial_{i}^{2}+m^{2}_{\phi}(t) \right)
\phi(x) =0\,,
\end{equation}
The vacuum causal and statistical propagators follow from
(\ref{causalpropagator}) and (\ref{statisticalpropagator}) where
$\hat{\rho}(t_{0})=|0\rangle\langle0|$ as:
\begin{subequations}
\label{propagatorvac}
\begin{eqnarray}
\imath\Delta^{c}_{\phi} (x;x^{\prime}) &=& \langle 0 |
 [\hat{\phi}(x),\hat{\phi}(x^{\prime})] |0 \rangle
 \label{causalpropagatorvac}\\
F_{\phi}(x;x') &=& \frac{1}{2} \langle 0 | \{ \hat{\phi}(x'),
\hat{\phi}(x) \} |0 \rangle \label{statisticalpropagatorvac} \,.
\end{eqnarray}
\end{subequations}
Let us quantise our fields in $D$-dimensions by making use of
creation and annihilation operators:
\begin{equation} \label{quantisationphi}
\hat{\phi}(x) = \int
\frac{\mathrm{d}^{\scriptscriptstyle{D}-1}\vec{k}}{(2\pi)^{\scriptscriptstyle{D}-1}}
\left( \hat{a}_{\vec{k}}\,\phi_{k}(t)e^{i\vec{k} \cdot \vec{x}} +
\hat{a}_{\vec{k}}^{\dag}\,\phi_{k}^{\ast}(t)e^{-i\vec{k} \cdot
\vec{x}} \right)\,.
\end{equation}
The annihilation operator acts as usual on the vacuum:
\begin{subequations}
\label{quantisationphi2}
\begin{equation}\label{quantisationphi2a}
\hat{a}_{\vec{k}}|0\rangle = 0\,,
\end{equation}
and we impose the following commutation relations:
\begin{equation}\label{quantisationphi2b}
[ \hat{a}_{\vec{k}}, \hat{a}_{\vec{k'}}^{\dag}]
=(2\pi)^{\scriptscriptstyle{D}-1}
\delta^{\scriptscriptstyle{D}-1}(\vec{k}-\vec{k'})\,.
\end{equation}
\end{subequations}
Hence the equation of motion for the mode functions $\phi_{k}(t)$
of $\phi(x)$, defined by relation (\ref{quantisationphi}), follows
straightforwardly as:
\begin{equation}\label{eom2}
\left(\partial_{t}^{2} + k^{2} + m^{2}_{\phi}(t) \right)
\phi_{k}(t) = 0 \,,
\end{equation}
where $k=\|\vec{k}\|$. The mode functions determine the causal and
statistical propagators from (\ref{propagatorvac}) completely:
\begin{subequations}
\label{propagatorF}
\begin{eqnarray}
\imath\Delta^{c}_{\phi} (k,t,t') &=&
\phi_{k}(t)\phi_{k}^{\ast}(t') - \phi_{k}(t')\phi_{k}^{\ast}(t)
 \label{causalpropagatorF}\\
F_{\phi}(k,t,t') &=& \frac{1}{2} \left\{
\phi_{k}(t')\phi_{k}^{\ast}(t) +
\phi_{k}(t)\phi_{k}^{\ast}(t')\right\}
\label{statisticalpropagatorF} \,.
\end{eqnarray}
\end{subequations}
Let us at this point for completeness calculate the constant mass
causal and statistical propagators in Fourier space. We just
insert a constant mass $m_{\phi}$, rather than a changing one as
in equation (\ref{masstanh}):
\begin{subequations}
\label{propagatorconstantmassF}
\begin{eqnarray}
\imath\Delta^{c}_{\phi} (k,t,t') &=& -
\frac{\imath}{\omega}\sin(\omega(t-t'))
 \label{causalpropagatorconstantmassF}\\
F_{\phi}(k,t,t') &=& \frac{1}{2 \omega}\cos(\omega(t-t'))
\label{statisticalpropagatorconstantmassF} \,,
\end{eqnarray}
\end{subequations}
where $\omega^{2}=m_{\phi}^{2}+k^{2}$. Let us now return to the
changing mass case. The physical picture is clear: we would like
to study reflection and transmission, i.e.: scattering, of an
incoming wave due to the changing mass. Before solving this
equation of motion exactly, let us first solve for the asymptotic
mode functions to gain intuitive understanding. In the asymptotic
past ($t \rightarrow - \infty$), equation (\ref{eom2}) is solved
by:
\begin{equation}\label{solin1}
\phi_{k}^{\mathrm{in}} (t) =
\frac{1}{\sqrt{2\omega_{\mathrm{in}}}} \exp\left[- \imath
\omega_{\mathrm{in}} t \right] \,,
\end{equation}
i.e.: one right-moving or incoming wave with frequency:
\begin{equation}\label{solomegain}
\omega_{\mathrm{in}} = \left( k^{2}+A-B \right)^{\frac{1}{2}}\,.
\end{equation}
In the infinite asymptotic future, the solution necessarily is an
appropriately normalised linear superposition of a left- and
right-moving wave:
\begin{equation} \label{solout1}
\phi_{k}^{\mathrm{out}}(t) = \alpha_{k}
\frac{1}{\sqrt{2\omega_{\mathrm{out}}}} \exp\left[- \imath
\omega_{\mathrm{out}} t \right] + \beta_{k}
\frac{1}{\sqrt{2\omega_{\mathrm{out}}}} \exp\left[ \imath
\omega_{\mathrm{out}} t \right] \,,
\end{equation}
where:
\begin{equation}\label{solomegaout}
\omega_{\mathrm{out}} = \left( k^{2}+A+B\right)^{\frac{1}{2}}\,,
\end{equation}
and where:
\begin{equation}\label{solnormout}
\|\alpha_{k}\|^{2} - \|\beta_{k}\|^{2} = 1 \,,
\end{equation}
for a consistent canonical quantisation. In both asymptotic
regions we can now immediately calculate the statistical
propagator from equation (\ref{statisticalpropagatorF}):
\begin{subequations}
\label{statpropsol1}
\begin{eqnarray}
F_{\mathrm{in}}(k,t,t') &=& \frac{1}{2 \omega_{\mathrm{in}}}
\cos(\omega_{\mathrm{in}}(t-t'))
\label{statpropsol1a} \\
F_{\mathrm{out}}(k,t,t') &=& \frac{1}{2
\omega_{\mathrm{out}}}\left[
(|\alpha_{k}|^{2}+|\beta_{k}|^{2})\cos(\omega_{\mathrm{out}}(t-t'))
+\alpha_{k}\beta_{k}^{\ast} e^{-\imath
\omega_{\mathrm{out}}(t+t')} +\alpha_{k}^{\ast}\beta_{k} e^{\imath
\omega_{\mathrm{out}}(t+t')}\right] \label{statpropsol1b} \,.
\end{eqnarray}
\end{subequations}
Using equation (\ref{deltaareainphasespace}), we can calculate the
area in phase space the in and out states occupy:
\begin{subequations}
\label{statDelta1}
\begin{eqnarray}
\Delta^{\mathrm{in}}_{k}(t) &=& 1
\label{statDelta1a} \\
\Delta^{\mathrm{out}}_{k}(t) &=& 1 \label{statDelta1b} \,.
\end{eqnarray}
\end{subequations}
Hence for the entropy we find:
\begin{equation}\label{solS}
S^{\mathrm{in}}_{k}(t) = 0 = S^{\mathrm{out}}_{k}(t) \,.
\end{equation}
We conclude that in both asymptotic regions the entropy is zero
and no entropy has been generated by changing the mass.

\begin{figure}[t!]
    \begin{minipage}[t]{.43\textwidth}
        \begin{center}
\includegraphics[width=\textwidth]{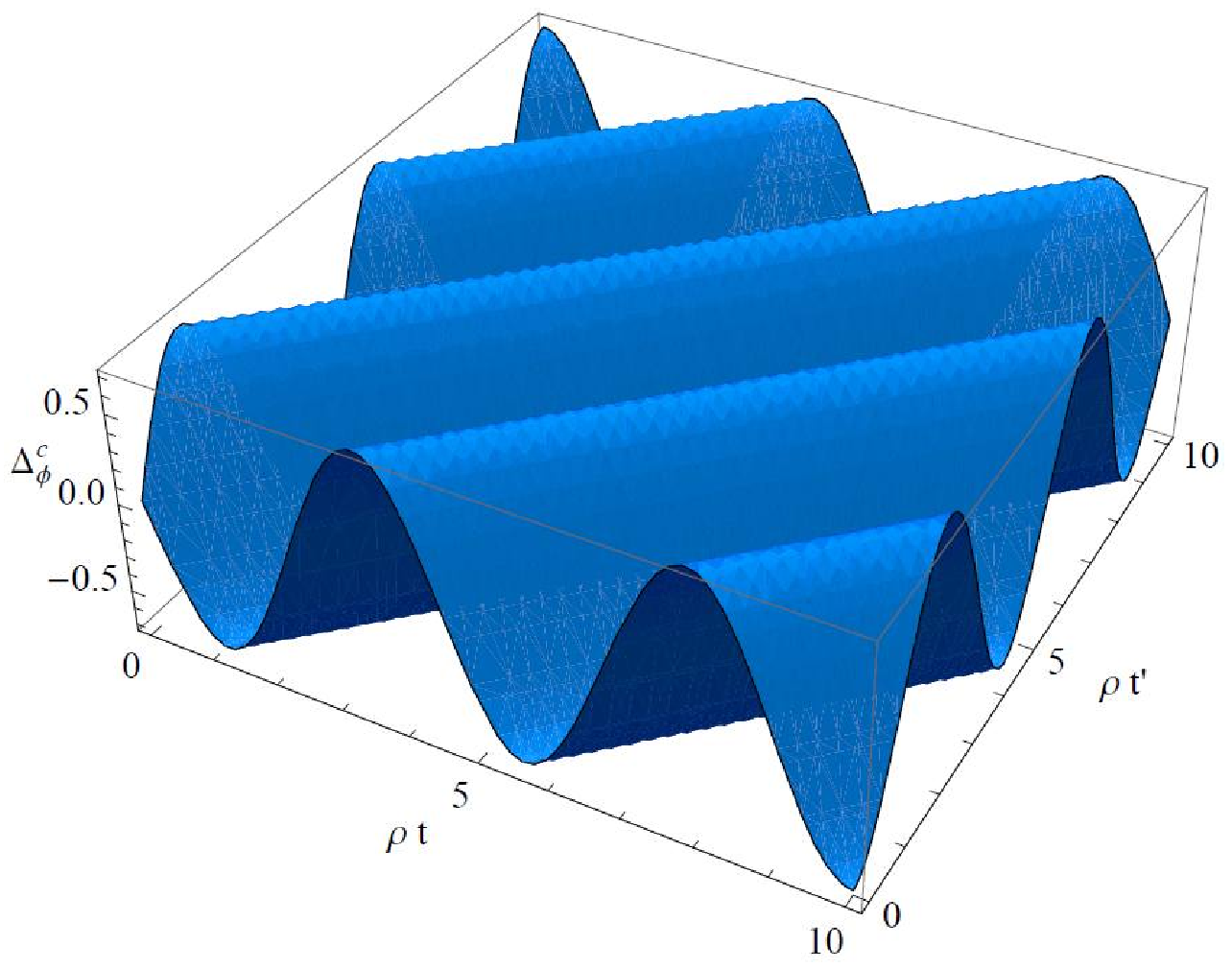}
   {\em \caption{Causal Propagator with a constant mass. Parameters: $k/\rho=1$, $m_{\phi}/\rho=1$. \label{fig:CausalPropagatorExact1} }}
\vskip 0.1cm
\includegraphics[width=\textwidth]{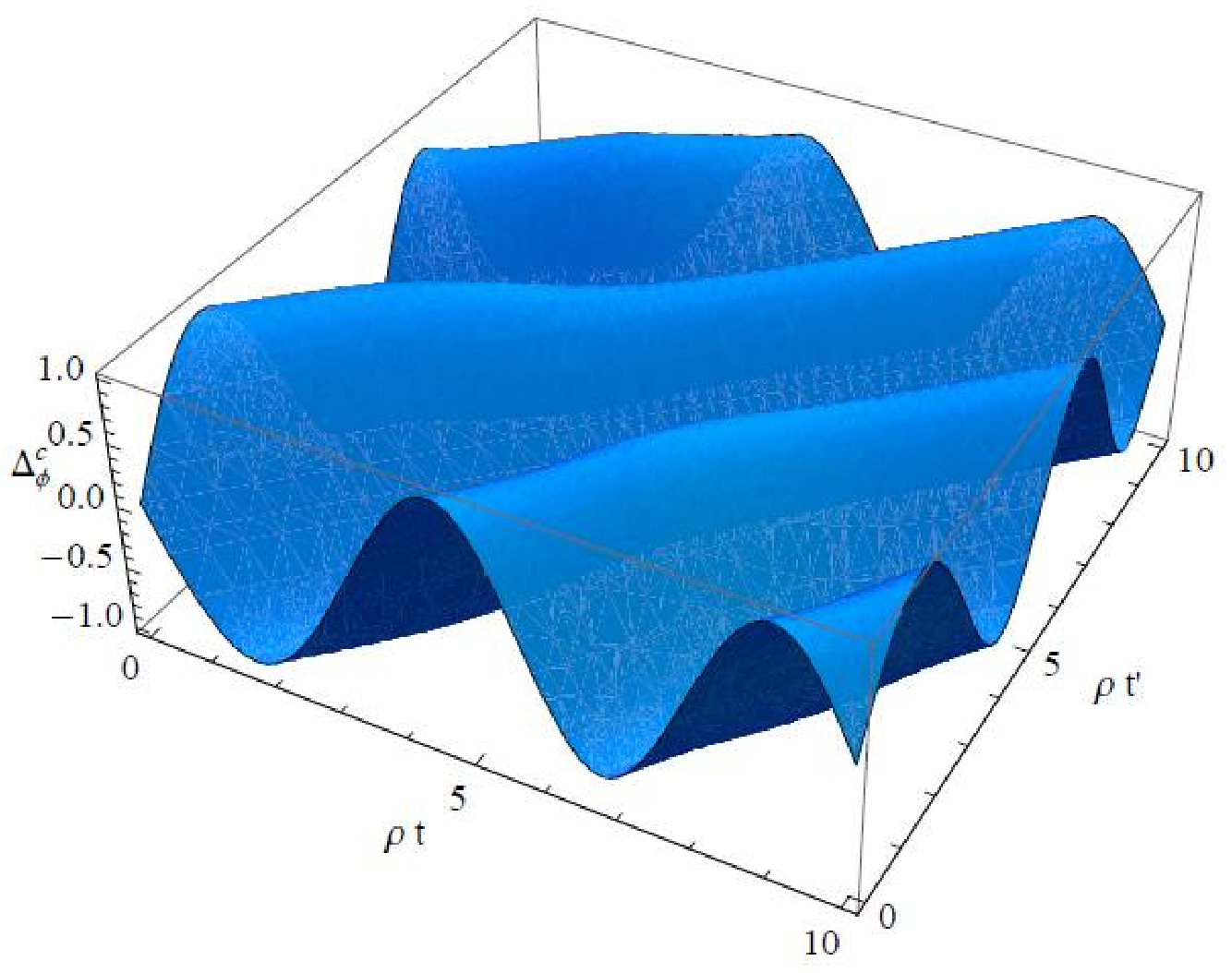}
   {\em \caption{Causal Propagator for a small change in the mass. Parameters:
   $k/\rho=1$, $A/\rho^{2}=B/\rho^{2}=1/2$.
   \label{fig:CausalPropagatorExact2} }}
\vskip 0.1cm
\includegraphics[width=\textwidth]{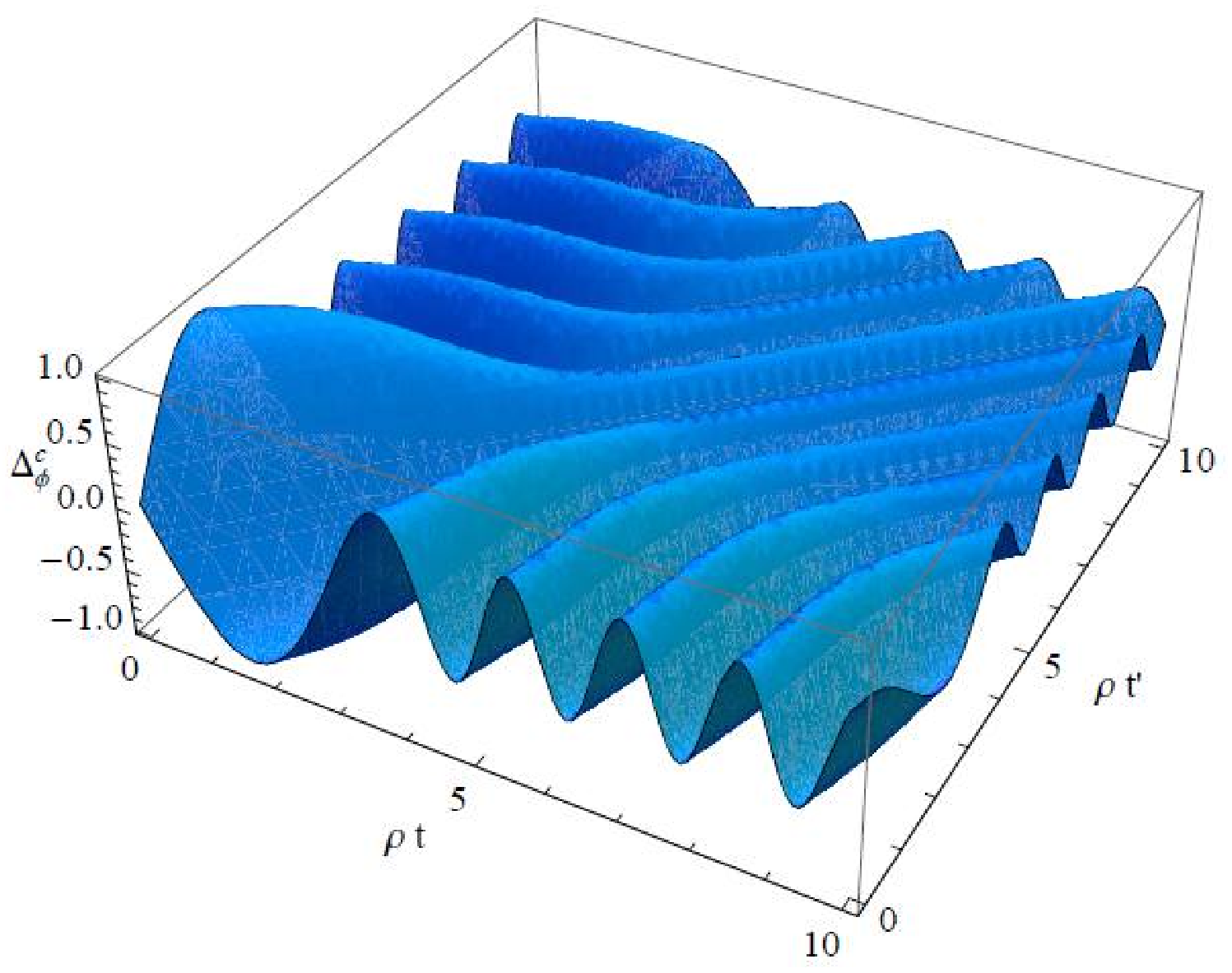}
   {\em \caption{Causal Propagator for a large change in the mass. Parameters:
   $k/\rho=1$, $A/\rho^{2}=B/\rho^{2}=2$.
   \label{fig:CausalPropagatorExact3} }}
        \end{center}
    \end{minipage}
\hfill
    \begin{minipage}[t]{.43\textwidth}
        \begin{center}
\includegraphics[width=\textwidth]{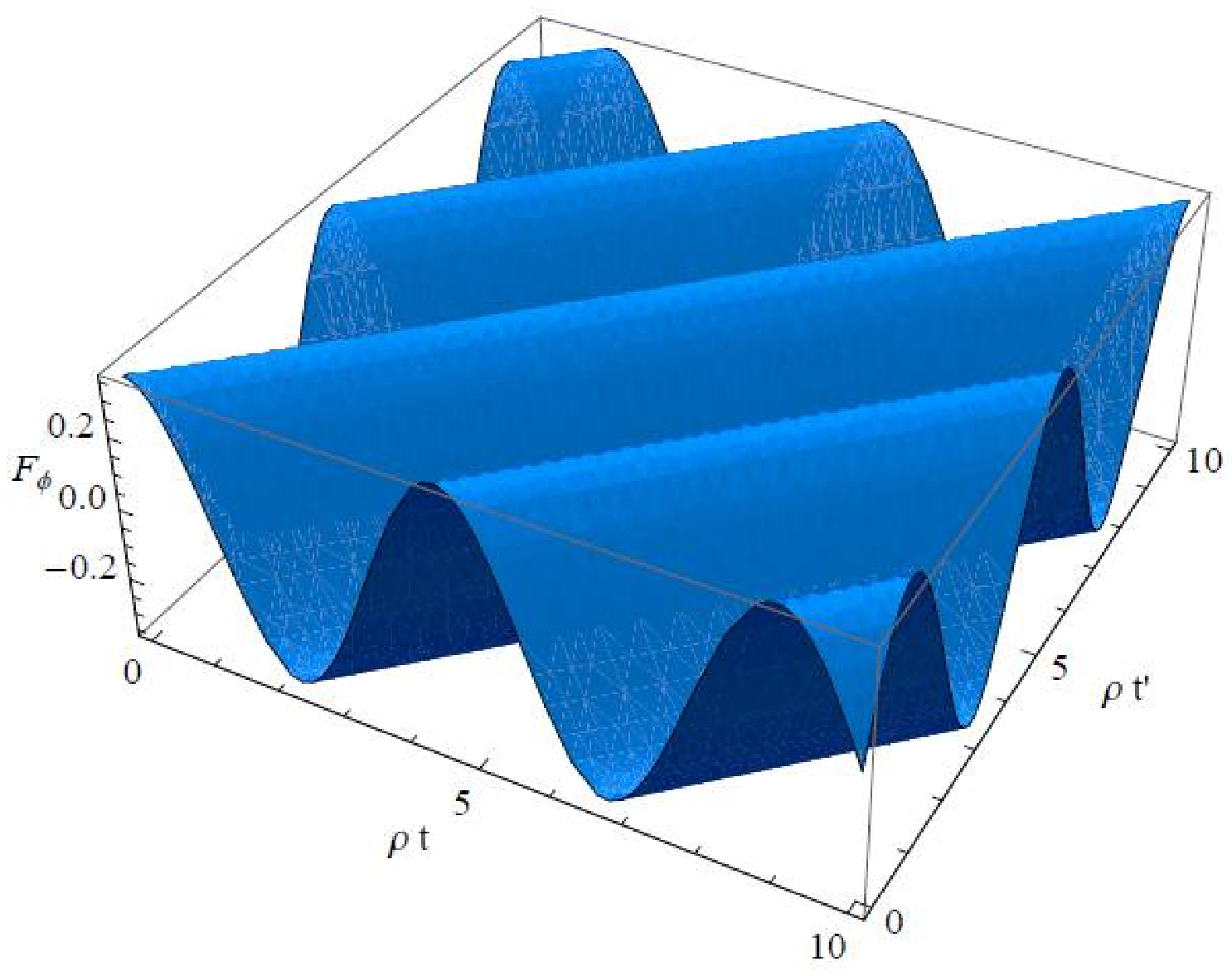}
   {\em \caption{Statistical Propagator with a constant mass. Parameters:
   $k/\rho=1$, $m_{\phi}/\rho=1$.
   \label{fig:StatisticalPropagatorExact1} }}
\vskip0.1cm
\includegraphics[width=\textwidth]{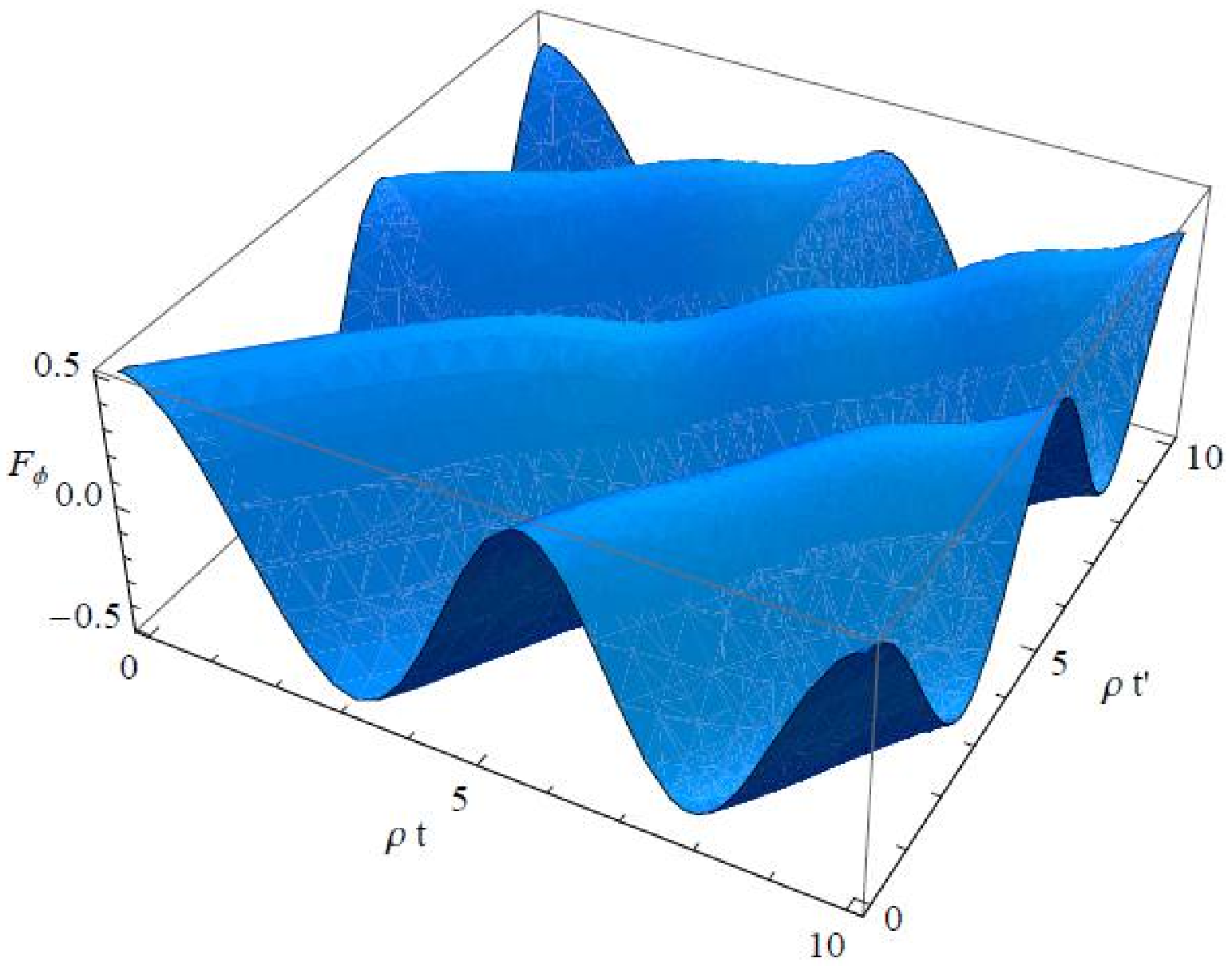}
   {\em \caption{Statistical Propagator for a small change in the mass. Parameters:
   $k/\rho=1$, $A/\rho^{2}=B/\rho^{2}=1/2$.
   \label{fig:StatisticalPropagatorExact2} }}
\vskip 0.1cm
\includegraphics[width=\textwidth]{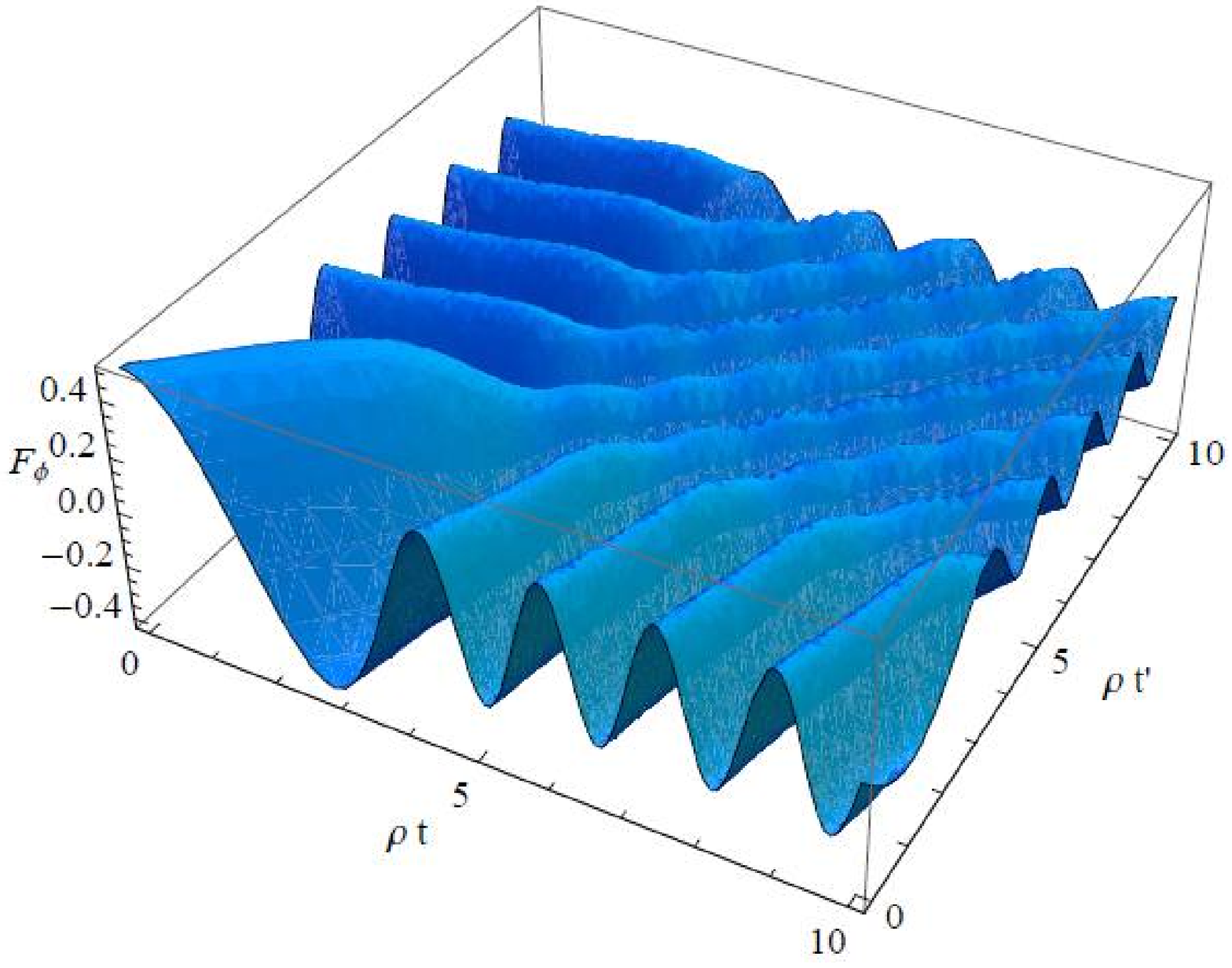}
   {\em \caption{Statistical Propagator for a large change in the mass. Parameters:
   $k/\rho=1$, $A/\rho^{2}=B/\rho^{2}=2$.
   \label{fig:StatisticalPropagatorExact3} }}
        \end{center}
    \end{minipage}
\end{figure}

However, we can do better than study the asymptotic behaviour
only. Birrell and Davies study cosmological particle creation in
section 3.4 of their book \cite{Birrell:1982ix} in a simple,
conveniently chosen cosmological setting. They consider a scale
factor as a function of conformal time $a(\eta)$ which behaves as:
\begin{equation}\label{BDscalefactor}
a^{2}(\eta) = A + B \tanh(\rho \eta)\,.
\end{equation}
This represents an asymptotically static universe with a smooth
expansion connecting these two asymptotic regions. Indeed, the
equation of motion (in conformal time) for the mode functions
Birrell and Davies consider coincides precisely with (\ref{eom2}).
The solution to (\ref{eom2}) which behaves as a positive frequency
mode in the asymptotic past ($t\rightarrow - \infty$) can be
expressed in terms of Gauss' hypergeometric function
$\phantom{1}_{2}F_{1}$:
\begin{equation} \label{modesolution1}
\phi_{k}^{\mathrm{in}}(t) = \frac{1}{\sqrt{2\omega_{\mathrm{in}}}}
\exp \left[- \imath \omega_{+} t - \imath
\frac{\omega_{-}}{\rho}\log\{2\cosh(\rho t )\}\right]
\phantom{1}_{2}F_{1}\left( 1+ \imath \frac{\omega_{-}}{\rho},
\imath \frac{\omega_{-}}{\rho}; 1- \imath
\frac{\omega_{\mathrm{in}}}{\rho};\frac{1}{2}\{1+\tanh(\rho
t)\}\right) \,,
\end{equation}
such that:
\begin{equation} \label{modesolution2}
\lim_{t\rightarrow - \infty} \phi_{k}^{\mathrm{in}}(t) =
\frac{1}{\sqrt{2\omega_{\mathrm{in}}}} \exp\left[- \imath
\omega_{\mathrm{in}} t \right] \,,
\end{equation}
where we defined $\omega_{\mathrm{in}}$ and
$\omega_{\mathrm{out}}$ in equations (\ref{solomegain}) and
(\ref{solomegaout}), respectively, and:
\begin{equation}
\omega_{\pm} = \frac{1}{2}( \omega_{\mathrm{out}} \pm
\omega_{\mathrm{in}}) \label{solasymptoticsc} \,.
\end{equation}
Alternatively, the modes which reduce to positive frequency modes
in the out region are given by:
\begin{equation} \label{modesolution4}
\phi_{k}^{\mathrm{out}}(t) =
\frac{1}{\sqrt{2\omega_{\mathrm{out}}}} \exp \left[- \imath
\omega_{+} t - \imath \frac{\omega_{-}}{\rho}\log\{2\cosh(\rho t
)\}\right] \phantom{1}_{2}F_{1}\left( 1+ \imath
\frac{\omega_{-}}{\rho}, \imath \frac{\omega_{-}}{\rho}; 1+ \imath
\frac{\omega_{\mathrm{out}}}{\rho};\frac{1}{2}\{1+\tanh(\rho
t)\}\right) \,,
\end{equation}
such that:
\begin{equation} \label{modesolution5}
\lim_{t\rightarrow \infty} \phi_{k}^{\mathrm{out}}(t) =
\frac{1}{\sqrt{2\omega_{\mathrm{out}}}} \exp\left[- \imath
\omega_{\mathrm{out}} t \right] \,.
\end{equation}
We can rewrite the hypergeometric functions using equations
(15.3.3) and (15.3.6) of \cite{Abramowitz} and identify:
\begin{equation} \label{modesolution6}
\phi_{k}^{\mathrm{in}}(t) = \alpha_{k} \phi_{k}^{\mathrm{out}}(t)
+ \beta_{k} \phi_{k}^{\mathrm{out\, \ast}}(t) \,,
\end{equation}
where
\begin{subequations}
\label{modesolution7}
\begin{eqnarray}
\alpha_{k} = \left(
\frac{\omega_{\mathrm{out}}}{\omega_{\mathrm{in}}}\right)^{\frac{1}{2}}
\frac{ \Gamma\left(1-\imath \omega_{\mathrm{in}}/\rho\right)
\Gamma\left(-\imath
\omega_{\mathrm{out}}/\rho\right)}{\Gamma\left(-\imath
\omega_{+}/\rho\right) \Gamma\left(1-\imath \omega_{+}/\rho\right)
} \label{modesolution7a}\\
\beta_{k} = \left(
\frac{\omega_{\mathrm{out}}}{\omega_{\mathrm{in}}}\right)^{\frac{1}{2}}
\frac{ \Gamma\left(1-\imath \omega_{\mathrm{in}}/\rho\right)
\Gamma\left(\imath
\omega_{\mathrm{out}}/\rho\right)}{\Gamma\left(\imath
\omega_{-}/\rho\right) \Gamma\left(1+\imath \omega_{-}/\rho\right)
} \label{modesolution7b} \,.
\end{eqnarray}
\end{subequations}
Having the mode functions at our disposal, we can find (the rather
cumbersome expressions for) the exact causal and statistical
propagators. The statistical and causal propagators can however
neatly be visualised. Figure \ref{fig:CausalPropagatorExact1}
shows the causal propagator with a constant mass from equation
(\ref{causalpropagatorconstantmassF}) for comparison to the
changing mass case. In figures \ref{fig:CausalPropagatorExact2}
and \ref{fig:CausalPropagatorExact3} we show the exact causal
propagator for a relatively small increase of the mass (from
$m_{\phi}/\rho=0$ to $m_{\phi}/\rho=1$) and a larger one (from
$m_{\phi}/\rho=0$ to $m_{\phi}/\rho=4$) for one particular Fourier
mode only ($k/\rho=1$). Figures
\ref{fig:StatisticalPropagatorExact1},
\ref{fig:StatisticalPropagatorExact2} and
\ref{fig:StatisticalPropagatorExact3} show the analogous
statistical propagators.

\begin{figure}[t!]
    \begin{minipage}[t]{.45\textwidth}
        \begin{center}
\includegraphics[width=\textwidth]{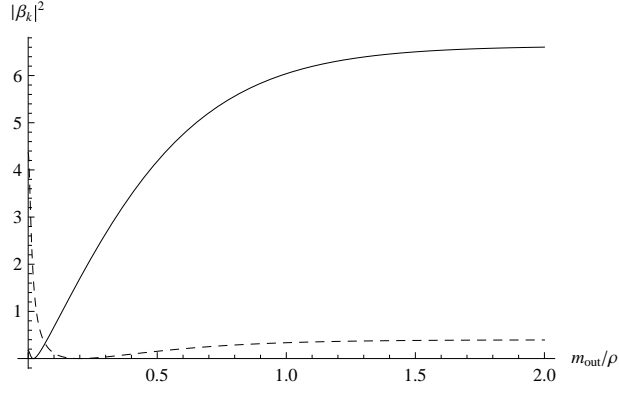}
   {\em \caption{This plot shows $|\beta_{k}|^{2}$ as a function
   of the final mass $m_{\mathrm{out}}/\rho$, for fixed
   $k/\rho=0.01$. The dashed line shows the adiabatic regime
   ($m_{\mathrm{in}}/\rho=0.2)$ whereas the solid line shows the
   non-adiabatic regime ($m_{\mathrm{in}}/\rho=0.02)$.
   \label{fig:adiabaticregimes} }}
        \end{center}
    \end{minipage}
\end{figure}

We can easily relate the statistical propagator to the phase space
area by making use of equation (\ref{deltaareainphasespace}). It
will not come as a surprise to the reader that we find:
\begin{equation} \label{statDelta2}
\Delta_{k}(t) = 1 \,,
\end{equation}
and hence:
\begin{equation}\label{solS2}
S_{k}(t) = 0 \,,
\end{equation}
also for all intermediate times. A final remark is in order. The
reader should not confuse $|\beta_{k}|^{2}$ calculated from
equation (\ref{modesolution7b}) with the phase space particle
number density or statistical number density
(\ref{particlenumber}). Although the mass is changing, the phase
space particle density remains zero but $|\beta_{k}|^{2}$, which
in the literature is often referred to as a particle number, can
change significantly as can be appreciated from figure
\ref{fig:adiabaticregimes}. This is just caused by the fact that
the in and out vacua differ. We plot the behaviour of
$|\beta_{k}|^{2}$ as a function of $m_{\mathrm{out}}/\rho$ in both
the adiabatic regime ($|\beta_{k}|^{2} \ll 1$) and non-adiabatic
regime ($|\beta_{k}|^{2} \gg 1$).

This simple example suggests the following: (i) the area in phase
space a state occupies is a good quantitative measure of the
entropy, (ii) the statistical propagator contains all the
information required to calculate this phase space area, and (iii)
a changing mass does not change the entropy for a free scalar
field. If we contrast this result with the calculations performed
in the main body of the paper, it is important to realise that
just a changing mass, in the absence of interactions, produces no
entropy, whereas we have shown that the entropy can change in the
interacting case.

\section{Evaluation of the Thermal Propagators}
\label{Evaluation of the Thermal Propagators}

In this appendix, we will for pedagogical reasons derive the
thermal propagators from first principles. The four thermal
propagators should solve the standard differential equation
(\ref{Keldysh propagator}). We start by summarising the conditions
the thermal propagators have to satisfy in position space:
\begin{subequations}
\label{propagatorrequirement1}
\begin{eqnarray}
\imath\Delta^{++}_{\chi}(x;x') + \imath\Delta^{--}_{\chi}(x;x')
&=& \imath \Delta^{-+}_{\chi}(x;x') +\imath
\Delta^{+-}_{\chi}(x;x') \label{propagatorrequirement1a}\\
\imath\Delta^{++}_{\chi}(x;x') - \imath\Delta^{--}_{\chi}(x;x')
&=& \mathrm{sgn}(t-t')(\imath \Delta^{-+}_{\chi}(x;x') - \imath
\Delta^{+-}_{\chi}(x;x')) \label{propagatoridentities1b} \\
\imath \Delta^{-+}_{\chi}(x;x')&=&\imath \Delta^{+-}_{\chi}(x';x)
\label{propagatoridentities1c} \\
\imath
\Delta^{-+}_{\chi}((t-\imath\beta,\bar{\mathbf{x}});x')&=&\imath
\Delta^{+-}_{\chi}(x';x) \label{propagatoridentities1d}
\\
\left. [ \chi(t,\bar{\mathbf{x}}) ,
\dot{\chi}(t,\bar{\mathbf{x}}')] =
\partial_{t'} \left\{ \imath\Delta^{-+}_{\chi}(x;x') -
\imath  \Delta^{+-}_{\chi}(x;x') \right\} \right|_{t=t'} &=&
\imath \delta^{(3)}(\bar{\mathbf{x}}-\bar{\mathbf{x}}')
\label{propagatoridentities1e} \,.
\end{eqnarray}
\end{subequations}
Here, the first condition (identical to
(\ref{propagatoridentitiesc})) and the second relate the sum and
the difference of the time ordered and anti-time ordered
propagators to the two Wightman functions, respectively. The third
condition is just identical to (\ref{propagatoridentitiesd}).
Condition (\ref{propagatoridentities1d}) is the well-known KMS
condition or Kubo-Martin-Schwinger condition, see
\cite{Kubo:1957mj,Martin:1959jp}. The KMS condition corresponds to
periodic boundary conditions in the imaginary time direction due
to assuming a thermal density matrix operator
$\hat{\rho}_{\mathrm{th}} \propto \exp[-\beta \hat{H}]$. The final
equation arises from requiring standard commutation relations.
Fourier transforming the equations above according to
(\ref{Fouriertransformdef2}) yields:
\begin{subequations}
\label{propagatorrequirement2}
\begin{eqnarray}
\imath\Delta^{++}_{\chi}(k^{\mu}) +
\imath\Delta^{--}_{\chi}(k^{\mu}) &=& \imath
\Delta^{-+}_{\chi}(k^{\mu}) +\imath \Delta^{+-}_{\chi}(k^{\mu})
\label{propagatorrequirement2a}
\\
\imath\Delta^{++}_{\chi}(k^{\mu}) -
\imath\Delta^{--}_{\chi}(k^{\mu}) &=&
\mathrm{P}\left[\frac{-2\imath}{k_{\mu}k^{\mu}+m^{2}_{\chi}}\right]\label{propagatorrequirement2b}
\\
\imath \Delta^{-+}_{\chi}(k^{\mu})&=&\imath \Delta^{+-}_{\chi}(-
k^{\mu})\label{propagatoridentities2c}
\\
\imath \Delta^{-+}_{\chi}(k^{\mu}) &=& e^{\beta k^{0}}  \imath
\Delta^{+-}_{\chi}(k^{\mu})\label{propagatoridentities2d}
\\
\imath \Delta^{-+}_{\chi}(k^{\mu}) -\imath
\Delta^{+-}_{\chi}(k^{\mu}) &=& 2\pi
\mathrm{sgn}(k^{0})\delta(k_{\mu}k^{\mu}+m^{2}_{\chi})
\label{propagatoridentities2e}\,.
\end{eqnarray}
\end{subequations}
To obtain the second relation (\ref{propagatorrequirement2b}), we
recall:
\begin{equation}\label{CauchyPrincipalValue}
\mathrm{sgn}(x) = \mathrm{P} \int_{-\infty}^{\infty} \mathrm{d}k
\frac{1}{\imath \pi k}e^{\imath k x}\,,
\end{equation}
where $\mathrm{P}$ denotes the Cauchy principal value. Relations
(\ref{propagatoridentities2d}) and (\ref{propagatoridentities2e})
trivially yield the two thermal Wightman functions:
\begin{subequations}
\label{Wightmanfunctions1}
\begin{eqnarray}
\imath \Delta^{+-}_{\chi}(k^{\mu}) &=& 2\pi\mathrm{sgn}(k^{0})
\delta(k_{\mu}k^{\mu}+m^{2}_{\chi}) n^{\mathrm{eq}}_{\chi}(k^{0})
\label{Wightmanfunctions1a}
\\
\imath \Delta^{-+}_{\chi}(k^{\mu}) &=& 2\pi\mathrm{sgn}(k^{0})
\delta(k_{\mu}k^{\mu}+m^{2}_{\chi})
(1+n^{\mathrm{eq}}_{\chi}(k^{0})) \label{Wightmanfunctions1b}\,,
\end{eqnarray}
\end{subequations}
where $n^{\mathrm{eq}}_{\chi}(k^{0})$ is the Bose-Einstein
distribution given by (\ref{BoseEinstein}). In order to solve for
the time ordered and anti-time ordered propagators, let us make
the following general ans\"atze:
\begin{subequations}
\label{ansatzthermalprops1}
\begin{eqnarray}
\imath \Delta^{++}_{\chi}(k^{\mu}) &=&
\frac{-\imath}{k_{\mu}k^{\mu}+m^{2}_{\chi} - \imath \epsilon}+
\delta(k_{\mu}k^{\mu}+m^{2}_{\chi})f(k^{0})
\label{ansatzthermalprops1a}
\\
\imath \Delta^{--}_{\chi}(k^{\mu}) &=&
\frac{\imath}{k_{\mu}k^{\mu}+m^{2}_{\chi} + \imath \epsilon}+
\delta(k_{\mu}k^{\mu}+m^{2}_{\chi})g(k^{0})
\label{ansatzthermalprops1b}\,,
\end{eqnarray}
\end{subequations}
The functions $f(k^{0})$ and $g(k^{0})$ do not depend on $k^{i}$
due to the delta function. We have already chosen the time ordered
and anti-time ordered pole prescription. This is particularly
convenient because, as we will appreciate in a moment, this allows
us to easily recover the familiar vacuum solutions when $T
\rightarrow 0$. We will return to this subtlety shortly. Condition
(\ref{propagatorrequirement2b}) immediately implies:
\begin{equation}\label{ansatzthermalprops2}
f(k^{0}) = g(k^{0})\,,
\end{equation}
where we have made use of the Dirac identity:
\begin{equation}\label{DiracIdentity}
\frac{1}{x+\imath \epsilon} = \mathrm{P}\frac{1}{x}-\imath \pi
\delta(x) \,.
\end{equation}
Because of the time ordering, $\imath\Delta^{++}(x;x')=
\imath\Delta^{++}(x';x)$ such that $\imath\Delta^{++}(k^{\mu})=
\imath\Delta^{++}(-k^{\mu})$. This consideration likewise applies
for the anti-time ordered propagator and suggests that the most
economic way of writing $f(k^{0})$ is in terms of $|k^{0}|$. We
observe:
\begin{equation}\label{BoseEinsteinIdentity}
\frac{1}{2}+n^{\mathrm{eq}}_{\chi}(k^{0}) =
\mathrm{sgn}(k^{0})\left(\frac{1}{2} +
n^{\mathrm{eq}}_{\chi}(|k^{0}|) \right) \,.
\end{equation}
Using the relation above and condition
(\ref{propagatorrequirement2a}):
\begin{equation}\label{ansatzthermalprops3}
f(k^{0}) = 2\pi n^{\mathrm{eq}}_{\chi}(|k^{0}|) \,.
\end{equation}
The thermal propagators are thus given by:
\begin{subequations}
\label{ThermalPropagatorApp}
\begin{eqnarray}
\imath\Delta_{\chi}^{++}(k^{\mu}) &=&
\frac{-\imath}{k_{\mu}k^{\mu}+m_{\chi}^{2} -\imath\epsilon} + 2\pi
\delta(k_{\mu}k^{\mu}+m_{\chi}^{2})
n_{\chi}^{\mathrm{eq}}(|k_{0}|) \label{ThermalPropagator++App}
\\
\imath\Delta_{\chi}^{--}(k^{\mu}) &=&
\frac{\imath}{k_{\mu}k^{\mu}+m_{\chi}^{2}+\imath\epsilon} +  2\pi
\delta(k_{\mu}k^{\mu}+m_{\chi}^{2}) n_{\chi}^{\mathrm{eq}}(|k^{0}|)\label{ThermalPropagator--App}\\
\imath\Delta_{\chi}^{+-}(k^{\mu}) &=& 2\pi
\delta(k_{\mu}k^{\mu}+m_{\chi}^{2}) \left[ \theta(-k^{0}) +
n_{\chi}^{\mathrm{eq}}(|k^{0}|)\right]
\label{ThermalPropagator+-App}
\\
\imath\Delta_{\chi}^{-+}(k^{\mu}) &=& 2\pi
\delta(k_{\mu}k^{\mu}+m_{\chi}^{2}) \left[ \theta(k^{0}) +
n_{\chi}^{\mathrm{eq}}(|k^{0}|)\right] \,.
\label{ThermalPropagator-+App}
\end{eqnarray}
\end{subequations}
If we now let $T\rightarrow 0$ to obtain the familiar vacuum
solutions, we find:
\begin{subequations}
\label{VacuumPropagatorApp}
\begin{eqnarray}
\imath\Delta_{\chi}^{++}(k^{\mu}) &=&
\frac{-\imath}{k_{\mu}k^{\mu}+m_{\chi}^{2} -\imath\epsilon}
\label{VacuumPropagator++App}
\\
\imath\Delta_{\chi}^{--}(k^{\mu}) &=&
\frac{\imath}{k_{\mu}k^{\mu}+m_{\chi}^{2}+\imath\epsilon} \label{VacuumPropagator--App}\\
\imath\Delta_{\chi}^{+-}(k^{\mu}) &=& 2\pi
\delta(k_{\mu}k^{\mu}+m_{\chi}^{2}) \theta(-k^{0})
\label{VacuumPropagator+-App}
\\
\imath\Delta_{\chi}^{-+}(k^{\mu}) &=& 2\pi
\delta(k_{\mu}k^{\mu}+m_{\chi}^{2}) \theta(k^{0}) \,.
\label{VacuumPropagator-+App}
\end{eqnarray}
\end{subequations}
Clearly, writing the thermal propagators in the form
(\ref{ThermalPropagatorApp}) above, facilitates obtaining the
vacuum solutions easily. The reason is that
$n_{\chi}^{\mathrm{eq}}(|k^{0}|)\rightarrow 0$ when $T\rightarrow
0$, whereas this statement does not hold for
$n_{\chi}^{\mathrm{eq}}(k^{0})$.

The freedom to choose a different pole prescription such as the
advanced or retarded pole prescription, is just an equivalent way
of writing the thermal propagators, which can easily be verified
by making use of the Dirac identity (\ref{DiracIdentity}). The
vacuum and thermal contributions to
$\imath\Delta_{\chi}^{++}(k^{\mu})$ and
$\imath\Delta_{\chi}^{--}(k^{\mu})$ separate only so neatly when
we use the time ordered and anti-time ordered contours to evaluate
these propagators, respectively.

\section{Alternative Method of Renormalising the Self-masses}
\label{Alternative Method of Renormalising the Self-masses}

In this appendix, we find the correctly renormalised self-masses
by means of an alternative Fourier space calculation. From
equations (\ref{selfMassa}) and (\ref{VacuumPropagator++}) we
immediately deduce:
\begin{eqnarray}\label{SelfMassFourier1}
\imath M_{\phi}^{++}(k^{\mu}) &=& - \frac{\imath h^{2}}{2} \int
\mathrm{d}^{\scriptscriptstyle{D}}(x-x') \left(
\imath \Delta_{\chi}^{++}(x;x')\right)^{2} e^{-\imath k(x-x')} \\
&=& \frac{\imath h^{2}}{2} \int
\frac{\mathrm{d}^{\scriptscriptstyle{D}}k'}{(2\pi)^{\scriptscriptstyle{D}}}
\frac{1}{k'_{\mu}k^{\prime\mu}+m^{2}_{\chi}-\imath \epsilon}
\frac{1}{(k_{\mu}-k'_{\mu})(k^{\mu}-k^{\prime \mu})+m^{2}_{\chi}-\imath \epsilon}\nonumber \\
&=& - \frac{h^{2}}{2} \int_{0}^{1} \mathrm{d}x \int
\frac{\mathrm{d}^{\scriptscriptstyle{D}}l}{(2\pi)^{\scriptscriptstyle{D}}}
\frac{1}{(l_{\mu}l^{\mu}+k_{\mu}k^{\mu}x(1-x)+m^{2}_{\chi}-\imath
\epsilon)^{2}}\nonumber \,,
\end{eqnarray}
where we used Feynman's trick (see e.g. \cite{Peskin:1995ev}),
performed a Wick rotation and we defined
$l_{\mu}=k'_{\mu}-xk_{\mu}$ as the new Euclideanised integration
variable. The integral can now straightforwardly be performed,
which yields when $m_{\chi} \rightarrow 0$:
\begin{eqnarray}\label{SelfMassFourier2}
\imath M_{\phi}^{++}(k^{\mu}) &=& \frac{ h^{2}
(D-2)(k_{\mu}k^{\mu}-\imath
\epsilon)^{\frac{\scriptscriptstyle{D}-4}{2}}
2^{1-2\scriptscriptstyle{D}}\pi^{\frac{3-D}{2}}\Gamma(\frac{D-2}{2})}{\Gamma(\frac{D}{2})
\Gamma(\frac{D-1}{2}) \sin(\frac{\pi D}{2})} \\
&=& \frac{h^{2}\mu^{4-\scriptscriptstyle{D}}}{16 \pi^{2} (D-4)} -
\frac{h^{2}}{32\pi^{2}}\left( 2-\gamma_{\mathrm{E}} -
\log\left(\frac{k_{\mu}k^{\mu}-\imath\epsilon}{4\pi
\mu^{2}}\right)\right) + \mathcal{O}(D-4) \nonumber\,,
\end{eqnarray}
where in the last line we have expanded around $D=4$ as usual and
we have again introduced a scale $\mu$ to make the argument of the
logarithm dimensionless. Observe that the (numerical value of the)
divergent term coincides with equation (\ref{SelfMassPosspace4})
as it should be. Note that for $\imath M_{\phi}^{--}(k^{\mu})$, we
would have to use the other Wick rotation (in order not to cross
the poles) giving us the desired minus sign difference just as in
the position space calculation. Finally, we need to perform the
$k^{0}$ integral in order to derive the self-mass in Fourier
space. The relevant integral is:
\begin{eqnarray}\label{SelfMassFourier3}
\imath M_{\phi,\mathrm{ren}}^{++}(k, t, t') &=&
\frac{h^{2}}{32\pi^{2}} \int_{-\infty}^{\infty}
\frac{\mathrm{d}k^{0}}{2\pi} e^{-\imath k^{0}\Delta t}\left(
\gamma_{\mathrm{E}}-2 +
\log\left(\frac{k_{\mu}k^{\mu}-\imath\epsilon }{4\pi\mu^{2}}\right) \right)\\
&=& - \frac{ h^{2}}{64\pi^{3}} \left[ 2\pi \delta(\Delta t)
\left\{\log\left(4\pi\mu^{2}\right)+ 2-\gamma_{\mathrm{E}}
\right\} - \int_{-\infty}^{\infty}\mathrm{d}k^{0} e^{-\imath
k^{0}\Delta t} \left\{ \log \left|k^{2}-(k^{0})^{2}\right| -\imath
\pi \theta ((k^{0})^{2}-k^{2})\right\} \right] \nonumber \,.
\end{eqnarray}
In order to make the inverse Fourier integral convergent, we
introduce $\epsilon$ regulators where appropriate. As an
intermediate result, we present:
\begin{eqnarray}\label{SelfMassFourier4}
\imath M_{\phi,\mathrm{ren}}^{++}(k, t, t') &=& - \frac{h^{2}}{32
\pi^{2}} \Bigg[ \delta(\Delta t)
\left\{\log\left(4\pi\mu^{2}\right)+ 2-\gamma_{\mathrm{E}}
\right\} + \Big[\gamma_{\mathrm{E}} + \log\left(-\imath \{|\Delta
t|+\imath\epsilon \} \right)\Big] \frac{\cos \left(k \{|\Delta
t|+\imath\epsilon \} \right)}{-\imath \pi (|\Delta
t|+\imath\epsilon
)} \\
&& \qquad + \Big[\gamma_{\mathrm{E}} + \log\left(\imath \{|\Delta
t|-\imath\epsilon \} \right)\Big] \frac{\cos \left(k \{|\Delta
t|-\imath\epsilon \} \right)}{\imath \pi (|\Delta
t|-\imath\epsilon )} + \frac{1}{2} \left(\frac{e^{-\imath k(\Delta
t|-\imath\epsilon)}}{|\Delta t|-\imath\epsilon} - \frac{e^{\imath
k(\Delta t|+\imath\epsilon)}}{|\Delta t|+\imath\epsilon} \right)
\Bigg] \nonumber\,.
\end{eqnarray}
Clearly, the $\epsilon$ regulators in the logarithms and exponents
are redundant and can be sent to zero. Using the Dirac rule in
equation (\ref{DiracIdentity}) once more, we finally arrive at:
\begin{equation}\label{SelfMassFinalFourier}
\imath M_{\phi,\mathrm{ren}}^{++}(k, t, t') = - \frac{h^{2}}{32
\pi^{2}} \left[ \delta(\Delta t)\left\{\gamma_{\mathrm{E}} + 2 +
\log\left( 4\pi \mu^{2} \Delta t^{2}\right)\right\} +
\frac{e^{-\imath k |\Delta t|}}{|\Delta t|-\imath
\epsilon}\right]\,.
\end{equation}
This result is divergent when we let $\Delta t \rightarrow 0$. As
already discussed in the paper in section \ref{Renormalising the
Kadanoff-Baym equations}, the correct way to deal with this is to
extract two time derivatives acting on e.g.
$Z_{\phi}^{++}(k,t,t')$. We then remain with a perfectly finite
result, which can be appreciated from equation
(\ref{SelfMassFourierSpFinalResult}).

To prove this, let us indeed evaluate the two time derivatives in
equation (\ref{SelfMassFourierSpFinalResult}). The result is:
\begin{eqnarray}\label{SelfMassFinalFourier2}
\imath M_{\phi,\mathrm{ren}}^{++}(k, t, t') &=& -\frac{h^{2}}{32
\pi^{2}} \left[ \delta(\Delta t)\left\{\log\left( 4 \mu^{2} \Delta
t^{2}\right)\right\} + \frac{e^{-\imath k |\Delta t|}}{|\Delta
t|-\imath \epsilon}\right] -\frac{h^{2}}{32 \pi^{2}}\delta(\Delta
t)\left[\gamma_{\mathrm{E}}+2+\log (\pi)\right] \\
&=& - \frac{h^{2}}{32 \pi^{2}} \left[ \delta(\Delta
t)\left\{\gamma_{\mathrm{E}} + 2 + \log\left( 4\pi \mu^{2} \Delta
t^{2}\right)\right\} + \frac{e^{-\imath k |\Delta t|}}{|\Delta
t|-\imath \epsilon}\right] \nonumber \,.
\end{eqnarray}
The first line contains two elements. The first is obtained
directly from evaluating the double time derivative in equation
(\ref{SelfMassFourierSpFinalResult}). The second contribution
originates from expanding the term multiplying the delta function
in equation (\ref{SelfMassPosspace4}) around $D=4$, corresponding
to the minimal subtraction renormalisation scheme. Indeed, the
second line of equation (\ref{SelfMassFinalFourier2}) is identical
to (\ref{SelfMassFinalFourier}) as it should be. This shows that the
position space and Fourier space calculations yield identical
results, however the former calculation proves to be superior to
the latter as the two extracted time derivatives appear naturally
in that case.

\section{Retarded Self-mass}
\label{Retarded Self-mass}

The retarded self-mass $\imath
M_{\phi,\mathrm{ren}}^{\mathrm{r}}(x;x')$ can be obtained by means
of an independent calculation by making use of equation
(\ref{SelfMassPosspace4Taylor}):
\begin{eqnarray}\label{RetardedSelfMass1}
\imath M_{\phi,\mathrm{ren}}^{\mathrm{r}}(x;x') \! &=& \! \imath
M_{\phi,\mathrm{ren}}^{++}(x;x') - \imath M_{\phi}^{+-}(x;x') =
\frac{\imath h^{2}}{128 \pi^{4}}
\partial^{2}\left[ \frac{\log(\mu^{2}\Delta x_{++}^{2}(x;x'))}{
\Delta x_{++}^{2}(x;x')} - \frac{\log(\mu^{2}\Delta
x_{+-}^{2}(x;x'))}{ \Delta x_{+-}^{2}(x;x')}
\right]\nonumber\\
\! &=& \! \frac{\imath h^{2}}{1024 \pi^{4}} \partial^{4}\! \left[
\log^{2}(\mu^{2}\! \Delta x_{++}^{2}\!(x;x'))\!-\! 2
\log(\mu^{2}\!\Delta x_{++}^{2}\!(x;x'))\! -\!
\log^{2}(\mu^{2}\!\Delta x_{+-}^{2}\!(x;x'))\!+\!2
\log(\mu^{2}\!\Delta
x_{+-}^{2}\!(x;x'))\! \right]\nonumber\\
\! &=&\! \frac{h^{2}}{256 \pi^{3}} \partial^{4}\left[\theta(\Delta
t^2-r^2)\theta(\Delta t)\left\{1 - \log\left(\mu^{2}(\Delta
t^{2}-r^{2})\right)\right\} \right]  \,,
\end{eqnarray}
where as before $r=\|\vec{x}-\vec{x}'\|$. In Fourier space we find
after some partial integrations:
\begin{eqnarray}\label{RetardedSelfMass2}
\imath M_{\phi,\mathrm{ren}}^{\mathrm{r}}(k, t, t') &=&
\frac{h^{2}}{256 \pi^{3}} (\partial^{2}_{t}+k^{2})^{2} \int
\mathrm{d}^{3}(\vec{x}-\vec{x}')\theta(\Delta
t^2-r^2)\theta(\Delta t) \left[1 - \log\left(\mu^{2}(\Delta
t^{2}-r^{2})\right) \right] e^{-\imath \vec{k}\cdot
(\vec{x}-\vec{x}')}\\
&=& \frac{h^{2}}{64 k \pi^{2}} (\partial^{2}_{t}+k^{2})^{2}
\theta(\Delta t) \Delta t^{2} \Bigg[  \frac{\sin(k \Delta t)-k
\Delta t \cos(k \Delta t)}{(k \Delta
t)^{2}}\left(1-\log(\mu^{2}\Delta
t^{2}) \right) \nonumber \\
&& \qquad\qquad\qquad\qquad\qquad\qquad - \int_{0}^{1} \mathrm{d}x
\, x \sin(k \Delta t x)\log\left(1-x^{2}\right) \Bigg] \nonumber
\,.
\end{eqnarray}
The last integral is performed in e.g. \cite{Prokopec:2002uw}:
\begin{eqnarray}\label{RetardedSelfMass3}
\imath M_{\phi,\mathrm{ren}}^{\mathrm{r}}(k, t, t') &=&
\frac{h^{2}}{64 k^{3} \pi^{2}} (\partial^{2}_{t}+k^{2})^{2}
\theta(\Delta t) \Bigg[  \left( k \Delta t \cos(k \Delta t) -
\sin(k \Delta t)\right) \left(\mathrm{ci}(2k\Delta t) - \gamma_{\mathrm{E}} - \log\left(\frac{k}{2\mu^{2}\Delta t}\right) -1 \right) \nonumber \\
&& \qquad\qquad\qquad\qquad\qquad\qquad +\left( \cos(k\Delta t)+
k\Delta t \sin(k\Delta
t)\right)\left(\frac{\pi}{2}+\mathrm{si}(2k\Delta
t)\right)-2\sin(k\Delta t) \Bigg] \,.
\end{eqnarray}
Since the term in square brackets is proportional to $(\Delta
t)^2$ as $\Delta t \rightarrow 0$, the $\theta(\Delta t)$ commutes
through one of the $(\partial^{2}_{t}+k^{2})$ operators.
Evaluating it further yields:
\begin{equation}\label{RetardedSelfMassFinalResult}
\imath M_{\phi,\mathrm{ren}}^{\mathrm{r}}(k, t, t') =
\frac{h^{2}}{32 k \pi^{2}} (\partial^{2}_{t}+k^{2})\theta(\Delta
t) \Bigg[  \cos(k \Delta t)
\left(\frac{\pi}{2}+\mathrm{si}(2k\Delta t)\right) - \sin(k \Delta
t) \left(\mathrm{ci}(2k\Delta t) - \gamma_{\mathrm{E}} -
\log\left(\frac{k}{2\mu^{2}\Delta t}\right)\! \right)\!\Bigg]\!
 \,.
\end{equation}
If we determine the retarded self-mass directly from equation
(\ref{SelfMassFourierSpFinalResult}), we find perfect agreement
with the result above, representing yet another consistency check
of (\ref{SelfMassFourierSpFinalResult}).

\section{The Statistical Propagator in Fourier Space}
\label{The Statistical Propagator in Fourier Space}

Let us now calculate the statistical propagator in Fourier space.
Starting point is the equation of motion (\ref{EOM4Fourier}) in
Fourier space. We can straightforwardly write the two Wightman
functions as:
\begin{subequations}
\label{FourierWightman}
\begin{eqnarray}
\imath \Delta^{-+}_{\phi}(k^{\mu}) &=& \frac{ - \imath
M^{-+}_{\phi}(k^{\mu}) \imath
\Delta^{\mathrm{a}}_{\phi}(k^{\mu})}{k_{\mu}k^{\mu}+m_{\phi}^{2}+\imath
M^{\mathrm{r}}_{\phi,\mathrm{ren}}(k^{\mu})}
 \label{FourierWightmana} \\
\imath \Delta^{+-}_{\phi}(k^{\mu}) &=& \frac{ - \imath
M^{+-}_{\phi}(k^{\mu}) \imath
\Delta^{\mathrm{a}}_{\phi}(k^{\mu})}{k_{\mu}k^{\mu}+m_{\phi}^{2}+\imath
M^{\mathrm{r}}_{\phi,\mathrm{ren}}(k^{\mu})}
\label{FourierWightmanb} \,,
\end{eqnarray}
\end{subequations}
where we have made use of the definition of the advanced
propagator (\ref{Delta:adv}) in Fourier space:
\begin{equation}
\imath \Delta^{\mathrm{a}}_{\phi}(k^{\mu}) = \frac{-\imath}{
k_{\mu}k^{\mu}+m_{\phi}^{2}+\imath
M^{\mathrm{a}}_{\phi,\mathrm{ren}}(k^{\mu})}
 \label{FourierAdvanced} \,.
\end{equation}
Moreover, we made use of:
\begin{subequations}
\label{FourierSelfMassRelation}
\begin{eqnarray}
\imath M^{\mathrm{r}}_{\phi,\mathrm{ren}}(k^{\mu}) &=& \imath
M^{++}_{\phi,\mathrm{ren}}(k^{\mu}) - \imath
M^{+-}_{\phi}(k^{\mu}) =  \imath
M^{-+}_{\phi}(k^{\mu}) - \imath M^{--}_{\phi,\mathrm{ren}}(k^{\mu})  \label{FourierSelfMassRelationa} \\
\imath M^{\mathrm{a}}_{\phi,\mathrm{ren}}(k^{\mu}) &=& \imath
M^{++}_{\phi,\mathrm{ren}}(k^{\mu}) - \imath
M^{-+}_{\phi}(k^{\mu}) =  \imath M^{+-}_{\phi}(k^{\mu}) - \imath
M^{--}_{\phi,\mathrm{ren}}(k^{\mu})
\label{FourierSelfMassRelationb} \,,
\end{eqnarray}
\end{subequations}
Clearly, we need to evaluate some self-masses in Fourier space.
The simplest method of determining e.g. $\imath
M^{+-}_{\phi}(k^{\mu})$ or $\imath M^{-+}_{\phi}(k^{\mu})$ is to
use the retarded self-mass in Fourier space and $\imath
M^{++}_{\phi,\mathrm{ren}}(k^{\mu})$ which we already derived
before in appendix \ref{Alternative Method of Renormalising the
Self-masses}. Using expressions (\ref{selfmasscausalret}) and
(\ref{RetardedSelfMassFinalResult}) we can derive:
\begin{eqnarray}
\imath M^{\mathrm{r}}_{\phi,\mathrm{ren}}(k^{\mu}) &=&
\int_{-\infty}^{\infty}\mathrm{d}\Delta t e^{\imath k^{0}\Delta t}
\imath M^{\mathrm{r}}_{\phi,\mathrm{ren}}(k,\Delta t) \label{Fourierretarded1} \\
&=& - \frac{h^{2}}{64k\pi^{2}} (-k_{0}^{2}+k^{2})
\int_{0}^{\infty}\mathrm{d}\Delta t \Bigg[
e^{\imath(k+k^{0})\Delta t}
\left\{-\imath\left(\mathrm{ci}(2k\Delta t)-\log(2k\Delta
t)-\gamma_{\mathrm{E}}\right) -\frac{\pi}{2}-\mathrm{si}(2k \Delta
t) \right\} \nonumber \\
&& \qquad\qquad\qquad\qquad\qquad\qquad\,\,\, +
e^{\imath(k^{0}-k)\Delta t}
\left\{\imath\left(\mathrm{ci}(2k\Delta t)-\log(2k\Delta
t)-\gamma_{\mathrm{E}}\right) -\frac{\pi}{2}-\mathrm{si}(2k \Delta
t) \right\} \nonumber\\
&&\qquad\qquad\qquad\qquad \qquad\qquad\qquad\qquad\qquad\quad -2
\imath \log\left(2\mu\Delta t
\right)\left(e^{\imath(k+k^{0}+\imath\epsilon)\Delta t} -
e^{\imath(k^{0}-k+\imath\epsilon)\Delta t}\right) \Bigg] \nonumber
\,,
\end{eqnarray}
where $k_{0}^{2} = (k^{0})^{2}$ and where we have used two partial
integrations and disposed ourselves of the boundary terms by
introducing an $\epsilon$ regulator where necessary. We can now
use:
\begin{equation}
\int_{0}^{\infty}\mathrm{d}x \log(\beta x) e^{\imath \alpha x} =
-\frac{\imath}{\alpha}\left[\log\left(\frac{-\imath \alpha +
\epsilon}{\beta}\right)+\gamma_{\mathrm{E}}\right]
 \label{Fourierretarded2} \,,
\end{equation}
and moreover we write:
\begin{equation}
e^{\imath(k^{0} \pm k)\Delta t} = \frac{-\imath}{k^{0} \pm
k}\partial_{t} e^{\imath(k^{0} \pm k)\Delta t}
\label{Fourierretarded3} \,,
\end{equation}
to prepare for another partial integration. Equation
(\ref{Fourierretarded1}) evaluates to:
\begin{eqnarray}
\imath M^{\mathrm{r}}_{\phi,\mathrm{ren}}(k^{\mu}) &=& -
\frac{h^{2}}{64k\pi^{2}} \Bigg[ 2(k^{0}-k)\left(
\log\left(\frac{-\imath(k+k^{0})+\epsilon}{2\mu}\right) +
\gamma_{\mathrm{E}}\right) - 2(k^{0}+k)\left(
\log\left(\frac{-\imath(k^{0}-k)+\epsilon}{2\mu}\right) +
\gamma_{\mathrm{E}}\right) \nonumber \\
&& \qquad\qquad + \int_{0}^{\infty}\mathrm{d}\Delta t \frac{2
k^{0}}{\Delta t}\left(e^{\imath(k^{0}+k)\Delta t} -
e^{\imath(k^{0}-k)\Delta t}\right)\Bigg] \,.
\label{Fourierretarded4}
\end{eqnarray}
For $\alpha , \beta \in \mathbb{R}$, we can use:
\begin{equation}
\lim_{z \downarrow 0} \int_{z}^{\infty}\mathrm{d}\Delta t\left[
\frac{\cos(\alpha\Delta t)-1}{\Delta t} - \frac{\cos(\beta\Delta
t)-1}{\Delta t}\right] = \log\left(\frac{|\beta|}{|\alpha|}\right)
\label{Fourierretarded5} \,,
\end{equation}
to evaluate the remaining integrals. The result is:
\begin{equation}
\imath M^{\mathrm{r}}_{\phi,\mathrm{ren}}(k^{\mu}) =
\frac{h^{2}}{32
\pi^{2}}\left[\log\left(\frac{-k_{0}^{2}+k^{2}-\imath
\mathrm{sgn}(k^{0})\epsilon}{4\mu^{2}}\right)+2\gamma_{\mathrm{E}}\right]
\label{Fourierretarded6}\,.
\end{equation}
From $\imath M^{\mathrm{r}}_{\phi,\mathrm{ren}}(k^{\mu}) = \imath
M^{++}_{\phi,\mathrm{ren}}(k^{\mu}) - \imath
M^{+-}_{\phi}(k^{\mu})$ we can immediately find the Wightman
self-masses. We take $\imath M^{++}_{\phi,\mathrm{ren}}(k^{\mu})$
from equation (\ref{SelfMassFourier2}), but we have to make sure
we use the same subtraction scheme as in our position space
calculation in equation (\ref{SelfMassPosspace4Taylor}). We
therefore modify equation (\ref{SelfMassFourier2}) slightly to:
\begin{equation}
\imath M^{++}_{\phi,\mathrm{ren}}(k^{\mu}) =
\frac{h^{2}}{32\pi^{2}}\left[
\log\left(\frac{-k_{0}^{2}+k^{2}-\imath\epsilon}{4\mu^{2}}\right)+2\gamma_{\mathrm{E}}\right]
\label{Fourier++} \,.
\end{equation}
We thus find:
\begin{subequations}
\label{FourierWightman2}
\begin{eqnarray}
\imath M^{+-}_{\phi}(k^{\mu}) &=& - \frac{\imath h^{2}}{16\pi}\theta(-k^{0}-k) \label{FourierWightman2a} \\
\imath M^{-+}_{\phi}(k^{\mu}) &=& - \frac{\imath
h^{2}}{16\pi}\theta(k^{0}-k) \label{FourierWightman2b} .
\end{eqnarray}
\end{subequations}
As a check, we consider the following relation that has to be
satisfied:
\begin{equation}
\imath M^{++}_{\phi,\mathrm{ren}}(k^{\mu})+ \imath
M^{--}_{\phi,\mathrm{ren}}(k^{\mu}) = \imath
M^{+-}_{\phi}(k^{\mu})+ \imath M^{-+}_{\phi}(k^{\mu})
 \label{SelfMassrelation} \,.
\end{equation}
Of course, $\imath M^{++}_{\phi,\mathrm{ren}}(k^{\mu})$ is given
in equation (\ref{Fourier++}) which also allows us to derive the
anti-time ordered self-mass:
\begin{equation}
\imath M^{--}_{\phi,\mathrm{ren}}(k^{\mu}) = -
\frac{h^{2}}{32\pi^{2}}\left[
\log\left(\frac{-k_{0}^{2}+k^{2}+\imath\epsilon}{4\mu^{2}}\right)+2\gamma_{\mathrm{E}}\right]
\label{Fourier--} \,.
\end{equation}
where $\imath M_{\phi,\mathrm{ren}}^{--}(k^{\mu})$ contains an
additional minus sign because of the Wick rotation (see appendix
\ref{Alternative Method of Renormalising the Self-masses} for
details). We thus find:
\begin{equation}
\imath M^{+-}_{\phi}(k^{\mu})+ \imath M^{-+}_{\phi}(k^{\mu}) = -
\frac{\imath h^{2}}{16\pi}\theta(k_{0}^{2}-k^{2})
 \label{SelfMassrelationcheck} \,.
\end{equation}
This is in perfect agreement with equation
(\ref{FourierWightman2}).

Using equations (\ref{FourierWightman}), (\ref{Fourier++}) and
(\ref{FourierWightman2}) we can derive our solutions for the two
Wightman functions in Fourier space:
\begin{subequations}
\label{FourierWightman3}
\begin{eqnarray}
\imath \Delta^{+-}_{\phi}(k^{\mu}) &=& \imath \theta(-k^{0}-k)
\Bigg[\frac{1}{k_{\mu}k^{\mu}+m_{\phi}^{2}+\frac{h^{2}}{32
\pi^{2}}\left(\log\left(\frac{|k_{\mu}k^{\mu}|}{4\mu^{2}}\right)+2\gamma_{\mathrm{E}}
\right)- \frac{\imath
h^{2}}{32\pi}\mathrm{sgn}(k^{0})\theta(k_{0}^{2}-k^{2})} \nonumber \\
&& \qquad\qquad\qquad -
\frac{1}{k_{\mu}k^{\mu}+m_{\phi}^{2}+\frac{h^{2}}{32
\pi^{2}}\left(\log\left(\frac{|k_{\mu}k^{\mu}|}{4\mu^{2}}\right)+2\gamma_{\mathrm{E}}
\right)+ \frac{\imath
h^{2}}{32\pi}\mathrm{sgn}(k^{0})\theta(k_{0}^{2}-k^{2})} \Bigg] \label{FourierWightman3a} \\
\imath \Delta^{-+}_{\phi}(k^{\mu}) &=& - \imath \theta(k^{0}-k)
\Bigg[\frac{1}{k_{\mu}k^{\mu}+m_{\phi}^{2}+\frac{h^{2}}{32
\pi^{2}}\left(\log\left(\frac{|k_{\mu}k^{\mu}|}{4\mu^{2}}\right)+2\gamma_{\mathrm{E}}
\right)- \frac{\imath
h^{2}}{32\pi}\mathrm{sgn}(k^{0})\theta(k_{0}^{2}-k^{2})} \nonumber \\
&& \qquad\qquad\qquad -
\frac{1}{k_{\mu}k^{\mu}+m_{\phi}^{2}+\frac{h^{2}}{32
\pi^{2}}\left(\log\left(\frac{|k_{\mu}k^{\mu}|}{4\mu^{2}}\right)+2\gamma_{\mathrm{E}}
\right)+ \frac{\imath
h^{2}}{32\pi}\mathrm{sgn}(k^{0})\theta(k_{0}^{2}-k^{2})} \Bigg]
\label{FourierWightman3b} .
\end{eqnarray}
\end{subequations}
The limit $h \rightarrow 0$ in the equations above nicely agrees
with the vacuum Wightman propagators in equations
(\ref{VacuumPropagator+-}) and (\ref{VacuumPropagator-+}). Hence
the statistical propagator in Fourier space reads:
\begin{eqnarray}
F_{\phi}(k^{\mu}) &=& - \frac{\imath}{2} \mathrm{sgn}(k^{0})
\theta(k_{0}^{2}-k^{2})
\Bigg[\frac{1}{k_{\mu}k^{\mu}+m_{\phi}^{2}+\frac{h^{2}}{32
\pi^{2}}\left(\log\left(\frac{|k_{\mu}k^{\mu}|}{4\mu^{2}}\right)+2\gamma_{\mathrm{E}}
\right)- \frac{\imath
h^{2}}{32\pi}\mathrm{sgn}(k^{0})\theta(k_{0}^{2}-k^{2})} \nonumber \\
&& \qquad\qquad\qquad\qquad\qquad -
\frac{1}{k_{\mu}k^{\mu}+m_{\phi}^{2}+\frac{h^{2}}{32
\pi^{2}}\left(\log\left(\frac{|k_{\mu}k^{\mu}|}{4\mu^{2}}\right)+2\gamma_{\mathrm{E}}
\right)+ \frac{\imath
h^{2}}{32\pi}\mathrm{sgn}(k^{0})\theta(k_{0}^{2}-k^{2})} \Bigg]
\label{FourierStatistical} \,.
\end{eqnarray}

\end{document}